\documentclass[iop, apj]{emulateapj}
\usepackage{graphics}      
\usepackage{amsmath}
\usepackage{amssymb}
\usepackage{morefloats}
\usepackage{natbib}
\usepackage{CJK}
\usepackage{soul}

\makeatletter

\newcommand{\Rmnum}[1]{\expandafter\@slowromancap\romannumeral #1@}
\makeatother

\def\Msol{M_\odot}
\def\Rsol{R_\odot}

\shorttitle{Global MHD Simulation of CV Disks}
\shortauthors{Ju et al.}

\begin{document}


\title{Global MHD Simulations of Accretion Disks in Cataclysmic Variables (CVs): {\Rmnum 2} The Relative Importance of MRI and Spiral Shocks}

\author{Wenhua Ju\altaffilmark{1}, James M. Stone\altaffilmark{1}, Zhaohuan Zhu\altaffilmark{1}}

\altaffiltext{1}{Dept.~of Astrophysical Sciences, Princeton University, Princeton, NJ 08544, USA}

\begin{abstract}
We perform global three-dimensional MHD simulations of unstratified accretion disks in cataclysmic variables (CVs). By including mass inflow via an accretion stream, we are able to evolve the disk to a steady state. We investigate the relative importance of spiral shocks and the magnetorotational instability (MRI) in driving angular momentum transport and how each depend on the geometry and strength of seed magnetic field and the Mach number of the disk (where Mach number is ratio of the azimuthal velocity and the sound speed of gas). We use a locally isothermal equation of state and adopt temperature profiles that are consistent with CV disk observations. Our results indicate that the relative importance of spiral shocks and MRI in driving angular momentum transport is controlled by the gas Mach number and the seed magnetic field strength. MRI and spiral shocks provide comparable efficiency of angular momentum transport when the disk Mach number is around 10 and the seed magnetic field has plasma $\beta=400$ (where $\beta$ is ratio of gas pressure and magnetic pressure). The MRI dominates whenever the seed field strength, or the disk Mach number, is increased. Among all of our simulations, the effective viscosity parameter $\alpha_{eff} \sim 0.016-0.1$ after MRI saturates and the disk reaches steady state. Larger values of $\alpha_{eff}$ are favored when the seed magnetic field has vertical components or the flow has stronger magnetization ($1/\beta$). Our models all indicate that the role of MRI in driving angular momentum transport thus mass accretion in CV disks is indispensable, especially in cool disks with weak spiral shocks. 
\end{abstract}


\section{Introduction}

Angular momentum transport in the accretion disks of cataclysmic variables (CVs) is key for driving the observed episodic outbursts in dwarf novae (DNe). Great success has been achieved in fitting the episodic outbursts with light curves computed using the \citet{1973Shakura} formalism in one-dimensional disk instability models (DIM, see \citealt{2001Lasota} for review). DIM was proposed by \citealt{1974Osaki}, and further developed by identifying a thermal instability arising from the outer part of the disk which changes the disk from radiative structure to convective structure \citep{1981Meyer, 1982smak, 1983Mineshige-Osaki, 1984Smak}. An anomalous kinematic viscosity $\nu= \alpha c_s H$ is assumed ($c_s$ and $H$ are sound speed and thermal scale height) to drive mass accretion and $\alpha$ is required to be $\sim 0.1-0.3$ during outburst state and $\sim 0.01$ during quiescence state \citep{ 1993Cannizzo, 1999Smak, 2012Cannizzo,2012Kotko}. However, perhaps the greatest limitation of these models is that the physical mechanisms accounting for angular momentum transport in CV disks during either quiescence or outburst states are not well understood yet \citep{2007King,2013King}. Possible mechanisms include spiral shocks which are excited by the non-axisymmetric gravitational potential and the magnetorotational instability (MRI). However, to date, no global MHD models of inviscid disks in close binaries where angular momentum transport is self-consistently calculated from first principles have been undertaken.

 One possible physical origin of episodic accretion in DNe disks may be that MRI turbulence is severely suppressed when the gas is predominantly neutral at low temperature during the quiescent state \citep{1998Gammie-Menou}. This implies that MRI may be the switch of DNe outbursts: during hot states when MRI and spiral shocks both contribute in angular momentum transport, efficient accretion thus outbursts are turned on, while during quiescence when the disk cold and predominantly neutral, only spiral shocks contribute in angular momentum transport and accretion rate is low. In the first paper of this series \citep{2016Ju}, we conducted a thorough study of the spiral shocks in CV disks using a series of 2D hydrodynamical simulations where angular momentum transport is driven by the deposition of negative angular momentum carried by the spiral density waves via shock dissipation. The effective viscosity parameter $\alpha_{eff}$ driven by spiral shocks is found to be $0.02-0.05$ which is consistent with the observed values in quiescence state of DNe (Note the Mach numbers used in our simulations are significantly lower than that expected from the quiescence state of DNe, but are similar to that during the DNe outburst state). However, during the hot outburst state where the ionized gas is well coupled to the magnetic field and MRI starts to operate in the disk, it is unclear how the dynamics of spiral shocks would change, or how spiral shocks and MRI interact, and what is the relative importance of spiral shocks and MRI in driving angular momentum transport. Therefore, the focus of this paper is to conduct global MHD models of inviscid CV disks where angular momentum transport is driven by spiral shocks and MRI self-consistently to investigate these issues.

Over the past two decades, it has been widely recognized that the magnetorotational instability (MRI) drives angular momentum transport in most disks \citep{1991Balbus-Hawley,1998Balbus-Hawley,2002Balbus-Hawley}. Subsequent numerical MHD simulations of the nonlinear regime of the MRI show that it saturates as turbulence, with a significant Maxwell stress that drives angular momentum transport (\citealt{1995Hawley}; \citealt{1996Stone}; \citealt{2012Sorathia}; \citealt{2013Hawley} see \citealt{2003Balbus} for a review). The Maxwell stress is greater than the Reynolds stress by a factor of $4-6$ in these simulations. The saturated stress-to-pressure ratio $\alpha$ varies widely ($\sim 0.01- 0.1$) depending on the net magnetic field strength \citep{1995Hawley,1999Hawley,2004Sano,2008Blackman,2009Guan-Gammie,2009Hawley,2010Davis,2011Guan, 2011Hawley, 2012Simon, 2016Shi}. 


A current challenge in accretion disk theory is to understand how the MRI acts on global scales \citep{2012Sorathia,2013Hawley} and determines the structure and evolution of disks on viscous times. Although local shearing box simulations have achieved great success in studying the local characteristics of well-resolved MRI , it is not clear whether they are representative of the global dynamics in systems such as CV disks. To date, most effort in this direction have focused on disks in both black holes \citep[e.g.][]{2009Hawley, 2010Pessah} and protostellar systems \citep[e.g.][]{2011Flock,2013Flock}. These studies begin with a rotationally supported torus of finite mass which then accretes as the MRI grows and saturates. However, the resulting accretion flow is ultimately transitory due to lack of mass supply. Moreover, it is becoming increasingly clear that some properties of the resulting flow depend on the initial assumed field geometry in the torus \citep{2010Penna,2012McKinney}. CV disks are ideal laboratories to understand the global physics of the MRI. The well-understood mass supply for the disk, e.g. gas streaming through the inner Lagrangian (L1) point at a rate of $\sim 10^{-9} - 10^{-10} M_\odot /yr$ \citep{2001Hellier} makes it possible for the disk to reach steady state. Moreover, the length and time scales in CV disks are much smaller than in the other accretion disk systems. For example, the radial range from the surface of the white dwarf to the inner Lagrangian point is only two orders of magnitude, to be compared with four orders of magnitude in protoplanetary disks and even larger in low-mass X-ray binary disks. In the context of CV disks, \citet{2012Latter} conducted local unstratified shearing box MHD simulation with radiative cooling and found bi-stability of thermal states. \citet{2014Potter} proposed that a limit cycle of disk stability could be achieved by assuming $\alpha$ is correlated with magnetic Prandtl number instead of a constant over the disk. But these investigations need to be extended to self-consistently evolving global models.

Controversy exists on whether MRI is sufficient to drive mass accretion during DNe outbursts. \citet{2007King} have suggested that even in the DNe outburst phase, MRI alone cannot explain observed accretion rates since some local shearing box MHD simulations with zero net vertical magnetic field ($B_z$) give $\alpha \le 0.02$ \citep[although see][]{2016Shi}. \citet{2012Kotko} reiterated this discrepancy using several methods to confirm the validity of $\alpha \sim 0.1 - 0.2$ from DNe outburst observations. \citet{2014Hirose} conducted local, vertically stratified, radiation MHD shearing box simulations with zero net vertical magnetic field, and found an enhanced stress to pressure ratio $\alpha$ is produced by strong thermal convection triggered due to large opacity near the temperature of $>10^4$ K. At higher temperature, thermal convection disappears and $\alpha$ decreases to quiescence values. \citet{2016Coleman} measured the values of MRI-driven $\alpha$ from these local simulations and applied them to one-dimensional disk instability models which successfully re-produced observed outburst and quiescence durations, as well as outburst amplitudes. However, these local models cannot model the spiral shocks which are important ingredients in driving the evolution of CV disks. To date, there are still no global MHD models where the angular momentum transport and cyclic outbursts are driven by spiral shocks and MRI turbulence self-consistently.

Motivated by these issues, we perform a series of global MHD simulations of unstratified CV disks with inviscid gas to explore angular momentum transport driven by spiral shocks and MRI. A constant supply of mass and seed magnetic field is provided through the L1 point. In \citet{2016Ju} we presented one such model with adiabatic equation of state and no cooling. In that case the disk was heated up to unrealistically high temperature, thus the spiral shocks totally dominate over MRI. Therefore, in this paper, we use locally isothermal equation of states to mimic steady-state thermodynamics, where the temperature profile is adopted from measurements through eclipse mapping of CV disks and the Mach number is adopted as high as we can computationally afford. We investigate the relative importance of spiral shocks and MRI in driving angular momentum transport and see how that changes with the following disk properties: seed magnetic field geometry, seed magnetic field strength, and disk Mach number. 

This paper is arranged as follows. We introduce our numerical methods and diagnostics in \S \ref{sec:method_mri}. We present the results of our fiducial model in \S \ref{sec:fiducial_model}. Then we compare the effects of seed field strength in \S \ref{sec:compare_beta}, the effects of seed field geometry in \S \ref{sec:compare_geometry}, and the effects of disk Mach numbers in \S \ref{sec:compare_machnumber}. Finally, further discussion and major conclusions are presented in \S \ref{sec:conclusion_mri}.

\section{Method and Diagnostics}
\label{sec:method_mri}

\subsection{Equations Solved}
We solve equations of ideal compressible MHD in cylindrical coordinates using {\it Athena++}, a recent extension of the grid-based Godunov code package {\it Athena} \citep{2008Stone}. The evolution of inviscid gas with locally isothermal equation of states follows
\begin{subequations}
    \begin{align}
    \frac{\partial \rho}{\partial t} + \nabla \cdot (\rho {\bf v}) &= 0,  
    \label{eq:massconservation-ch-MRI} \\
    \frac{\partial (\rho {\bf v})}{\partial t} + \nabla \cdot (\rho {\bf vv} - {\bf BB} + P^*{\bf I}) &= -\rho \nabla \Phi_{tot} + {\bf F}_{Cori}, \label{eq:momentumcons-ch-MRI} \\
    P = \rho c_s(R)^2, & \\
    \frac{\partial B}{\partial t} - \nabla \times ({\bf v \times B}) &= 0,
    \end{align}
\end{subequations}
where $\rho$ is the gas density; $\bf v$ is the velocity vector; $\bf B$ is the magnetic field vector; $P^* = P + |{\bf B}|^2/2$ is the total pressure consisting of magnetic and gas pressure; $c_s(R)$ is the sound speed at radius $R$ which we will elaborate in \S \ref{subsec:local-iso-eos}.

We solve the equations in a frame of reference which is centered on the WD and corotates with the the donor star such that the donor star stays static in this frame. The position and velocity vectors in this frame are ${\bf r} = (R, \phi, z)$ and ${\bf v} = (v_R, v_\phi, v_z)$ respectively. Therefore, a Coriolis force and an indirect gravitational force must be included to account for the non-inertial frame \citep{1987Binney-Tremaine}. The total gravitational potential in this frame is 
\begin{equation}
\Phi_{tot} = - \frac{G M_1}{|{\bf r}|_{z=0}} - \frac{G M_2}{|{\bf r} - {\bf R_2}|_{z=0}} + G M_2 \frac{ ({\bf R_2} \cdot {\bf r})_{z=0} }{R_2^3}, 
\label{eq:phitot-ch-MRI}
\end{equation}
which consists of the potentials of the WD and donor star and the indirect term due to movement of the origin. $M_1$ and $M_2$ are the masses of the WD and the donor star respectively, and $R_2$ is the position of the donor star. We define the mass ratio of the binary to be $q = M_2 / M_1$. Throughout this work, we adopt $q=0.3$ which is a typical value for dwarf novae \citep{2001Hellier}. Since we are studying unstratified disks as in \citet{2001Hawley} and \citet{2012Sorathia}, the gravitational potential does not include the vertical component of gravity. 

The Coriolis force writes
\begin{equation}
    {\bf F}_{Cori} = -2 \rho \Omega_0 {\bf z} \times {\bf v},
\end{equation}
where $\Omega_0 = \sqrt{G(M_1+M_2) / R_2^3}$ is the orbital frequency of the binary stars which is also the rotating frequency of the frame we adopt.

\subsection{Locally Isothermal Equation of States}
\label{subsec:local-iso-eos}
We adopt locally isothermal equation of states with a power-law radial profile for the isothermal sound speed $c_s(R)$
\begin{equation}
\label{eq:cs}
    c_s(R) = c_s(R_{min}) * (R/R_{min})^{-3/8},
\end{equation}
where $R_{min}$ is the inner boundary of the disk. 

Such a power-law radial profile for $c_s$ has both theoretical and observational origins. On the theoretical side, the effective temperature in a standard thin $\alpha$ disk \citep{1973Shakura} is
\begin{equation}
T_{eff}^4 = \frac{3 G M \dot{M}}{8 \pi \sigma R^3} (1 - \sqrt{R_{min} / R})
\end{equation}
where $G$ and $\sigma$ are Newton's and Stefan's constants, $M$ and $R_{min}$ are the mass and size of the central object, $\dot{M}$ is the mass accretion rate, $R$ is the radius of interest. When $\dot{M}$ is constant, this yields to a scaling of $T \propto R^{-3/4}$ and thus $c_s \propto R^{-3/8}$. Observationally, the radial profiles of surface brightness temperature of CV disks are able to be reconstructed using the eclipse mapping method \citep{1985Horne-Cook, 1986Wood, 1989Wood, 1992Rutten,1992Rutten-a, 1991Baptista, 2001Baptista, 2002Vrielmann, 2003Vrielmann, 2005Borges, 2006Shafter, 2007Baptista, 2015Baptista}. Those work has found the radial temperature profiles of the disks can be fit using the power law $T \propto R^{-3/4}$ at phases close the peaks of outbursts. Although the temperature profile is much flatter than $T \propto R^{-3/4}$ at other phases of the limit cycle when the disk is not in steady state, we do not consider those cases in this paper. Therefore, we adopt the $T \propto R^{-3/4}$ thus $c_s \propto R^{-3/8}$ power law which stays constant over time in our simulations. Assuming a Keplerian disk, the Mach number of the disk $\mathcal{M} = v_K / c_s$ scales with $R^{-1/8}$.

\begin{figure}
\centering
\includegraphics[width=0.48\textwidth]{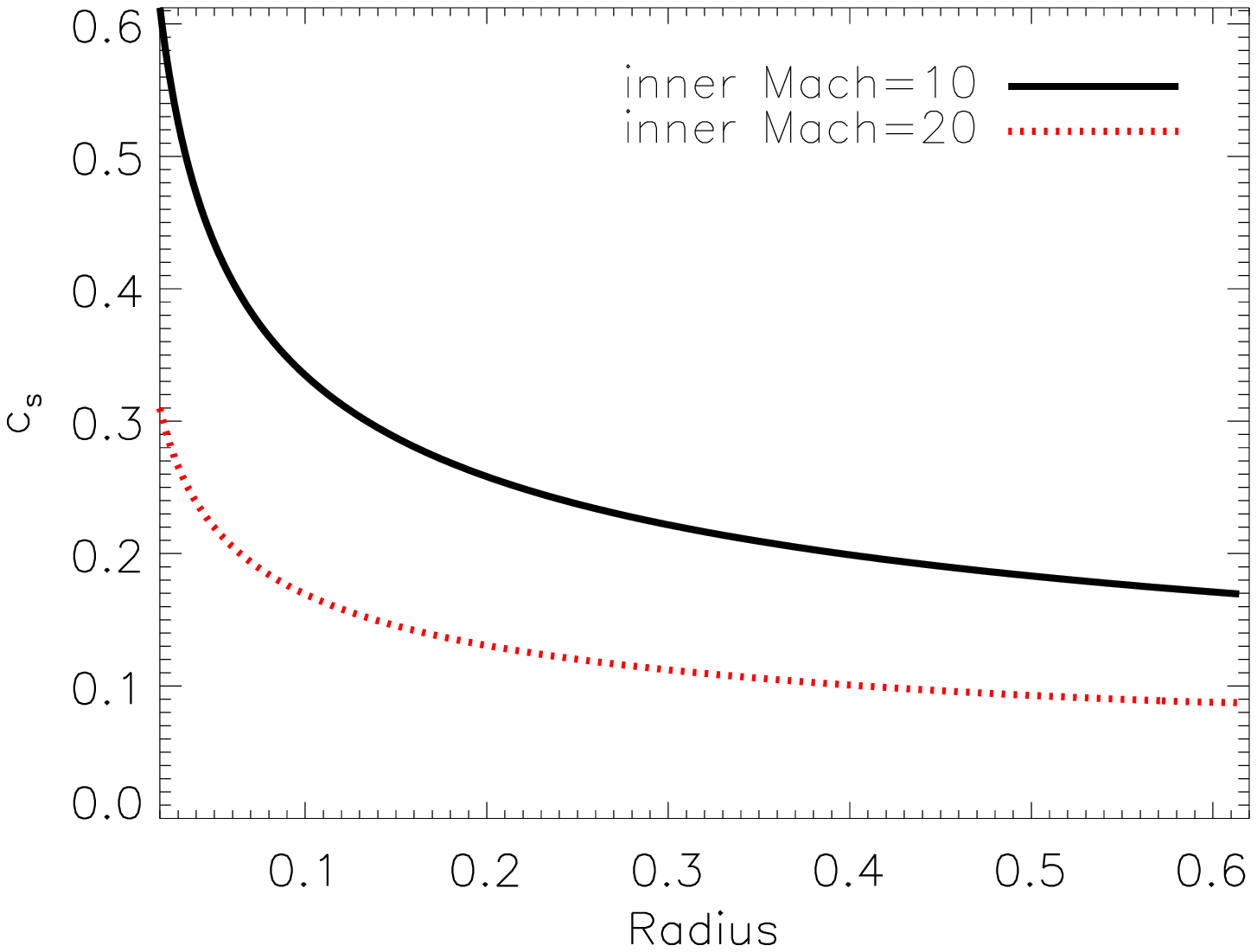}
\includegraphics[width=0.48\textwidth]{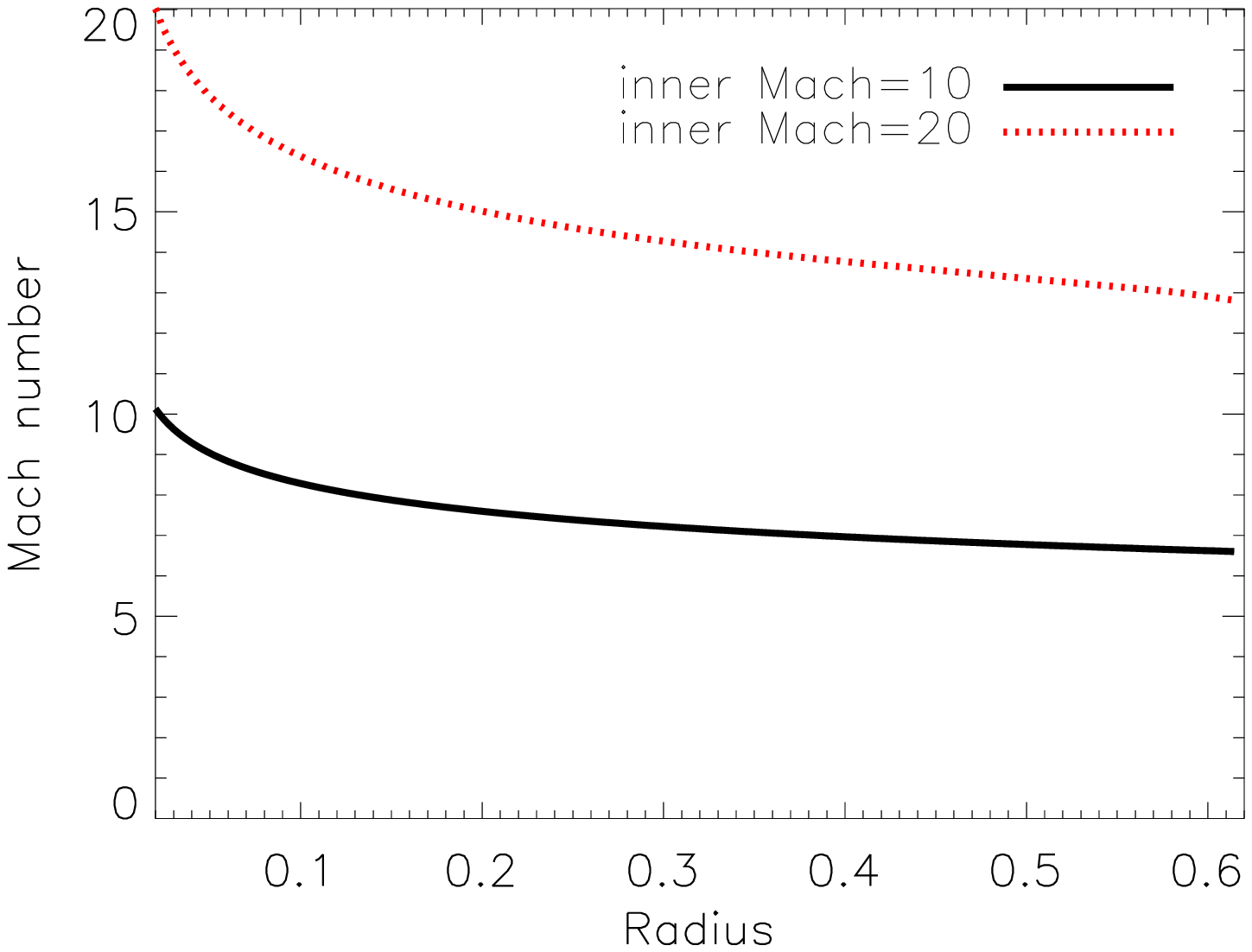}
\caption{Radial profiles of the sound speed (left panel) and the Mach numbers (right panel) used in the locally isothermal equation of states (Eq.\ref{eq:cs}). The slope of the profiles is adopted from the standard thin disk theory $T \propto R^{-3/4}$ which is also consistent with CV disk observations. Two scaling factors are chosen: the hotter disk has $\mathcal{M} (R_{min}) = 10$ (solid black line), and the cooler disk has $\mathcal{M} (R_{min}) = 20$ (dotted red line).}
\label{fig:soundspeed}
\end{figure}

The scaling factor $c_s(R_{min})$ is set such that the Mach number $\mathcal{M}$ at $R_{min}$ is 10 or 20 in this work. The radial profiles of sound speed and $\mathcal{M}$ are shown in Figure \ref{fig:soundspeed}. Hereafter we denote the sound speed profile with $\mathcal{M} (R_{min}) = 10$ with ``Mach10", and the profile with $\mathcal{M}(R_{min}) = 20$ with ``Mach20". These $\mathcal{M}$ values may be lower than estimated from CV observations: according to brightness map reconstructed by eclipse mapping, the surface temperature of CV disks spans from 8000K (outer radius) to 40000K (inner radius) during outbursts which corresponds to $\mathcal{M} \sim 50 - 200$ \citep{1985Horne-Cook,1986Wood, 1992Rutten, 1998Baptista, 2015Baptista}. Such extreme Mach numbers are very challenging for numerical simulations. However, it is worth noting that the mid-plane temperature that governs gas dynamics in our simulations may be a few times higher than the surface brightness temperature measured from observations \citep{2014Hirose,2016Coleman}.
The temperature values from eclipse maps should be taken with caution before being applied to numerical models directly.

\subsection{Dimensionless Units}
Equations are solved in dimensionless form in this work. The dimensionless units are identical to \citet{2016Ju} (see their \S 2.2). We define the unit of mass such that $G M_0 = GM_1 + GM_2 =1$, and the unit of length to be the separation of the binary $a_0=1$. Therefore, the orbital frequency of the binary $\Omega_0 = \sqrt{GM_0 / a_0^3} = 1$.  The unit of time is the inverse of the angular frequency of the binary orbit $t_0 = \Omega_0^{-1}=1$ which makes one orbital period of the binary $P/t_0 = 2\pi$. The donor star orbits on a unit circle at frequency $\Omega_0$, so ${\bf R_2} = (cos(t), sin(t), 0)$.

To set up initial conditions that are in general agreement with the properties of typical CV systems, we list the properties of a well-studied CV system: SS Cygni. This system has a 1.2 $\Msol$ WD and a 0.7 $\Msol$ companion star in orbit with a period of 6.6 hr, which implies a binary separation of 2.24 $\Rsol$. The mean interval time between outbursts of SS Cygni is 40 days, with typical time of decay from outbursts being 2.4 days. The mass inflow rate from the companion star is of order $10^{-9} \Msol /$ yr \citep{1993Cannizzo}. According to thermal limit models \citep[e.g.][]{1993Cannizzo}, at a radius of $2 \times 10^{10}$ cm from the WD (around the mid-radius area of the disk), the typical surface density of the WD accretion disk is $\approx 200$ g cm$^{-2}$, and the midplane temperature is $\approx 3000$ K which implies the local Mach number of $\mathcal{M} \sim 280$ during the quiescence state (where $\mathcal{M}$ is ratio of kinematic velocity and sound speed of the gas). While there is a wide variation of these parameters among CV systems, we take the SS Cygni system as a reference system. 

If our internal units were scaled to the SS Cygni system, then our unit of length is $a_0 = 2.24 R_\odot$, our unit of mass is $M_0 \sim 2.44 \times 10^{-9} \Msol$ and our unit of time is 1.05 hr which yields a binary orbit of 6.6 hrs. If we express the properties of SS Cygni in our internal units, then the binary stars are separated by $1 R_0$, the mean interval time between outbursts is 914 $t_0$, and outbursts decay on the timescale of 54 $t_0$; the mass inflow rate from the companion star is $4.9 \times 10^{-5} M_0 / t_0$; the reference radius $2 \times 10^{10}$ cm is about 0.13 $R_0$, and at this radius, the surface density of the disk is around 1 $M_0/R_0^2$.

\subsection{Initial and Boundary Conditions}
\label{sec:IC_BC}

\begin{deluxetable*}{cccccc}
\tablecolumns{6} 
\tabletypesize{\scriptsize}
\tablewidth{0pt}
\tablecaption{Model Parameters \label{tab:parameter}}
\tablehead{ 
\colhead{Model Name} & \colhead{Seed field geometry} & \colhead{Seed field $\beta$}  & \colhead{Inner $\mathcal{M}$} & \colhead{Resolution} & \colhead{Comments}\\
\colhead{} & \colhead{} & \colhead{}  & \colhead{} & \colhead{($R \times \phi \times z$)} & \colhead{}
}
\startdata
Bz-Mach10-$\beta$100-res1 & Bz & 100  & 10 &  $384 \times 704 \times 32$ & fiducial model \\
\hline 
Bz-Mach10-$\beta$400-res1 & Bz & 400  & 10 &  $384 \times 704 \times 32$ & compare field strength \\
Loop-Mach10-$\beta$400-res1 & loop & 400  & 10 &  $384 \times 704 \times 32$ & compare field geometry\\
Hydro-Mach10-res1 & no B & 400  & 10 &  $384 \times 704 \times 1$ & \\
Bz-Mach20-$\beta$100-res2 & Bz & 100  & 20 &  $768 \times 704 \times 64$ & compare Mach number \\
Bz-Mach10-$\beta$400-res2 & Bz & 400  & 10 &  $768 \times 704 \times 64$ & convergence study
\enddata
\tablecomments{
\scriptsize \\
(1) Seed field geometry: geometry of magnetic field in the initial disk and in the L1 gas inflow.\\
(2) Seed field $\beta$: plasma $\beta$ of the magnetic field in the L1 inflow, defined as ratio of gas pressure and magnetic pressure.\\
(3) Inner $\mathcal{M}$: Mach number at the inner boundary at $R_{min}=0.02$.
}
\end{deluxetable*}

The goal of this paper is to explore how the relative importance of MRI and spiral shocks change with disk properties. Therefore, we design a series of simulations for comparison by varying the following parameters: the Mach number of the disk $\mathcal{M}$, the geometry and strength of the seed magnetic field at the Roche lobe overflow. We summarize the parameters of all models in this work in Table \ref{tab:parameter} and elaborate them in this section.

The WD resides at the origin and the donor star resides at $(R, \phi, z) = (1, 0, 0)$ throughout the simulation time. In cylindrical coordinates, our computational domain spans $(R, \phi, z) \in [0.02, 0.62] \times [0, 2\pi] \times [-0.02,0.02]$ for 3D MHD simulations and $(R, \phi) \in [0.02, 0.62] \times [0, 2\pi]$ for 2D hydro simulations. Our fiducial simulations have  $384 \times 704 \times 32$ cells for 3D MHD models and $384 \times 704$ cells for 2D hydro cases. The outer radial boundary at $R_{max}=0.62$ is the radius of the first Lagrangian point $L_1$ for a binary with mass ratio $q=0.3$, and the inner radial boundary $R_{min}=0.02$ is approximately four times the surface radii of the WD. Considering the temperature profile we adopt in Eq. \ref{eq:cs} which implies that the Mach number is $\mathcal{M} = 10 (R/R_{min})^{-1/8}$ in our fiducial model, the thermal scale height is $H=R/\mathcal{M}=0.002 (R/R_{min})^{9/8}$. The vertical range $[-0.02,0.02]$ thus contains about 2 scale heights at $R=0.15$ ( As a reference for this location in the disk, the radial size of CV disks in both MHD and hydrodynamical models in \citet{2016Ju} is about 0.3). This yields that there are $3-75$ grid cells per scale height depending on radius in the vertical direction with 16 grid cells per scale height at $R=0.15$. However, note that the thermal scale height does not have a physical meaning except an indicator of the local temperature in unstratified disks because there is no vertical structure.

As in \citet{2016Ju}, we use logarithmic grid spacing in the radial direction ($\Delta R \propto R$) and uniform grids in the azimuthal and vertical directions. Spiral arms excited in binary systems are nearly self-similar \citep{1987Spruit}, so that the radial spacing between shocks becomes smaller at small radii. By using logarithmic grid spacing in radius, we can resolve the spiral shocks equally well at any radii. Moreover, with a logarithmic grid, we are able to ensure $\Delta R /R = \Delta \phi = const$, i.e. the grid cells are square in physical size at all radii, which reduces numerical diffusion due to highly distorted grid cells.

The initial density of the disk is uniform $\rho=1$ and the initial velocity is Keplerian where $v_R=v_z=0$, $v_{\phi} = \sqrt{GM_1/R^3}$. The initial sound speed profile is set as Eq. \ref{eq:cs} defines, where the scaling factor $c_s(R_{min})$ is adopted such that the Mach number at inner boundary $R_{min}$ is 10 or 20. The hotter cases with $\mathcal{M}(R_{min}) = 10$ are marked as ``Mach10" in the model names (see Table \ref{tab:parameter}), and the cooler cases with $\mathcal{M}(R_{min}) = 20$ are marked as ``Mach20". With these two Mach number profiles, we compare how disk temperature affects the relative importance of spiral shocks and MRI.

The initial magnetic field in the disk has two geometries. The first is vertical magnetic field which is denoted with ``Bz" in the model names in Table \ref{tab:parameter}. In this case, the radial and azimuthal components of ${\bf B}$ are zero $B_R = B_\phi=0$ and the vertical component is set such that wavelength of the fastest growing unstable mode of MRI is equal to a quarter of the vertical size of our computational domain $\lambda_{MRI} \equiv 2\pi \sqrt{16/15} v_{A,z} / \Omega = 0.1$ where $v_{A,z} = B_z / \sqrt{\rho}$ is the vertical Alfv{\'e}n speed. Thus
\begin{equation}
B_z =  \sqrt{\frac{15}{16}} \frac{0.1 \Omega \sqrt{\rho} }{2\pi} .
\end{equation}
The second initial magnetic field geometry is toroidal field loops which is denoted with ``Loop" in the model names in Table \ref{tab:parameter}. In this case, the radial and vertical components are zero $B_R = B_z = 0$ and the azimuthal component is set such that the critical azimuthal wavenumber of the toroidal field $M_c = R \Omega / v_{A,\phi}=20$ where $v_{A,\phi} = B_\phi / \sqrt{\rho}$. This leads to
\begin{equation}
B_\phi = \sqrt{\rho} \frac{R \Omega}{20}.
\end{equation}

We inject gas stream from the L1 point at a constant rate to represent the Roche lobe overflow. The L1 region spans $(\phi,z) \in [-0.1, 0.1] \times [-0.02,0.02]$ at the outer boundary and consists of 10 (azimuthal) $\times$ 32 (vertical) cells with the fiducial resolution. At the L1 region ghost cells, the gas density is $\rho_0=1$, radial velocity is $v_R=-0.01$, azimuthal velocity is $v_\phi=0$ since the L1 point is static in the rotating frame. Considering the area of the L1 region is $R_{max} \Delta \phi \Delta z = 0.62 \times 0.2 \times 0.04$, the mass inflow rate from the companion star through the L1 region is $\rho_0 (-v_R) R_{max} \Delta \phi \Delta z \sim 5 \times 10^{-5}$ in internal units and $\sim 1 \times 10^{-9} M_\odot / yr$. The gas sound speed is set to be $c_s(R_{max})$ as defined in Eq.\ref{eq:cs}, i.e., $c_s \simeq 0.18$ in the ``Mach10" cases, or $c_s \simeq 0.09$ in the ``Mach20" cases. All the variables are constant over time at the L1 ghost region. At all other radial boundary regions, we use "free outflow, no inflow" boundary condition, which copies variables from the last active cells to the ghost cells except radial velocities $v_R$ and azimuthal velocities $v_\phi$. Azimuthal velocities in ghost zones are set to their local Keplerian values in the rotating frame $v_K  - \Omega_0 R$. For radial velocities, if $v_R$ of the last active cell is outflowing (i.e. $v_R>0$ at the outer boundary, or $v_R<0$ at the inner boundary), we copy its value to the ghost cells; otherwise, we set $v_R=0$ at the ghost cells to avoid unwanted mass inflow. We use periodic boundary conditions in the azimuthal and vertical directions.

Seed magnetic field is also injected along with the gas inflow at the L1 region. Corresponding to the ``Bz" and ``Loop" cases in the initial conditions, there are also two types of field geometry at the L1 boundary. In ``Bz" cases, $B_R=B_\phi = 0$ at the L1 ghost cells and the vertical magnetic field is set as 
\begin{equation}
B_z = \sqrt{\frac{2 P_g}{\beta}} \cos(\frac{\pi}{2} \frac{\phi}{l_{L1}}),
\end{equation}
where $\beta = P_g/P_B$ is the ratio of gas pressure and magnetic pressure, $l_{L1}=0.1$ is half the azimuthal range of the L1 region. The cosine term is to provide a continuous transition between the L1 and the non-L1 ghost cells. In this work, we adopt vertical seed field ``Bz" with $\beta = 100$ and ``Mach10" as our fiducial model. To study the effects of magnetic field strength, we add another model ``Bz-Mach10-$\beta400$" which is similar to the fiducial model except the seed filed has $\beta=400$. 

In ``Loop" cases, magnetic field loops are set by a magnetic vector potential 
\begin{equation}
\label{eq:seed_loop}
{\bf A}({\bf R}) = \sqrt{\frac{2P_{gas}}{\beta}} \frac{2R_{loop}}{\pi} \cos(\frac{\pi}{2} \frac{|{\bf R}-{\bf R}_{ref}(t)|}{R_{loop}}) {\bf \hat{z}},
\end{equation}
where $\beta$ is magnetic pressure to gas pressure ratio, ${\bf R}_{ref}(t) = (R=R_0 + v_Rt, \phi=\Omega_0 t, z)$ is a reference point that advects inward at the same velocity with the gas inflow at L1 point. The size of magnetic loops is $R_{loop} \sim 0.06$. The magnetic field components can then be calculated by
\begin{equation}
{\bf B} = \nabla \times {\bf A},
\end{equation}
which forms field loops in the disk plane advecting with the background inflow at L1 region. In this work, this field geometry is adpoted in model ``Loop-Mach10-$\beta400$" to be compared with the ``Bz-Mach10-$\beta400$" model for effects of seed field geometry. At non-L1 regions, we simply copy the magnetic field components of the last active cells to ghost cells.

Note that the MHD run with a cooler disk ``Bz-Mach20-$\beta100$-res2" has double the resolution in the radial and vertical direction (Table \ref{tab:parameter}). The reason is that the spiral shocks are more tightly wound in a cooler disk \citep{2016Ju} and thus harder to resolve. Therefore, when we use fiducial resolution for the ``Mach20" runs, the gas piles up at $R=0.2-0.3$ forming a ring because the spiral waves are not well resolved and thus cannot propagate further inward. 
Doubling the radial resolution better resolves the spiral waves, and doubling the vertical resolution better resolves MRI. We do not double resolution in the azimuthal direction to save computational cost since there is not much structure along the azimuthal direction. 

\subsection{Diagnostics}

In order to evaluate the efficiency of angular momentum transport driven by MRI and spiral shocks in our simulations, we define the effective viscosity parameter $\alpha_{eff}$ such that the corresponding kinetic viscosity under the standard $\alpha$ theory \citep{1973Shakura} induces the same mass accretion rate observed in our simulations,
\begin{equation}
\alpha_{eff} = \frac{\dot{M}}{3\pi \Sigma c_s H},
\label{eq:alpha_eff}
\end{equation}
where $\dot{M}$ is the mass accretion rate (the total amount of mass passing through the radius R per unit time), $c_s$ is the sound speed and $H = c_s/\Omega$ is the thermal scale height, and $\Sigma$ $=$ $(1/2\pi)$ $\int_z \int_{0}^{2\pi} \rho d\phi dz$ is the azimuthally averaged surface density.

To compare the relative importance of MRI and spiral shocks in driving angular momentum transport, we split the expression of $\alpha_{eff}$ into several terms that represent the contribution from MRI and spiral shocks respectively. Such splitting can be derived from angular momentum conservation equation. We did similar analysis in \S 2.4 and \S 4.3.1 of \citealt{2016Ju} where we analyzed the angular momentum budget of the disk. For reader's convenience, we repeat some of the derivation here.

For inviscid gas, the angular momentum conservation equation in cylindrical coordinates is
\begin{eqnarray}
\label{eq:angmom0}
 \partial_t <\rho R v_\phi&>&= -\frac{1}{R} \partial_R (R^2<\rho v_R v_\phi> \nonumber \\
&-& R^2 <B_R B_\phi> ) + <\bf{R} \times \bf{F}_{ext}>,
\end{eqnarray}
where $\bf{F}_{ext} = -\rho \nabla \Phi_{tot} + {\bf F}_{Cori}$ is the total external force including the gravitational force from the companion star and the Coriolis force (see the momentum equation Eq.\ref{eq:momentumcons-ch-MRI}). The notation $<X>$ represents the vertical and azimuthal integration of variable $X$ within a ring of radial width $\Delta R$ at radius $R$: 
\begin{eqnarray}
<X> &=& \int_{z_{min}}^{z_{max}} \int_{0}^{2\pi} X d\phi dz \Delta R \nonumber \\
&=& \sum_k \sum_j X_{i,j,k}  \Delta \phi_j \Delta z_k \Delta R_i.
\end{eqnarray}

Once the disk reaches steady state, it is almost Keplerian throughout the simulation time, so the evolution can be more clearly seen in the perturbed angular momentum $\rho R \delta v_\phi = \rho R (v_\phi-v_K)$ where $v_K$ is the Keplerian azimuthal velocity. Substituting $v_\phi = v_K + \delta v_\phi$ and multiplying $R$ on both sides, equation \ref{eq:angmom0} becomes
\begin{eqnarray}
&& \partial_t <\rho> (R v_K) R +  \partial_t <\rho R \delta v_\phi>R \nonumber \\
&=& - \partial_R <R\rho v_R> (R v_K)  - <R\rho v_R> \partial_R (R v_K) \nonumber \\
&& - \partial_R (R^2<\rho v_R \delta v_\phi> ) +  \partial_R (R^2<B_R B_\phi> ) \nonumber \\
&&+  <\mathbf{R} \times \mathbf{F_{ext}}>R .
\end{eqnarray}
Note that the first terms on the left and right side cancel due to mass conservation equation (Eq.\ref{eq:massconservation-ch-MRI}). The conservation equation for perturbed angular momentum becomes
\begin{eqnarray}
\label{eq:angmom}
 &&\partial_t <\rho R \delta v_\phi> R \nonumber \\
 &=& - <R\rho v_R> \partial_R (R v_K) - \partial_R (R^2<\rho v_R \delta v_\phi> ) \nonumber \\
&&  + \partial_R (R^2<B_R B_\phi> ) + <\mathbf{R} \times \mathbf{F}_{ext}>R \nonumber \\
&=& \dot{M} \partial_R (R v_K) - \partial_R (R^2<\rho v_R \delta v_\phi> ) \nonumber \\
&&  + \partial_R (R^2<B_R B_\phi> ) + <\mathbf{R} \times \mathbf{F}_{ext}>R,
\end{eqnarray}
where $\dot{M}=- <R\rho v_R>$ is the mass accretion rate at radius $R$. At this point, it is clear to see the angular momentum budget of the accretion disk in CVs: the time change of angular momentum of a local ring at $R$ in the disk ($\partial_t <\rho R \delta v_\phi> R$) is caused by the accreted gas carrying differential angular momentum ( $\dot{M} \partial_R (R v_K)$ ), the radial gradient of Reynolds stress ($T_R = <\rho v_R \delta v_\phi>$) driven by spiral arms and Maxwell stress ($T_M = -<B_R B_\phi>$) driven by MRI, and lastly the external torque exerted on the disk ($<\mathbf{R} \times \mathbf{F}_{ext}>R$). If we integrate this equation over the whole radial range of the disk,  the time change of the total angular momentum of the disk would be caused by the gas which carries angular momentum entering (e.g. inflow from L1 point) or leaving the disk (e.g. outflow at outer boundary and accretion into the inner boundary), the radial advection of angular momentum driven by Reynolds stress and Maxwell stress, and the external torque exerted on the disk.

In a steady-state MHD disk, the time derivative term on the left hand side is zero, which gives
\begin{eqnarray}
 &&\frac{\dot{M}}{R} \partial_R (R v_K) \nonumber \\
 &=& \frac{1}{R} \partial_R (R^2<\rho v_R \delta v_\phi> - R^2 <B_R B_\phi> ) \nonumber \\
 &&- <\bf{R} \times \bf{F}_{ext}>.
\end{eqnarray}
Integrating this equation in the radial direction gives
\begin{align}
\dot{M}  =& \frac{1}{R v_K} \left( R^2 <\rho v_R \delta v_\phi> - R^2 <B_R B_\phi> \right. \nonumber \\
&  - \int R \left. <\mathbf{R} \times \mathbf{F}_{ext}> dR +  C \right)
\end{align}
where we assume $\dot{M}$ is a constant over radius in steady state. Plugging this formula into the expression for $\alpha_{eff}$ we get
\begin{eqnarray}
\label{eq:effalpha-ch-MRI}
\alpha_{eff} &=&  \frac{2}{3} \frac{T_R}{P} + \frac{2}{3} \frac{T_M}{P} \nonumber \\
&& + \frac{ -\int R <\mathbf{R} \times \mathbf{F}_{ext}> dR + C}{3\pi \Sigma c_s H R v_K}
\end{eqnarray}
where 
\begin{equation}
T_R = <\rho v_R \delta v_\phi>/2\pi L_z,
\end{equation} 
is the azimuthally- and vertically averaged Reynolds stress, and
\begin{equation}
T_M = - <B_R B_\phi> /2\pi L_z
\end{equation}
is the azimuthally- and vertically averaged Maxwell stress. The constant $C$ is set by the boundary condition. 

From this equation, we can clearly see the mechanisms that drive mass accretion in CV disks: the Reynolds stress from the spiral shocks, the Maxwell stress from the MRI, and the external torques. While the external torque exerted by the companion star plays an important role at the large radii where tidal effects are strong, the stresses induced by spiral shocks and MRI dominate at the small radii where most of the disk resides. Therefore, we measure the time-, azimuthal- and vertical- averaged $T_R$ and $T_M$ to compare the relative importance of the two major mechanisms in driving angular momentum transport: spiral shocks and MRI.

\section{The Fiducial Model}
\label{sec:fiducial_model}

\begin{figure*}
\centering
\includegraphics[width=0.4\textwidth]{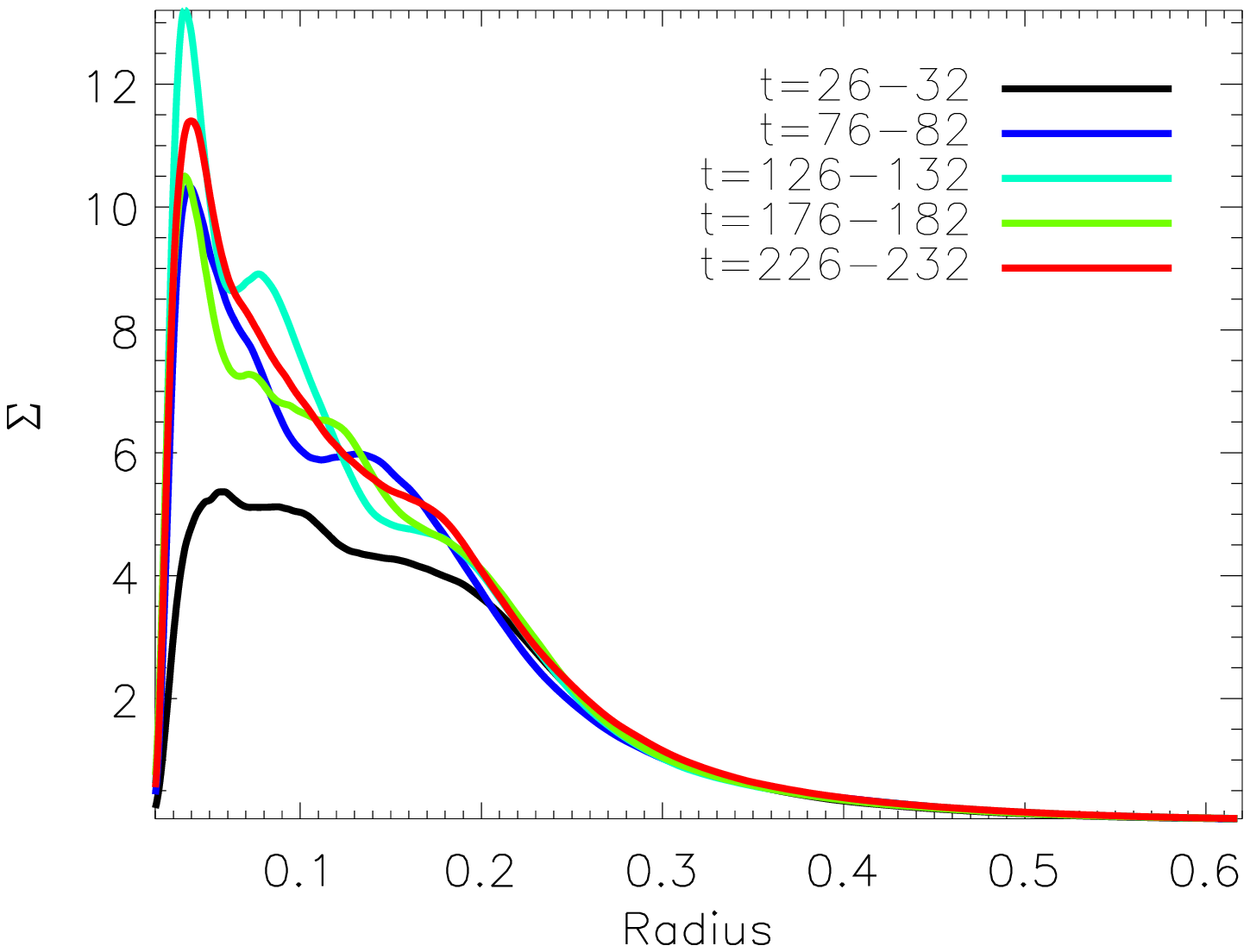}
\includegraphics[width=0.4\textwidth]{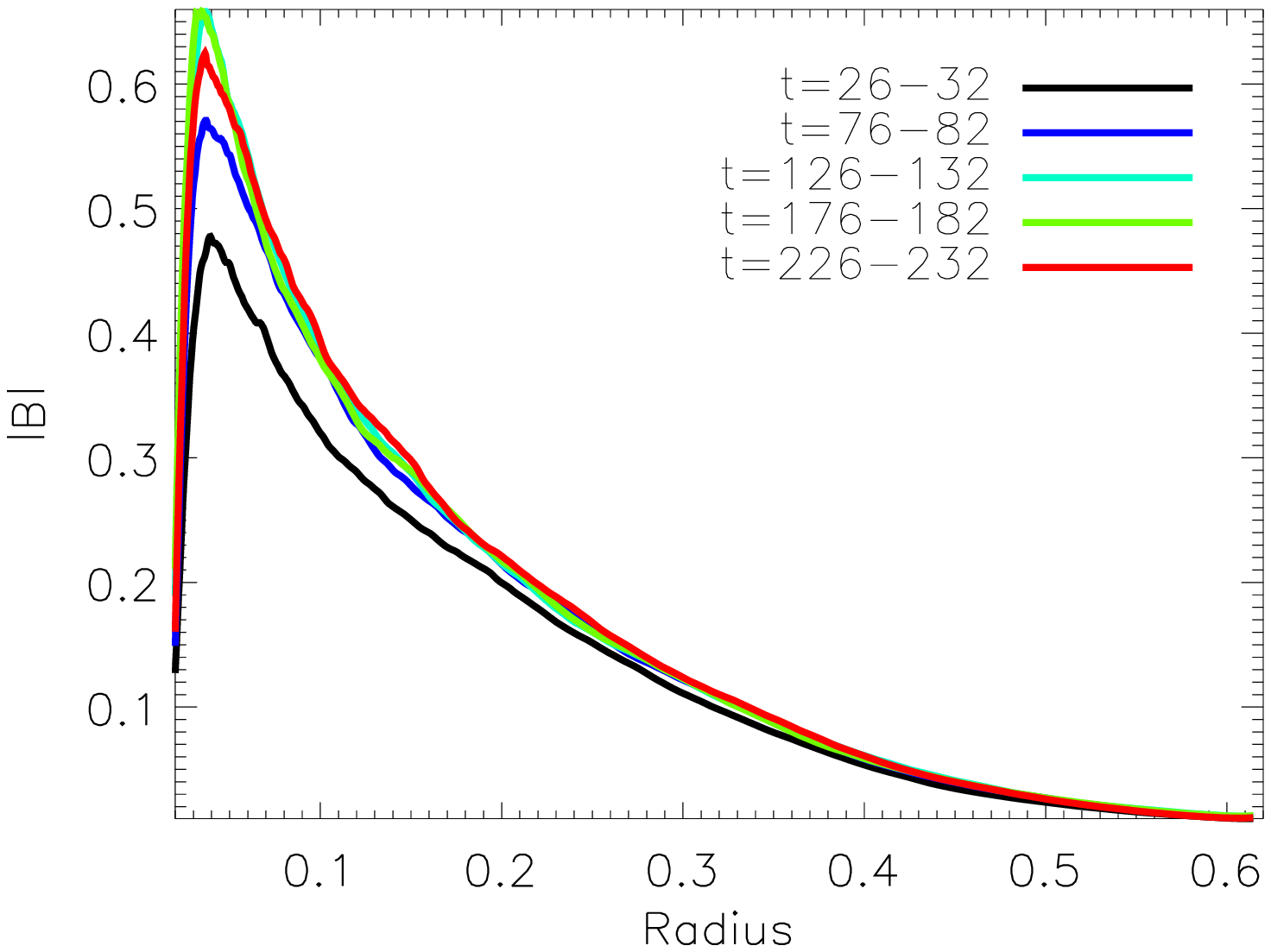}
\includegraphics[width=0.4\textwidth]{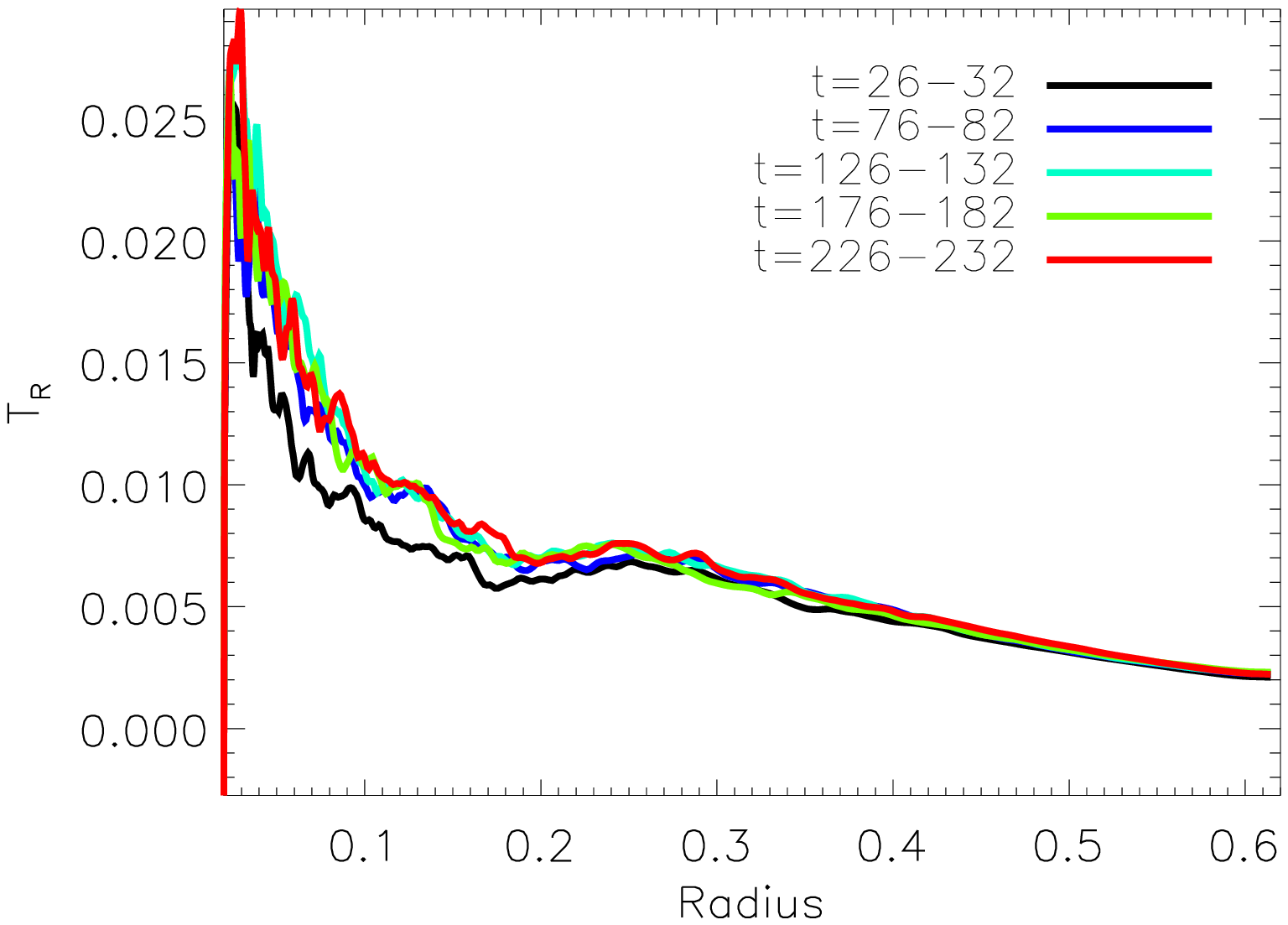}
\includegraphics[width=0.4\textwidth]{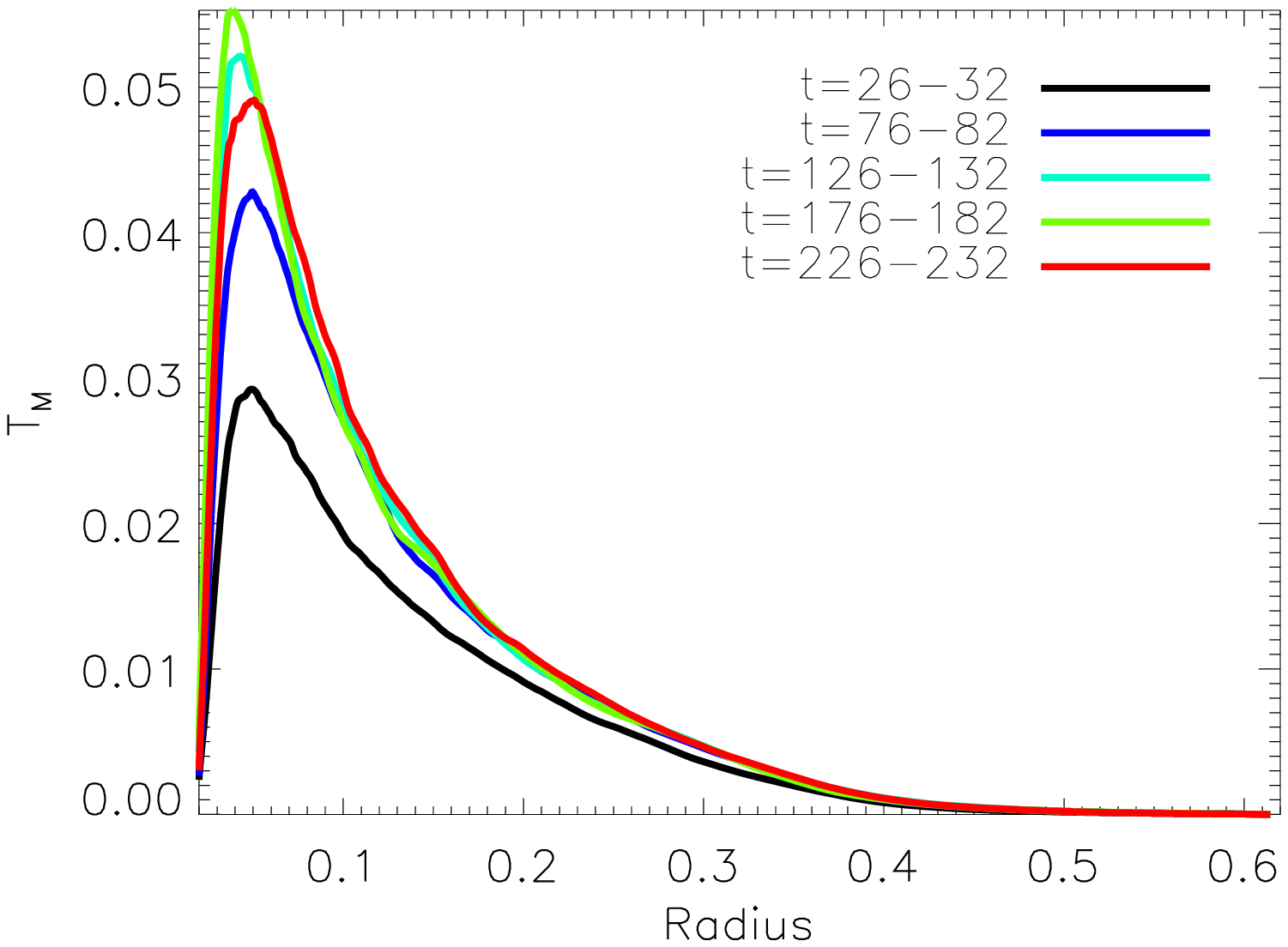}
\includegraphics[width=0.4\textwidth]{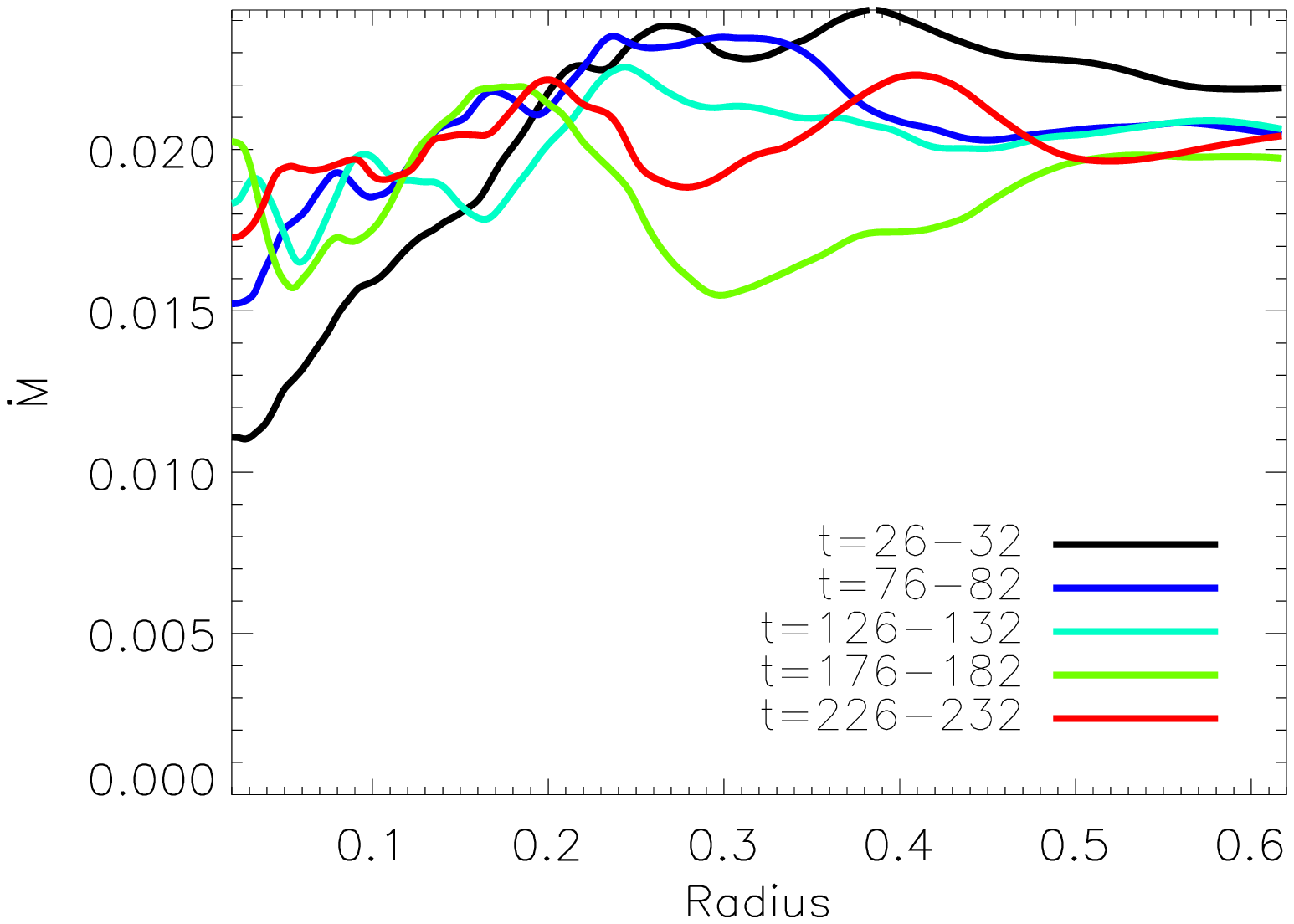}
\includegraphics[width=0.4\textwidth]{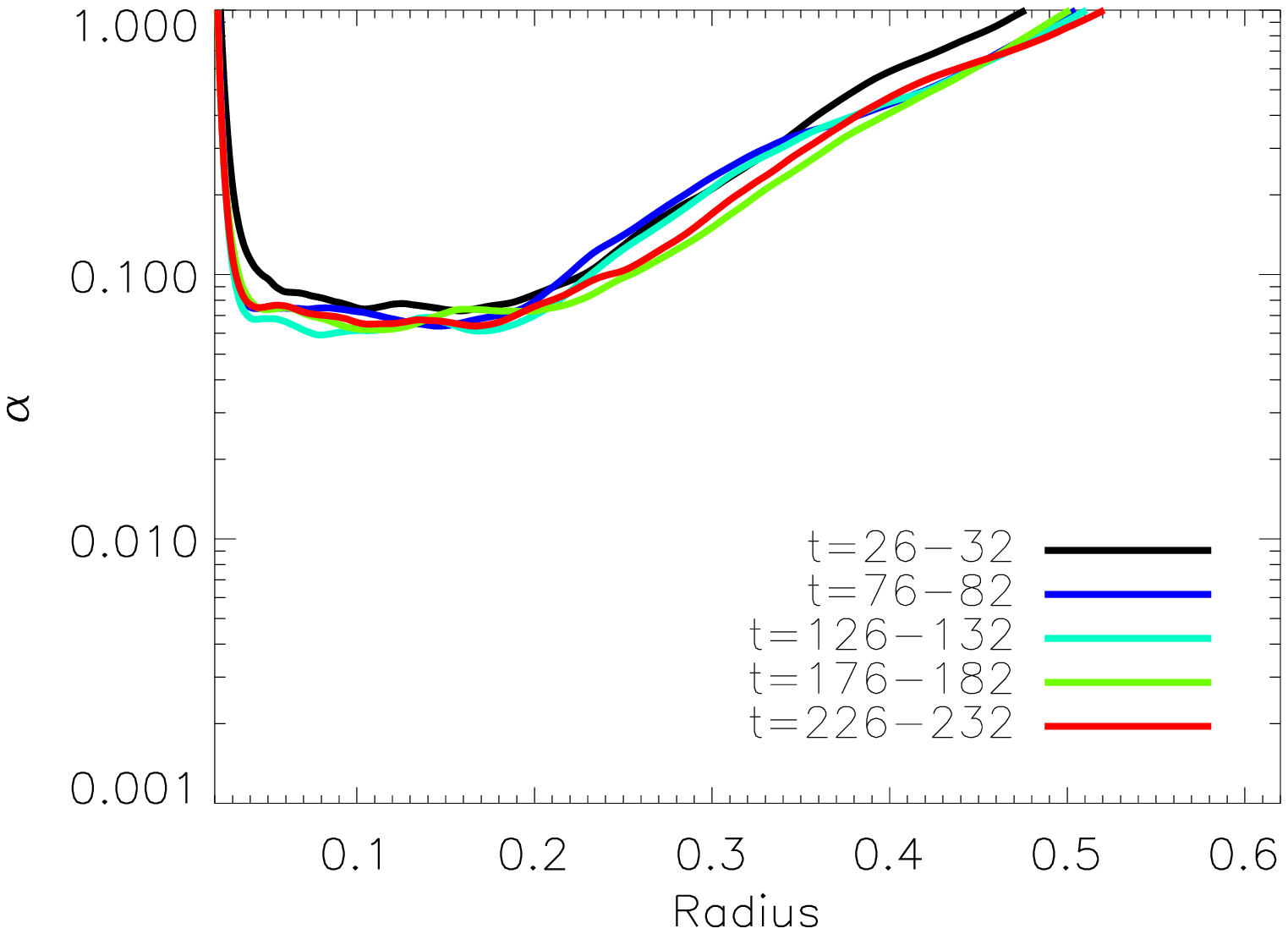}
\includegraphics[width=0.4\textwidth]{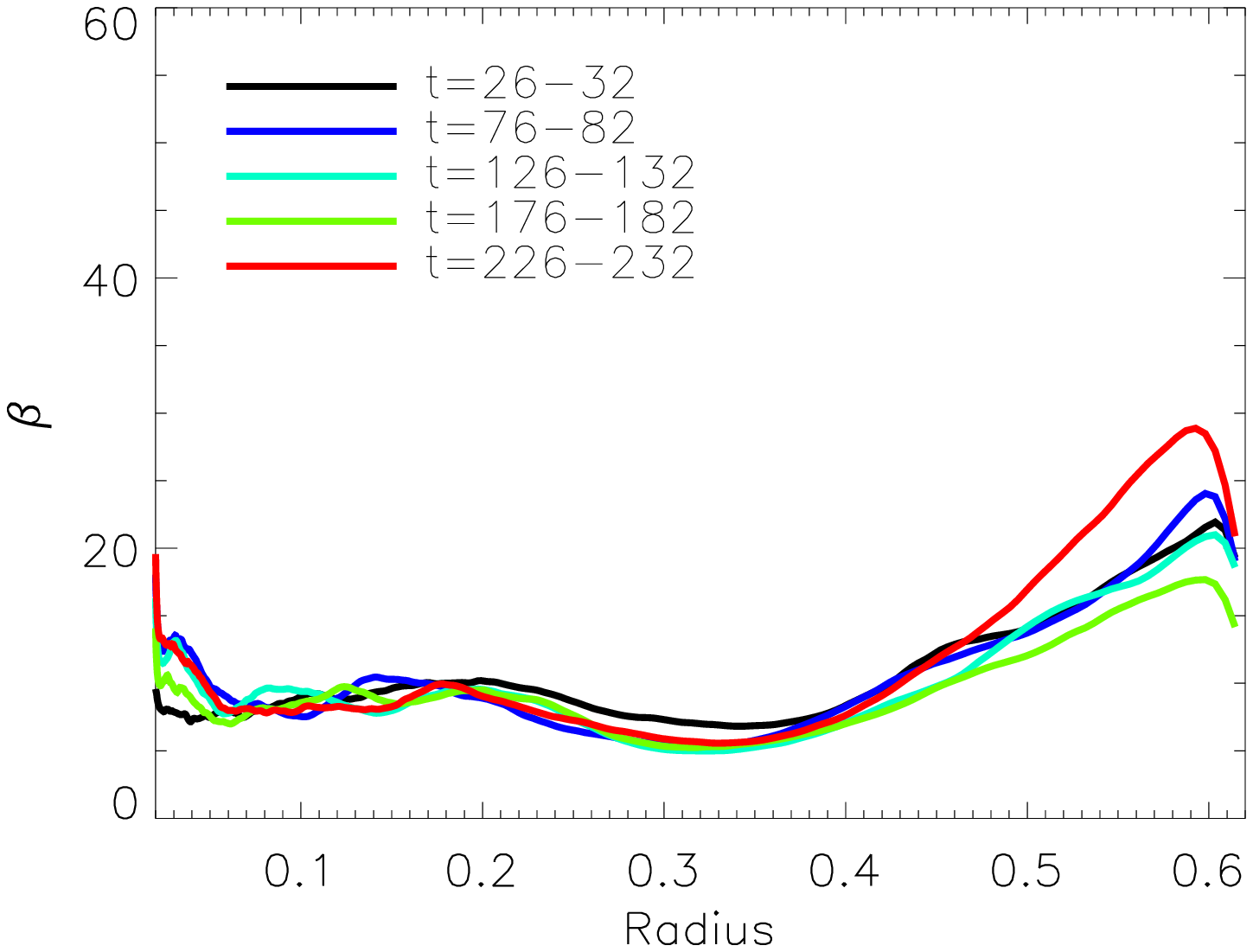}
\includegraphics[width=0.4\textwidth]{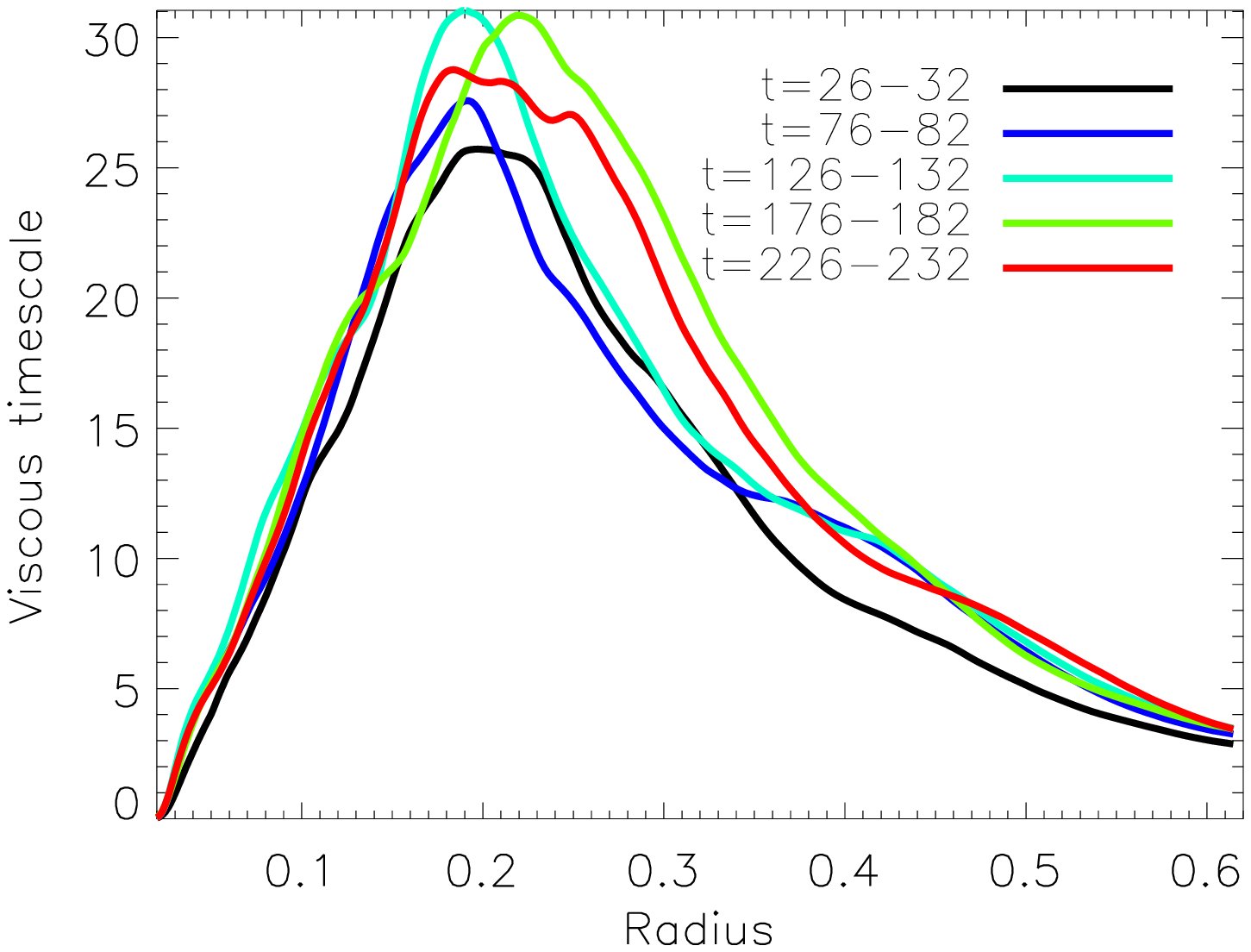}
\caption{Time evolution of characteristic properties of the disk in the fiducial model ``Bz-Mach10-$\beta100$-res1": the surface density $\Sigma$ (first row left), the magnitude of the magnetic field $|{\bf B}|$ (first row right), the Reynolds stress $T_R$ (second row left), the Maxwell stress $T_M$ (second row right), the mass accretion rate $\dot{M}$ (third row left), the effective viscosity parameter $\alpha_{eff}$ (third row right), the plasma $\beta$ (last row left), and the viscous timescales (last row right). All quantities are volume averaged and time averaged over the following time periods: [26, 32], [76, 82], [126, 132], [176, 182], [226, 232].}
\label{fig:Bz_Mach10_res2_time_evolution}
\end{figure*}

In order to study the effects of varying the geometry and strength of injected magnetic field as well as the Mach number of gas, we first consider a standard model as a reference for the comparison studies. We take the model ``Bz-Mach10-$\beta$100-res1" as our fiducial model. The parameters of this model are listed in Table \ref{tab:parameter}.

\begin{figure*}
\centering
\includegraphics[width=0.48\textwidth]{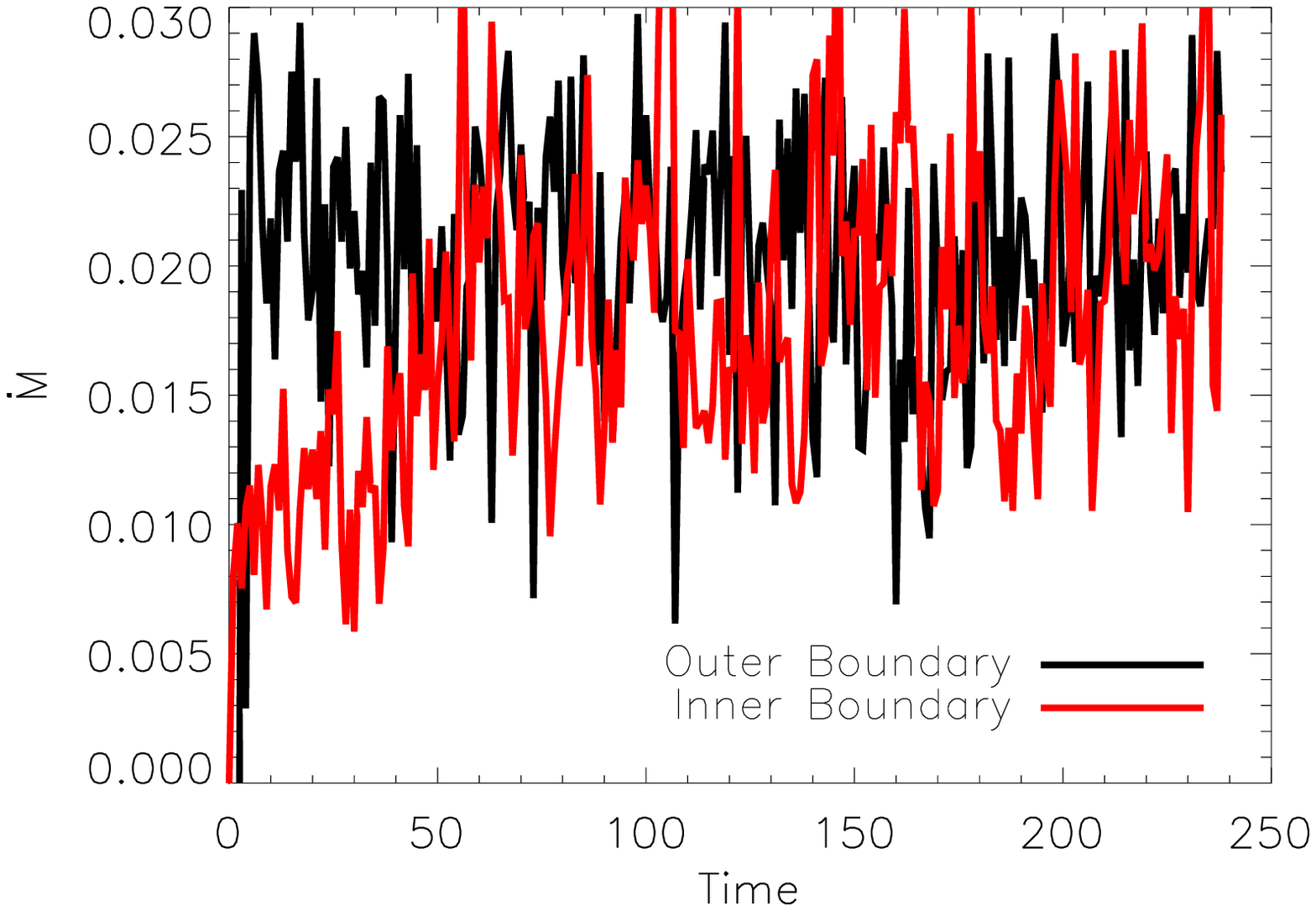}
\includegraphics[width=0.48\textwidth]{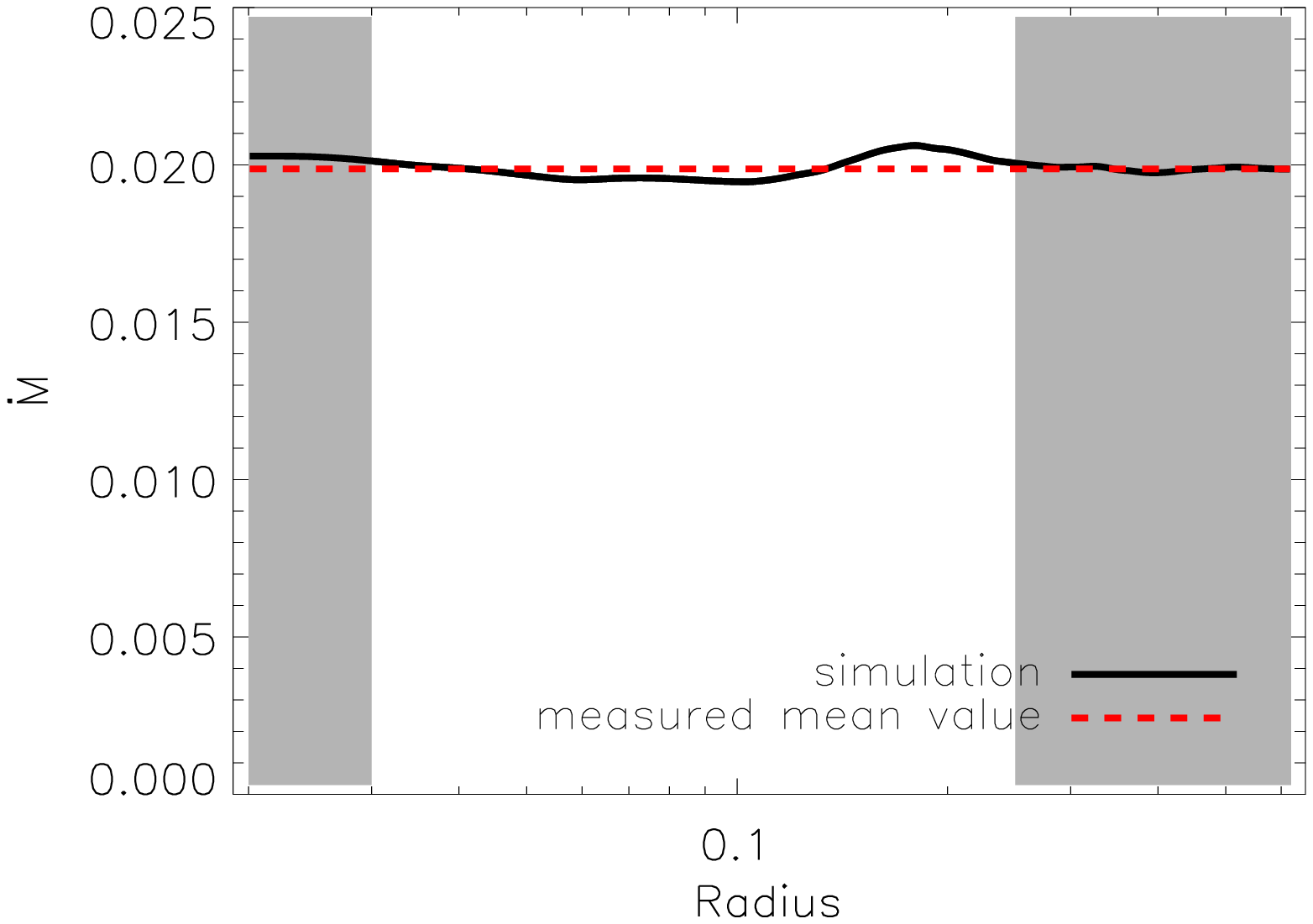}
\includegraphics[width=0.48\textwidth]{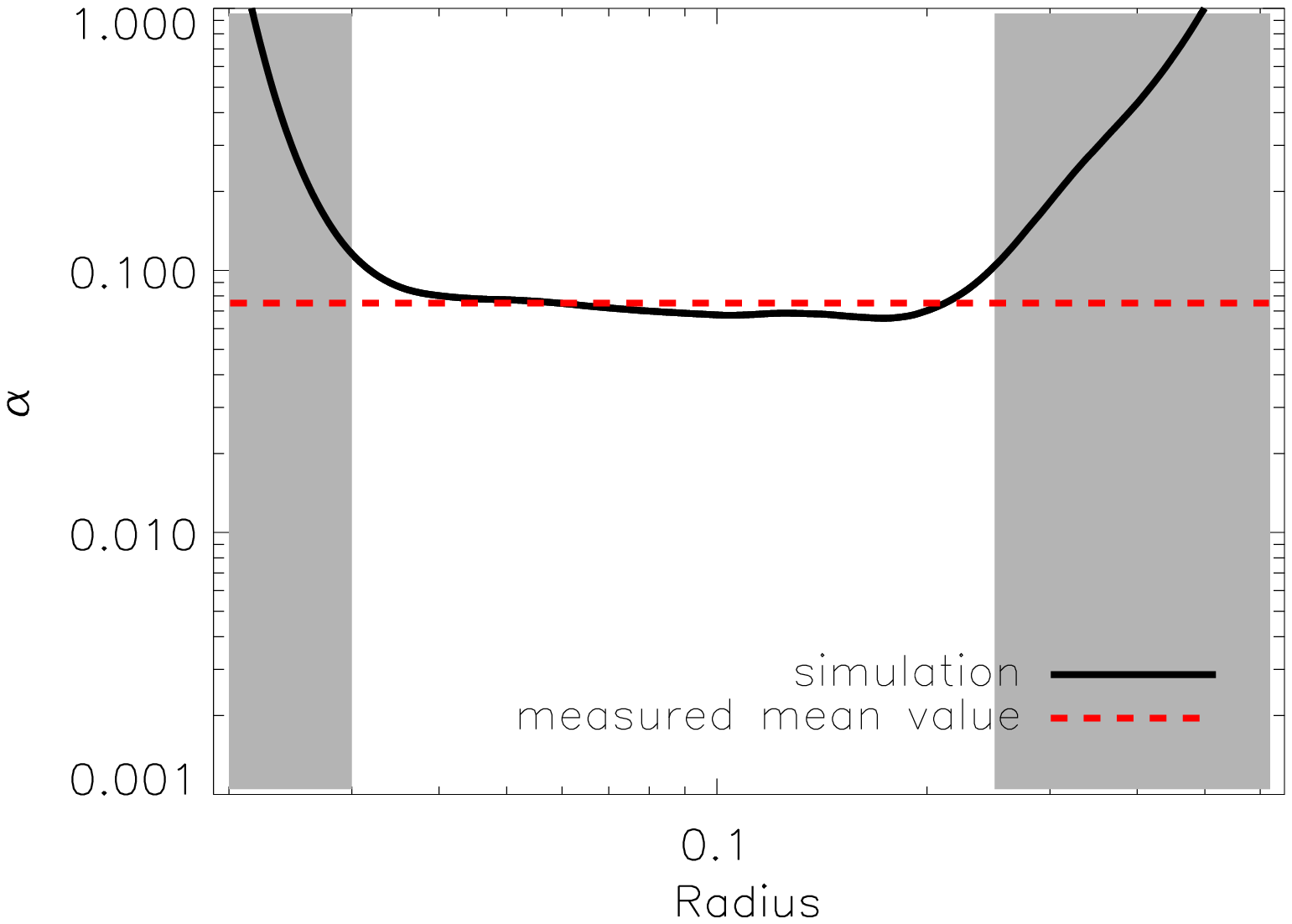}
\includegraphics[width=0.48\textwidth]{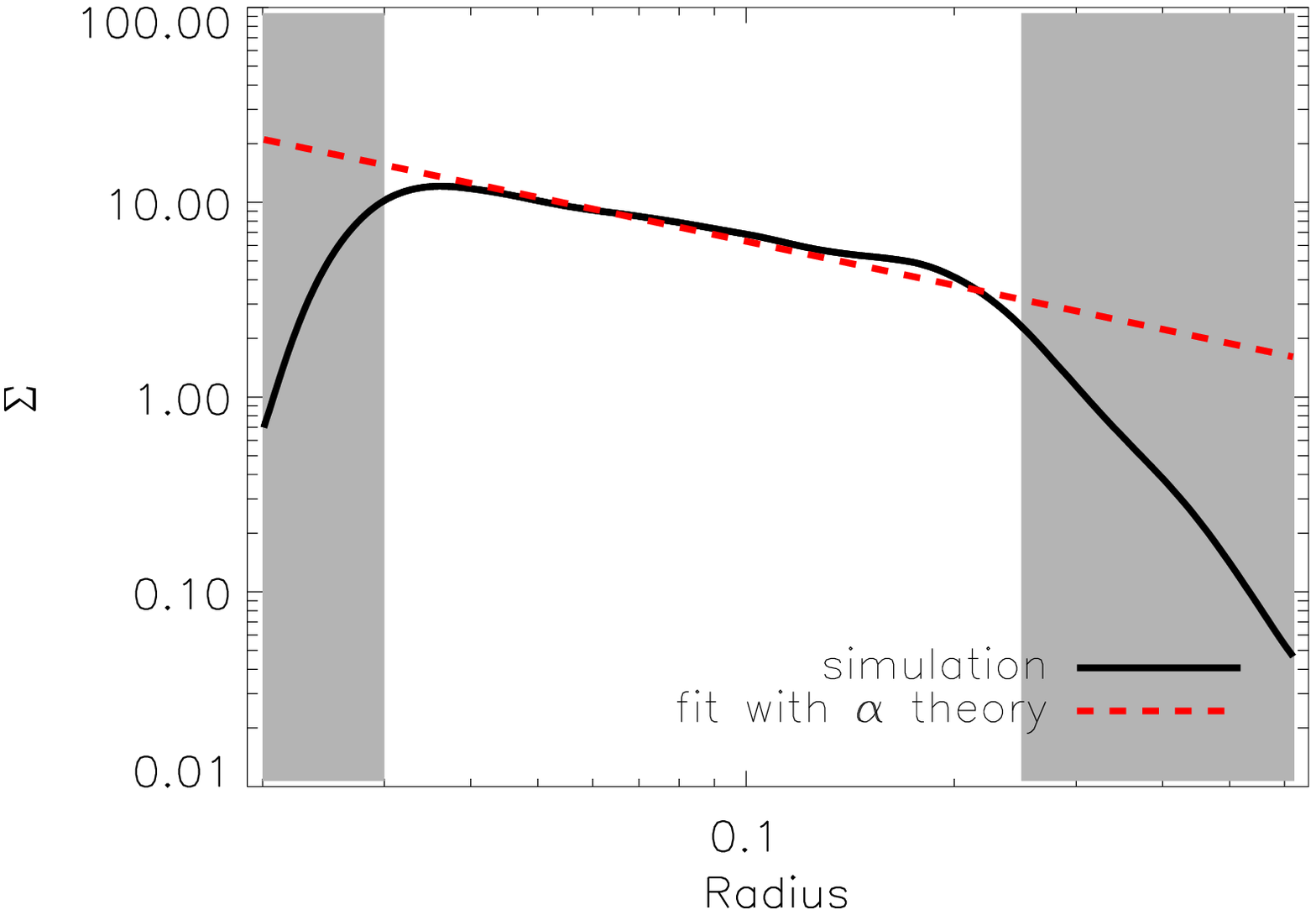}
\caption{ {\it Top left}: time evolution of mass accretion rates at inner (red line) and outer (black line) boundaries for the fiducial model ``Bz-Mach10-$\beta100$-res1". {\it The rest three panels}: fitting the fiducial model ``Bz-Mach10-$\beta100$-res1" using the steady-state $\alpha$ disk theory. The black lines are the radial profiles of the mass accretion rate $\dot{M}$ (top right), $\alpha_{eff}$ (bottom left), and the surface density $\Sigma$ (bottom right), all of which are volume- and time- averaged over $t=200-238$. The area at $R>0.25$ is dominated by the inflow stream and the area at $R<0.03$ is affected by the inner boundary conditions, so they are both shaded for less attention. Mean values of $\dot{M}$ and $\alpha_{eff}$ (dashed red lines) are measured within the unshaded area $0.03<R<0.25$ where the majority of the disk resides. The predicted profile of $\Sigma$ according standard $\alpha$ disk theory is calculated as $\Sigma = \overline{\dot{M}} / 3 \pi \nu$ ($\nu=\overline{\alpha_{eff}} c_s H$ is the kinematic viscosity), which is the dashed red line in the bottom right panel.}
\label{fig:Bz_Mach10_res2_fit_with_alpha_theory}
\end{figure*}

We begin by describing the evolution of this model. In Figure \ref{fig:Bz_Mach10_res2_time_evolution} we show the evolution of characteristic properties of the disk: the surface density $\Sigma$, the magnitude of the magnetic field $|{\bf B}|$, the Reynolds stress $T_R$, the Maxwell stress $T_M$, the mass accretion rate $\dot{M}$, the effective $\alpha$, the plasma $\beta$ which is the ratio of the gas pressure to the magnetic pressure, and lastly the viscous timescales $t_s=\frac{1}{3(2-\gamma)^2} \frac{R^2}{\nu}$ (where $\gamma$ is the specific heat ratio, $\nu=\alpha c_s H$ is the kinematic viscosity). These quantities are all volume averaged in azimuthal and vertical directions and time averaged over several time periods: [26, 32], [76, 82], [126, 132], [176, 182], [226, 232]. Each period is 6 time units in length which is about 1 binary orbit (one binary orbit = $2\pi$ time units). The disk starts with uniform surface density and vertical magnetic field, so spiral arms and MRI are quickly developed. Meanwhile, the gas stream, as well as vertical magnetic field, flows in through the L1 point. At the beginning of the simulation, the surface density keeps increasing since the mass accretion rate at the inner boundary induced by spiral shocks and MRI is smaller than the mass inflow rate at the L1 point. Consequently, the spiral shocks become stronger thus the Reynolds stress $T_R$ increases over time.  Likewise, the vertical magnetic field piles up in the disk due to the higher injection rate of magnetic flux at the outer boundary and therefore $|{\bf B}|$ increases over time. MRI becomes stronger thus the Maxwell stress $T_M$ increases. As $T_R$ and $T_M$ both increase, the induced mass accretion rate $\dot{M}$ near the inner boundary increases and eventually catches up with the mass inflow rate at L1 point. The flat radial profiles of $\dot{M}$ indicates a steady state is reached. In Figure \ref{fig:Bz_Mach10_res2_fit_with_alpha_theory} (upper left panel), we show the time history of the mass accretion rate at the outer boundary (black line) and at the inner boundary (red line). The disk reaches steady state after $t\sim 50$. The whole simulation runs to $t=240$ which covers $\sim 580$ Keplerian orbits at $R= 0.15$ (middle area of the disk), $\sim 60$ binary orbits, or $\sim 8$ viscous timescales.


As a consistency check, we compare our model with the steady-state $\alpha$ disk theory in Fig.\ref{fig:Bz_Mach10_res2_fit_with_alpha_theory}. The black solid lines are the radial profiles of mass accretion rate $\dot{M}$, $\alpha_{eff}$ and surface density $\Sigma$ averaged over time $200-238$ from our simulation. In a steady-state $\alpha$ disk, $\dot{M}$ and $\alpha$ are constant in the disk. Therefore, we measure the mean values of $\dot{M}$ and $\alpha_{eff}$ within the radial range $R\in [0.03, 0.25]$ where the majority of the disk resides, and mark them as red dashed lines. The measurement gives $\overline{\dot{M}} = 0.02$ and $\overline{\alpha_{eff}}=0.08$. Using these mean values, we predict the radial surface density profile under the standard $\alpha$ disk theory $\Sigma = \overline{\dot{M}} / 3 \pi \nu$ where $\nu=\overline{\alpha_{eff}} c_s H$ is the kinematic viscosity. The predicted profile for $\Sigma$ is marked as the dashed red line in the lower right panel of Fig.\ref{fig:Bz_Mach10_res2_fit_with_alpha_theory}, which well fits the surface density profile from our simulation data shown in black line. Therefore, our model is consistent with a standard $\alpha$ disk with $\alpha=0.08$ and $\Sigma \propto R^{-4/3}$. 

\begin{figure}
\centering
\includegraphics[width=0.46\textwidth]{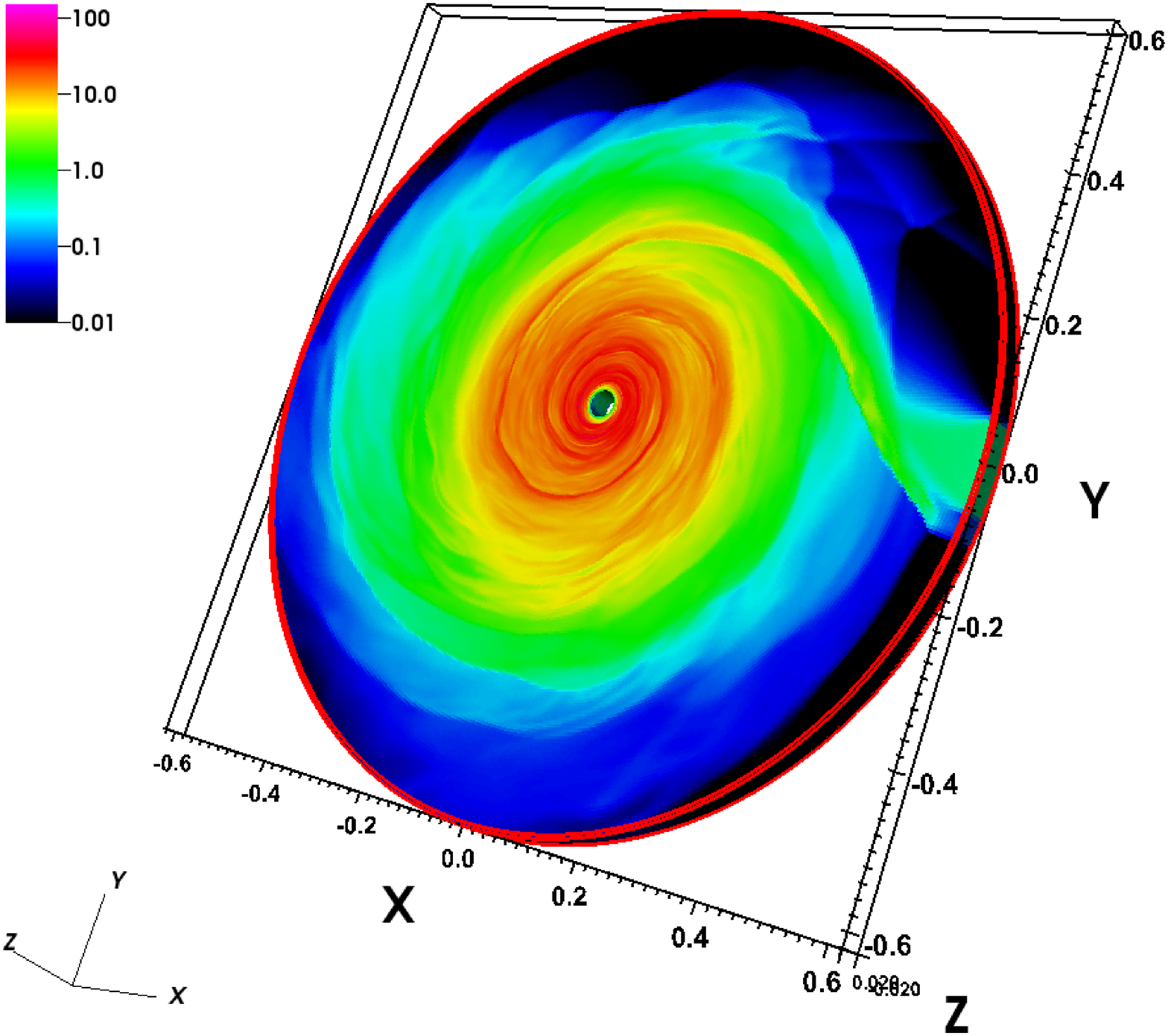}
\includegraphics[width=0.46\textwidth]{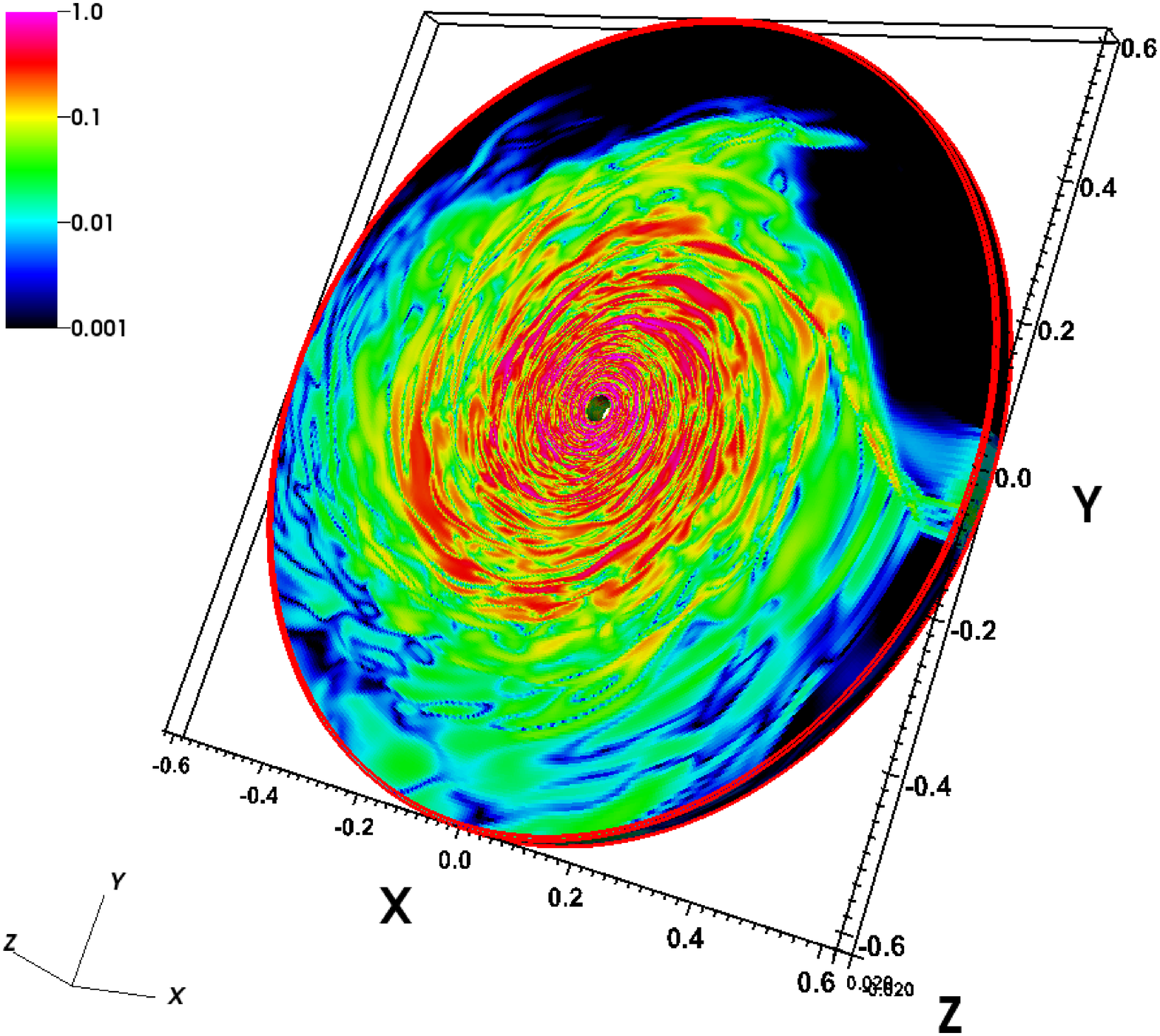}
\caption{Snapshots of the gas density (upper panel) and the strength of magnetic field $|{\bf B}|$ (lower panel) at $t=238$ in the fiducial model ``Bz-Mach10-$\beta100$-res1".}
\label{fig:Bz_Mach10_res2_3Dsnapshot_t48}
\end{figure}

\begin{figure}
\centering
\includegraphics[width=0.48\textwidth]{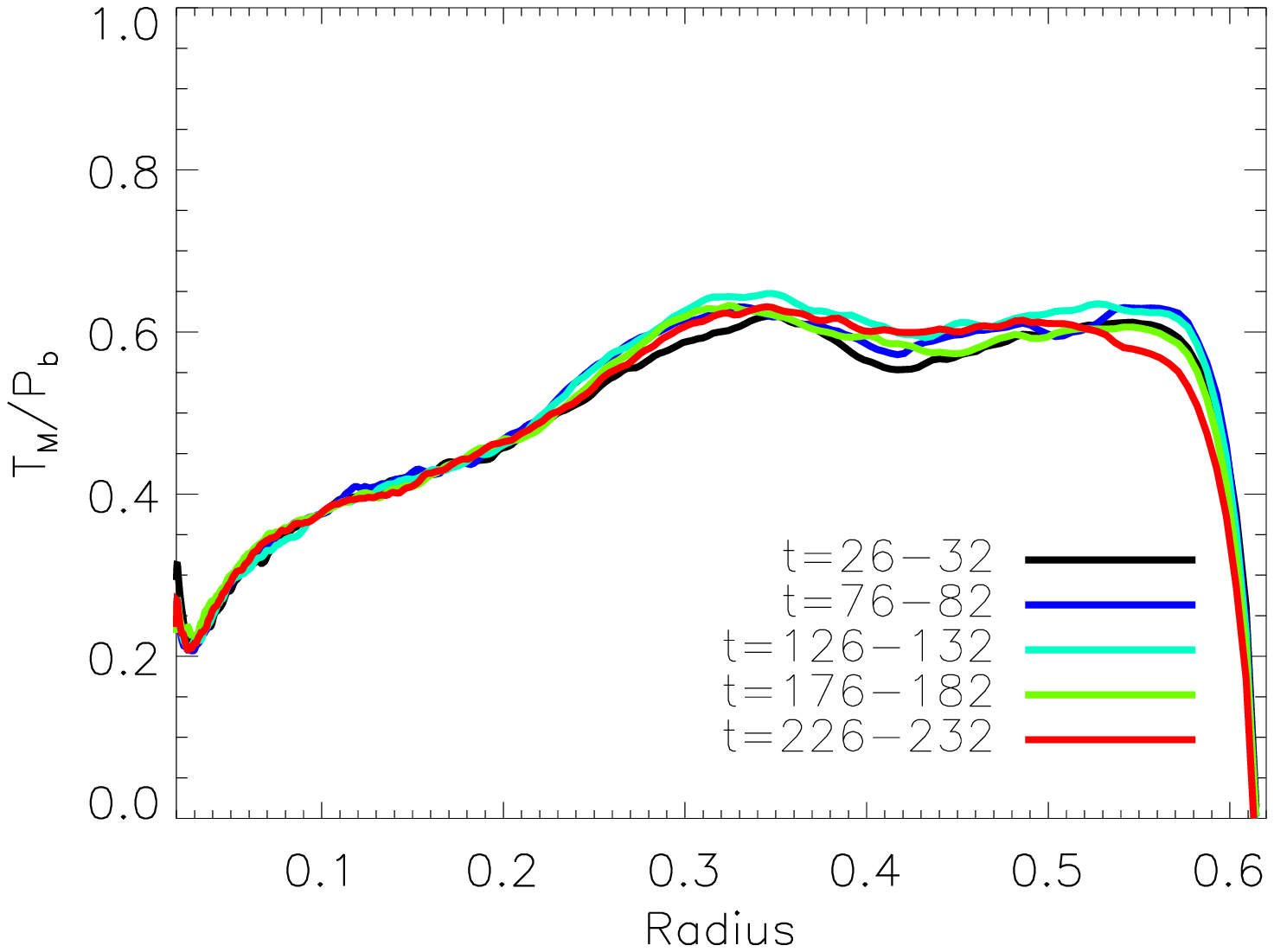}
\caption{MRI saturation predictor: the ratio of the Maxwell stress $T_M$ to the magnetic pressure $P_b=(1/2) |{\bf B}|^2$ in the fiducial model ``Bz-Mach10-$\beta100$-res1". The values of $T_M$ and $P_b$ are volume- and time- averaged over the following time periods: [26, 32], [76, 82], [126, 132], [176, 182], [226, 232].}
\label{fig:Bz_Mach10_res2_ratio_TM_to_Pb}
\end{figure}

This is the first global steady-state MHD model of CV disk with relatively realistic temperature profile. 
In Figure \ref{fig:Bz_Mach10_res2_3Dsnapshot_t48} we show the snapshots of gas density (upper panel) and magnitude of magnetic field $|{\bf B}|$ (lower panel) of the steady-state CV disk at $t=238$. The two-armed spiral structure excited by the binary potential is clearly seen in the density field (see \citealt{2016Ju} for more discussions on spiral shocks). During the whole evolution, the spiral patterns are static in the rotating frame. As mass is injected via L1 point, the density of the spiral arms increases everywhere with an increasing concentration at inner radius due to the outward angular momentum transport driven by spiral shocks and MRI. The ratio of density across the shocks is $\sim 1.5 - 3$ and remains in the same range as the disk becomes more massive. Meanwhile, the magnetic field is turbulent throughout the disk indicating the presence of MRI, which also induces turbulence in the density field. The magnetic field is amplified due to MRI: the steady-state disk has $\beta = 10$ (see radial profiles of $\beta$ in Figure \ref{fig:Bz_Mach10_res2_time_evolution}) while the seed magnetic field at L1 point has $\beta = 100$. Another important diagnostic of well-resolved MRI-driven turbulence that is independent of seed magnetic field topology or strength is the ratio of the Maxwell stress to the magnetic pressure $\alpha_{mag} = T_M / P_b \approx 0.45$ \citep{1995Hawley,2011Hawley,2013Hawley} or equivalently the magnetic tilt angle $\theta_B = \arcsin(\alpha_{mag})/2 \approx 13^{\circ}$ \citep{2009Guan-Gammie}. \citet{2013Hawley} did a series of global 3D MHD simulations of stratified disks and reported that $\alpha_{mag}$ falls in the range of $[0.35, 0.55]$ with a mean of 0.45 in MRI-resolved models. When the vertical resolution decreases, they found the value of $\alpha_{mag}$ tends to decrease. In Figure \ref{fig:Bz_Mach10_res2_ratio_TM_to_Pb} we plot the radial profiles of $\alpha_{mag}$ of our fiducial model that are averaged at different times of the simulation. The profile is mostly time invariant and the values of $\alpha_{mag}$ almost linearly increase from $\sim 0.2$ at the inner boundary to $\sim 0.6$ at outer boundary. This is because the thermal scale height $H=R/\mathcal{M}=0.002 (R/R_{min})^{9/8}$ which increases with radius while the vertical resolution is fixed at all radii. Therefore, there are increasing grid cells per scale height as radius increases. This is unavoidable for disk simulations in cylindrical coordinates. The way to solve this problem is to have a varying vertical resolution that is adaptive to the thermal scale height profile.

\begin{figure}
\centering
\includegraphics[width=0.48\textwidth]{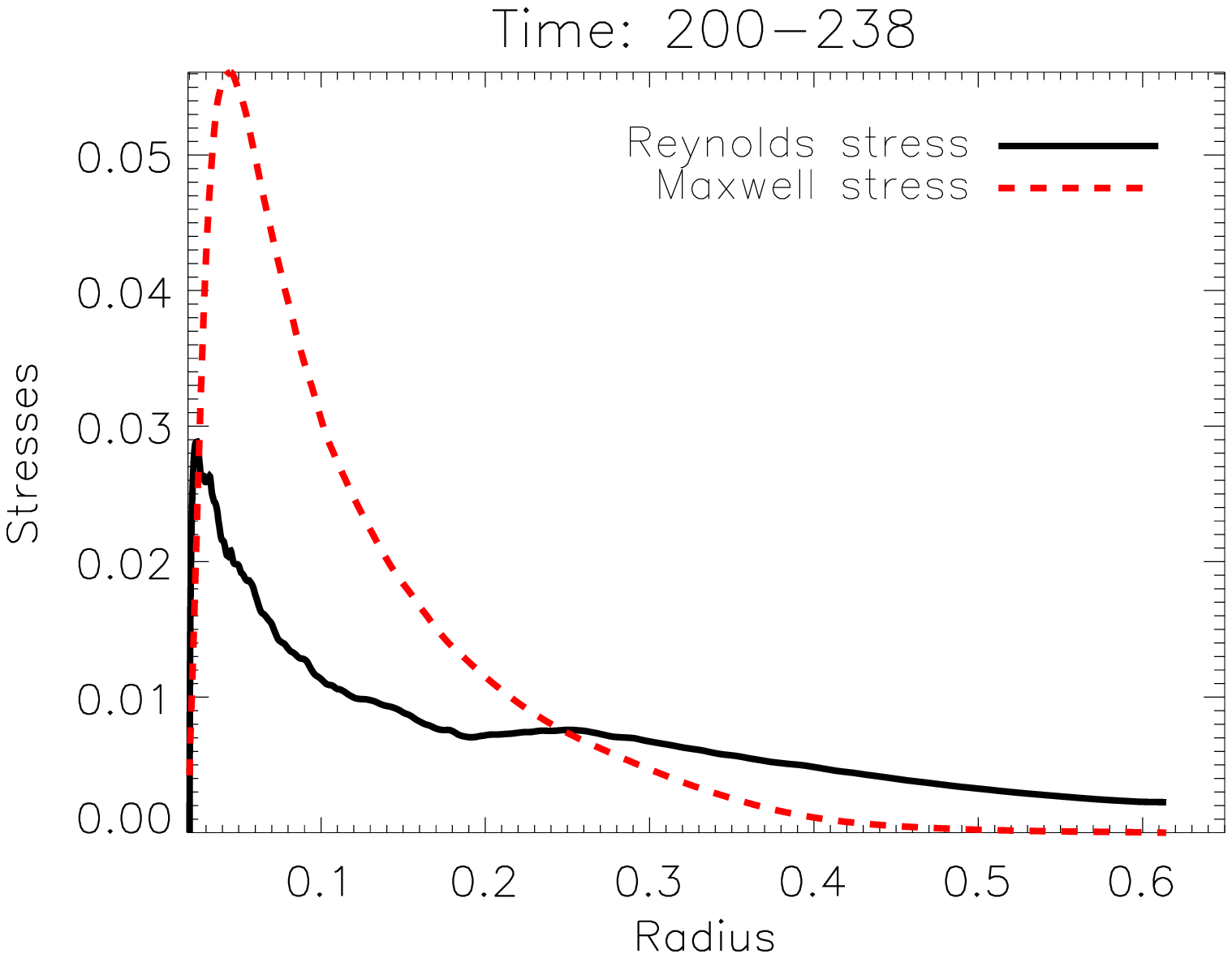}
\caption{Comparison of the Reynolds stress and the Maxwell stress in the fiducial model ``Bz-Mach10-$\beta$100-res1" as an indicator of the relative importance of spiral shocks and MRI in driving angular momentum transport. Both quantities are volume- and time- averaged over $t=200-238$.}
\label{fig:Bz_Mach10_res2_stress}
\end{figure}

To compare the relative importance of spiral shock and MRI in driving angular momentum transport, we compare the Reynolds stress and the Maxwell stress in Figure \ref{fig:Bz_Mach10_res2_stress}. The profiles are volume-averaged in azimuthal and vertical directions and time-averaged over $t=200-238$. Both stresses are peaked near the inner edge due to the more concentrated density and magnetic field there. Within $R<0.25$ where the majority of the disk resides, the Maxwell stress is greater than the Reynolds stress with a ratio $T_M/T_R \sim 3$. Previous investigations about MRI in isolated disks around a single gravitational source have found the Maxwell stress always dominates the Reynolds stress where the Reynolds stress is from the turbulence generated by MRI. The ratio of Maxwell stress to Reynolds stress in previous simulations is $4-6$ \citep{1995Hawley,1996Stone,1999Hawley,2008Blackman,2012Sorathia,2016Shi}. In our CV disk models, the Reynold stress also comes from the spiral shocks, which makes it closer to the Maxwell stress than conventional MHD models. This indicates that the role of spiral shocks in driving angular momentum transport in CV disks cannot be neglected.

\section{Effects of Seed Field Strength}
\label{sec:compare_beta}

The strength of seed magnetic field or the magnetization of the disk is a potential factor to affect angular momentum transport in the disk driven by MRI. The magnetization is usually measured with $1/\beta$ where $\beta = \rho c_s^2 / (1/2 B^2)$ is the ratio of gas pressure to magnetic pressure. Previous MHD simulations found that the initial onset of MRI is faster when $\beta$ is smaller because the growth rate of MRI at a specific wavelength longer than $\lambda_{MRI}$ is proportional to the Alfv{\'e}n speed $v_A$ \citep{2013Hawley}. They also found the effective viscosity parameter $\alpha_{eff}$ during quasi-steady states is proportional to the magnetization $1/\beta$. In CV systems, observational data about the strength of magnetic field is still rare. Therefore, it is worth exploring the effects of varying seed magnetic field strength in CV disk simulations. 

To compare with the fiducial model, we conduct another model ``Bz-Mach10-$\beta400$-res1" which has all the same numerical parameters as the fiducial model ``Bz-Mach10-$\beta100$-res1" except that $\beta$ of the seed magnetic field in the inflow at L1 point in this model is four times of that in the fiducial model. In other words, the seed field in this model has half of the strength of that in the fiducial model. 

\begin{figure*}
\centering
\includegraphics[width=0.49\textwidth]{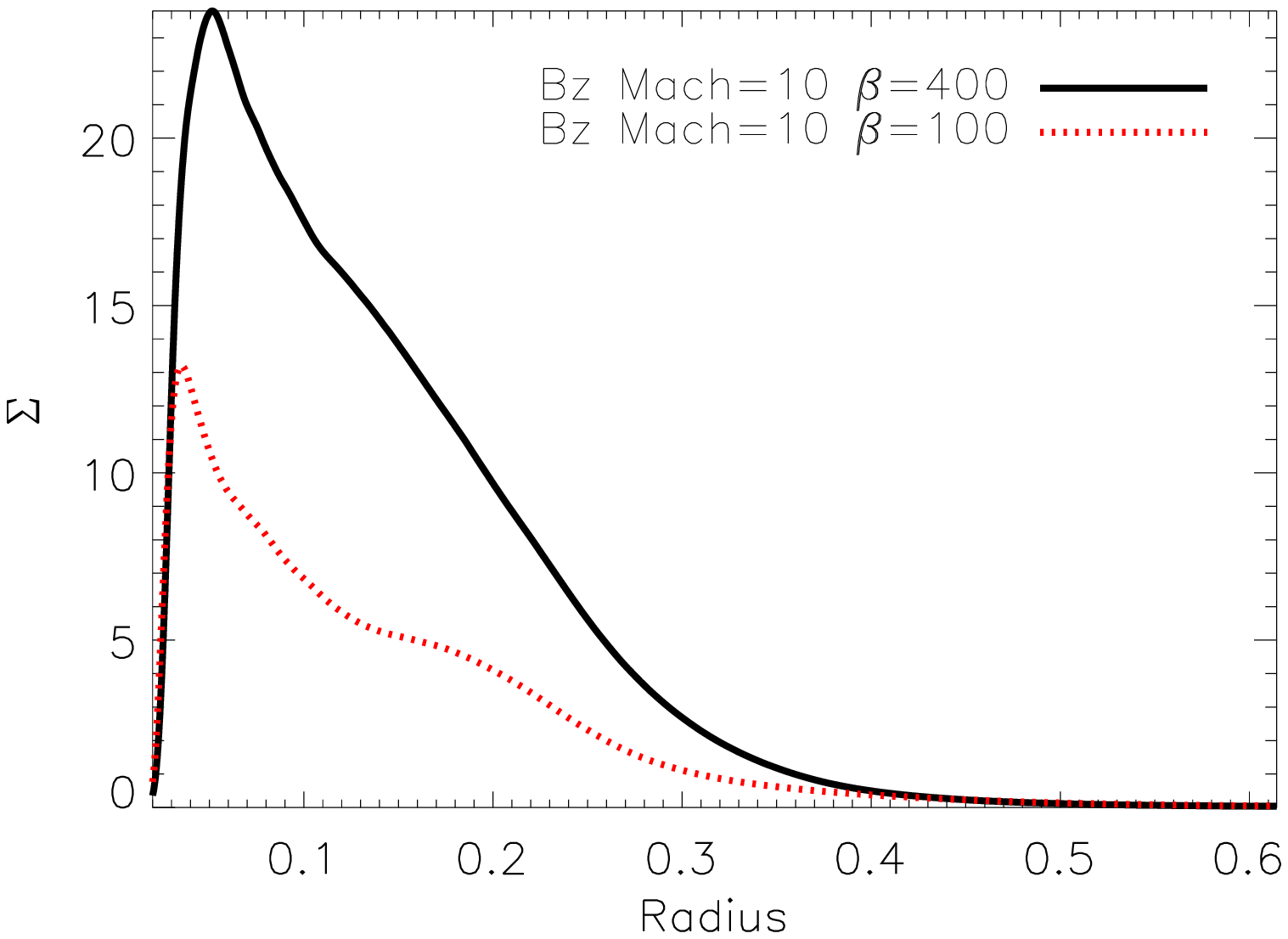}
\includegraphics[width=0.49\textwidth]{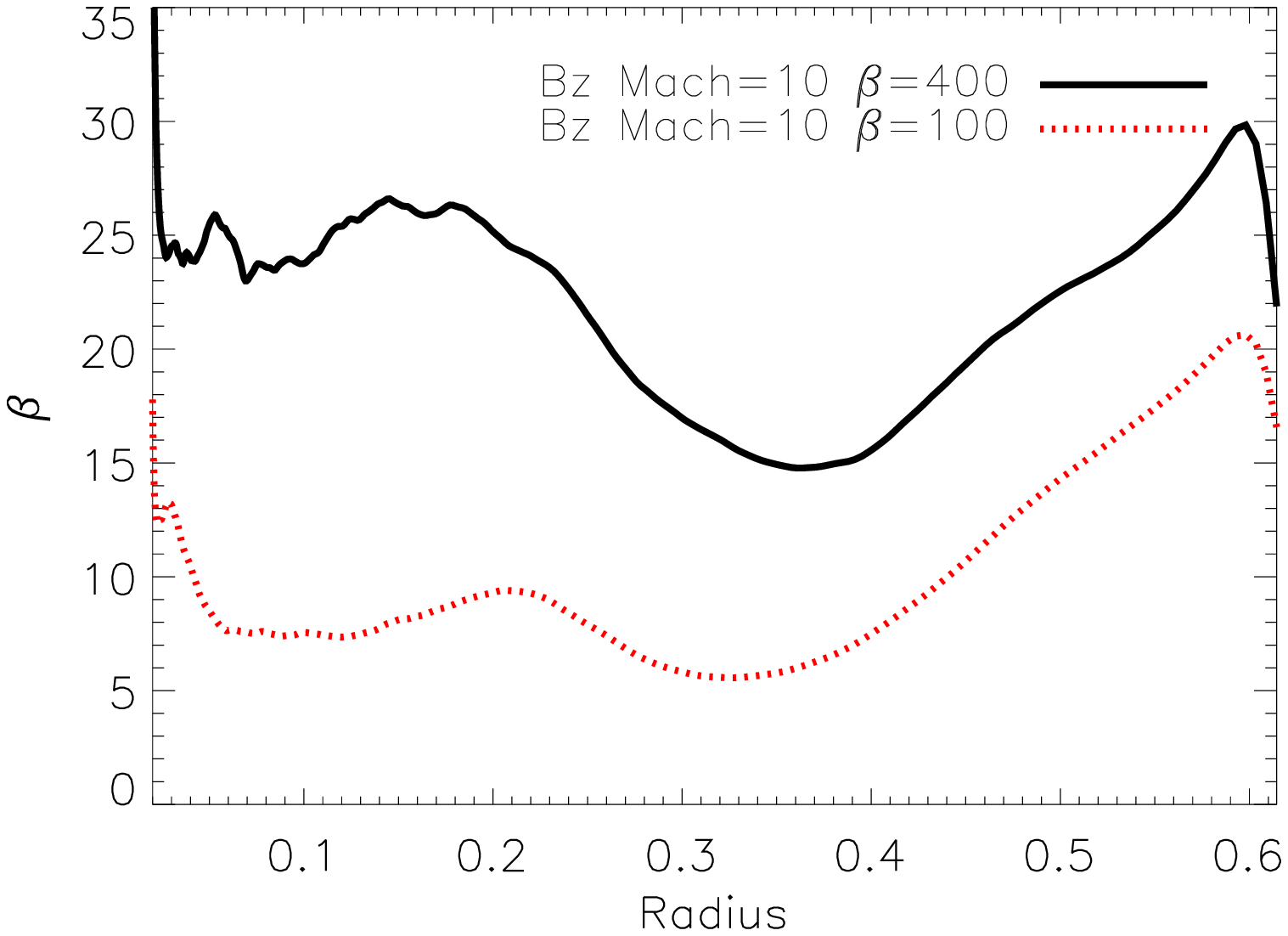}
\includegraphics[width=0.49\textwidth]{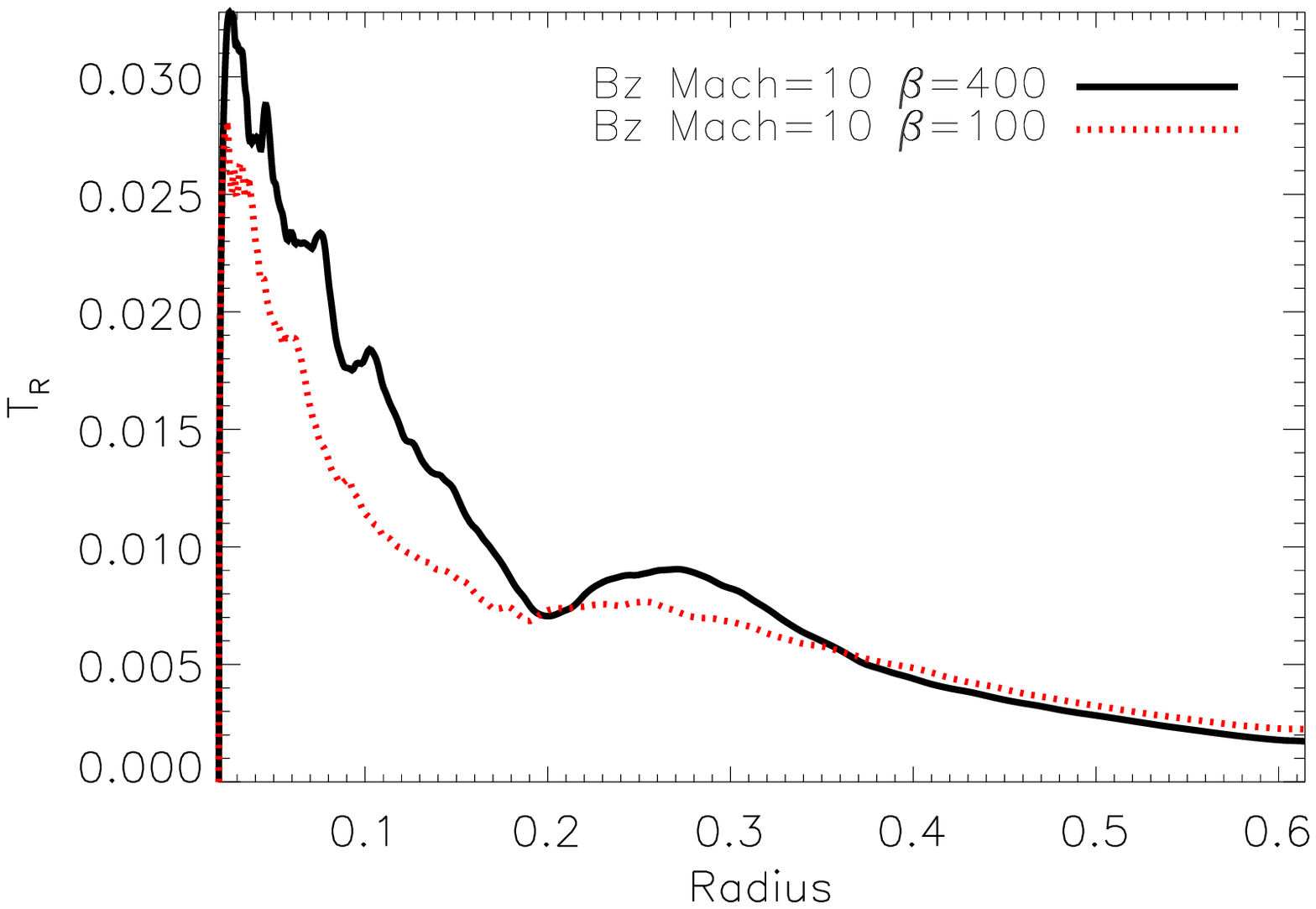}
\includegraphics[width=0.49\textwidth]{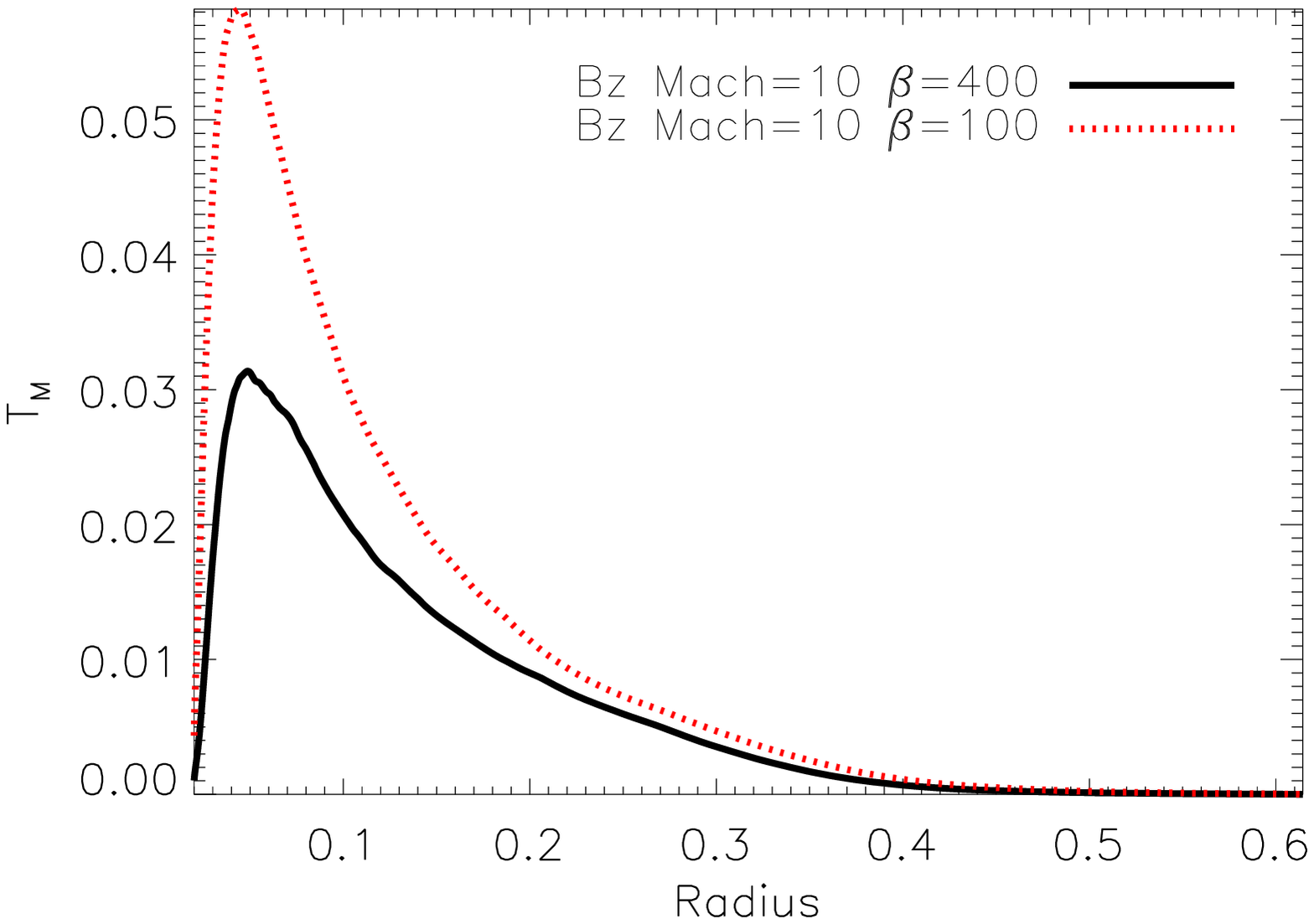}
\includegraphics[width=0.49\textwidth]{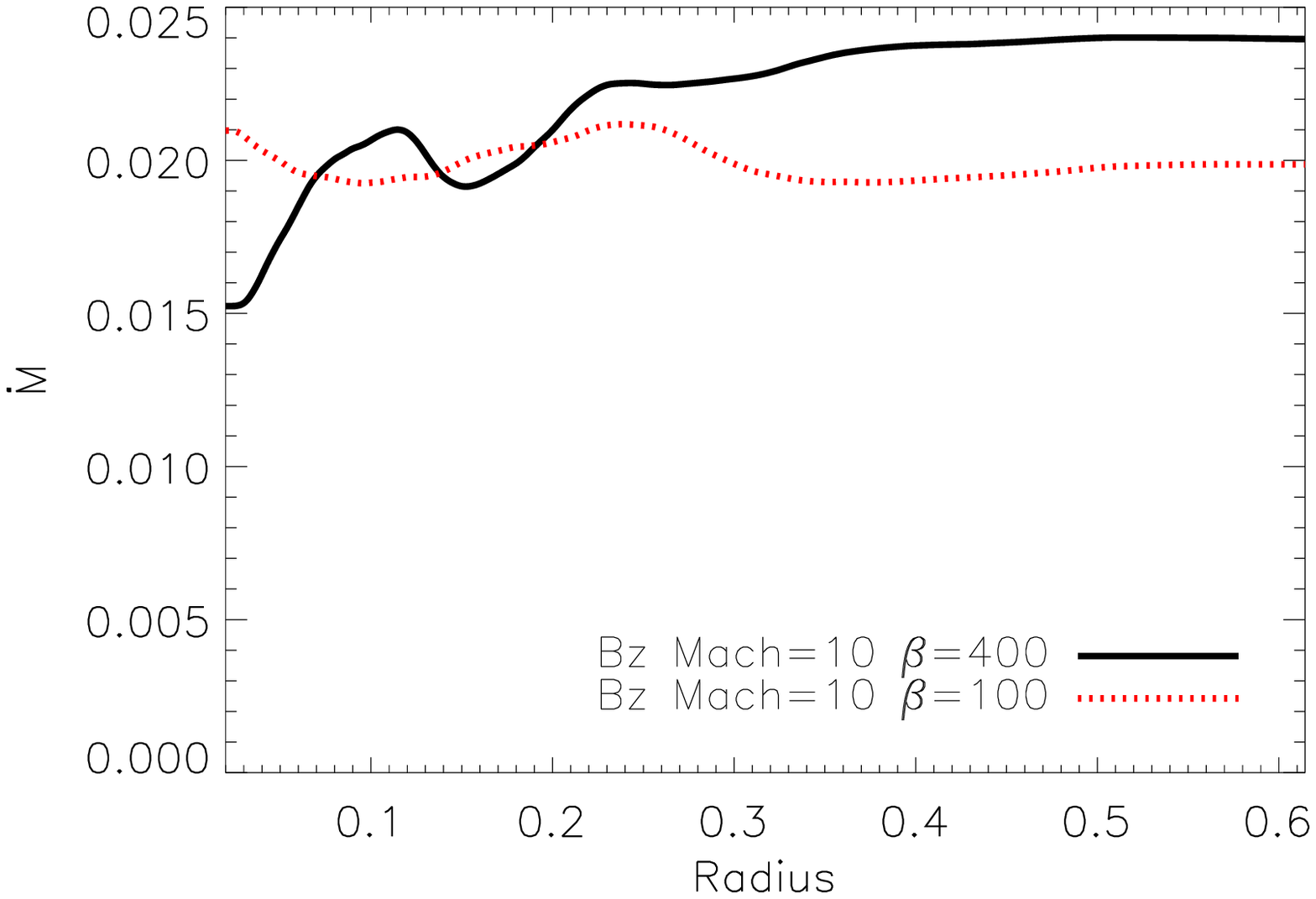}
\includegraphics[width=0.49\textwidth]{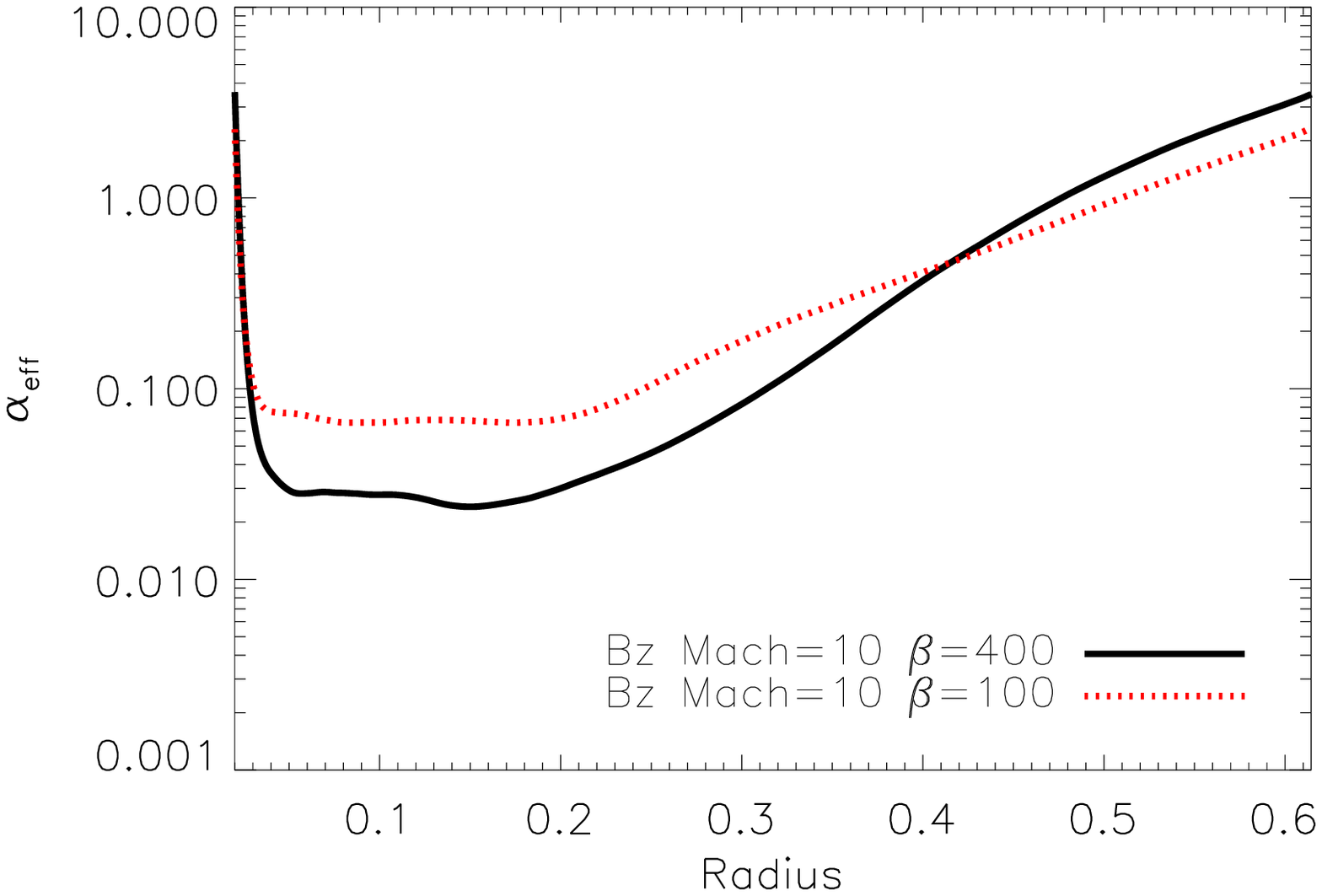}
\caption{Comparison of effect of seed magnetic field strength. Characteristic properties of the disk in the fiducial model ``Bz-Mach10-$\beta400$-res1" (solid black lines) and the model ``Bz-Mach10-$\beta100$-res1" (dotted red lines): the surface density $\Sigma$ (top row left), the plasma $\beta$ (top row right), the Reynolds stress $T_R$ (middle row left), the Maxwell stress $T_M$ (middle row right), the mass accretion rate $\dot{M}$ (bottom row left) and the effective viscosity parameter $\alpha_{eff}$ (bottom row right). All quantities are volume averaged and time averaged over $t=200-238$.}
\label{fig:M10_compare_beta}
\end{figure*}

In Figure \ref{fig:M10_compare_beta} we compare the characteristic diagnostics of the ``Bz-Mach10-$\beta100$-res1" model and the ``Bz-Mach10-$\beta400$-res1" model, the surface density $\Sigma$, the plasma $\beta$, the Reynolds stress $T_R$, the Maxwell stress $T_M$, the mass accretion rate $\dot{M}$ and the effective viscosity parameter $\alpha_{eff}$, all of which are volume- and time- averaged over $t=200-238$. During this time period, both models have reached steady state. The magnetic field of the model ``Bz-Mach10-$\beta400$-res1" is amplified from $\beta = 400$ in the L1 inflow stream to $\beta \sim 25$ in the steady-state disk, which is to be compared with the fiducial model ``Bz-Mach10-$\beta100$-res1" where the magnetic field is amplified from $\beta = 100$ in the L1 inflow stream to $\beta \sim 8$ in the steady-state disk. Correspondingly, the Maxwell stress $T_M$ in the ``Bz-Mach10-$\beta100$-res1" model is greater than the the ``Bz-Mach10-$\beta400$-res1" model by a factor of $1.5-2$ which indicates more efficient angular momentum transport driven by MRI. The surface density of the steady-state disk in the ``Bz-Mach10-$\beta100$-res1" model is lower than that in the fiducial model by a factor of $\sim 1.5$ due to the different accretion history: MRI in the ``Bz-Mach10-$\beta100$-res1" model has a faster growth rate, so the disk reaches steady state earlier, whereas the ``Bz-Mach10-$\beta400$-res1" model has a longer phase of mass pile-up before reaching steady state. The Reynolds stresses of the two models are comparable with the stress in the ``Bz-Mach10-$\beta100$-res1" model being slightly smaller due to the smaller surface density. The mass accretion rates of the two models are comparable especially in the disk area within $R<0.25$. The $\dot{M}$ value of the ``Bz-Mach10-$\beta100$-res1" model at the outer boundary is slightly lower because the stronger MRI in this model may drive more mass loss as winds through the outer boundary. The effective viscosity parameter is $\alpha_{eff} \sim 0.1$ in the ``Bz-Mach10-$\beta100$-res1" model which is larger than that in the ``Bz-Mach10-$\beta400$-res1" model by a factor of 1.5 because this model has the same mass accretion rate as the fiducial model but lower surface density.

\begin{figure}
\centering
\includegraphics[width=0.48\textwidth]{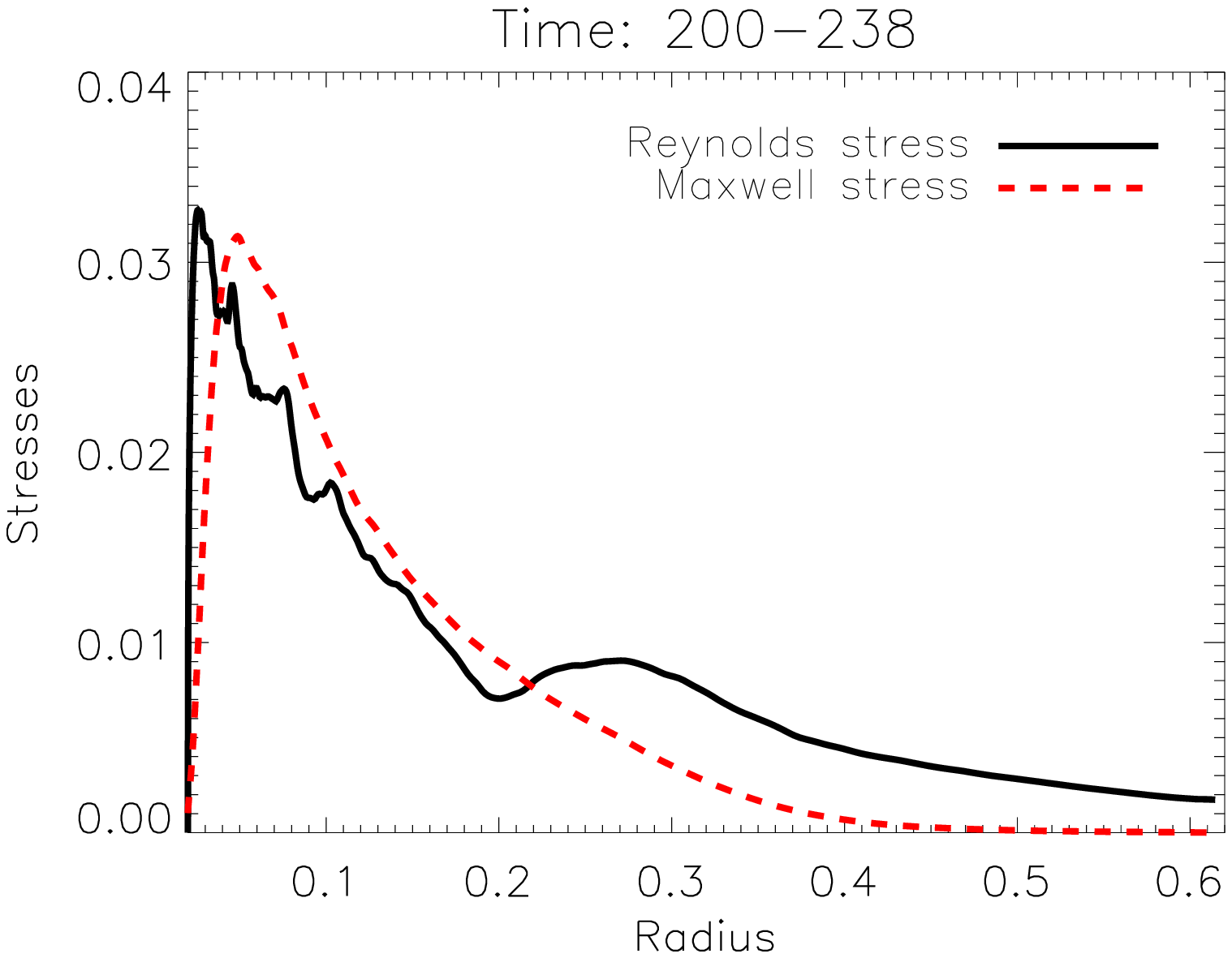}
\caption{Comparison of the Reynolds stress and the Maxwell stress in the model ``Bz-Mach10-$\beta$400-res1" as an indicator of the relative importance of spiral shocks and MRI in driving angular momentum transport. Both quantities are volume- and time- averaged over $t=200-238$.}
\label{fig:Bz_Mach10_beta400_stress}
\end{figure}

In Figure \ref{fig:Bz_Mach10_beta400_stress} we compare the Reynolds stress and the Maxwell stress in the ``Bz-Mach10-$\beta$400-res1" model. In this model, the two stresses are comparable in the majority of the disk which indicates that the spiral shocks and the MRI have comparable importance in driving angular momentum transport when the seed filed has $\beta \sim 400$ in a ``Mach10" disk. Compared with the fiducial ``Bz-Mach10-$\beta$100-res1" model, since MRI is stronger and drives more efficient angular momentum transport when $\beta$ of the seed magnetic field is smaller, i.e., the inflow gas is more strongly magnetized, MRI plays a more important role than spiral shocks when the seed field has $\beta < 400$. 


What is a plausible range for the value of $\beta$ in the plasma flowing from the donor star in CV systems? The closest analogy to the surface activity of donor stars in CVs is the surface activity of the Sun. The plasma-$\beta$ of the solar photosphere is found to have a wide range from $\sim 1$ at a sunspot to a few$\times 10^2$ in a plage region (\citealt{2001Gary}; see \citealt{2014Wiegelmann} for review). Recently, \citet{2015Wiegelmann} conducted a magneto-static magnetic field model for the solar atmosphere using high-resolution photospheric magnetic field measurements from SUNRISE/IMaX as boundary condition, and found the horizontally averaged $\beta \sim 1-8 \times 10^2$ in the photosphere of the Sun (see their Figure 3).

\section{Effects of Seed Field Geometry}
\label{sec:compare_geometry}


\citet{2007King} have questioned whether MRI can explain angular momentum transport in accretion disks at all, given some local shearing-box MHD simulations gave $\alpha_{eff} \lesssim 0.01$ while observations of dwarf novae outbursts require $\alpha_{eff} \sim 0.1-0.3$ during outbursts. They especially selected shearing-box simulations with zero net vertical magnetic flux due to the concern that simulations with a superimposed net vertical field tend to yield estimates of $\alpha_{eff}$ larger by an order of magnitude than those which do not according to the local simulations by \citet{1995Hawley}. Actually, more recent global MHD simulations have reported that $\alpha_{eff}$ could be of order $0.1$ even for models with zero net vertical flux although it can only be achieved during the transient phase of MRI growth \citep{2001Hawley-Krolik,2012Sorathia,2013Hawley}. The general conclusions from the  various MHD simulations, both local and global, that studied the effects of seed field geometry to MRI \citep{1995Hawley,1996Stone,2004Sano,2012Sorathia,2013Hawley,2016Shi} are: the models with initial toroidal field have a longer initial MRI growth phase before MRI turbulence saturates than the models with initial vertical field; the ratio of the Maxwell stress to the gas pressure, i.e. the efficacy of angular momentum transport driven by MRI, is up to one order of magnitude larger in the initial vertical field models than in the initial toroidal field models upon saturation, but has much smaller differences in long-term average of the quasi-steady states of global MHD models \citep{2012Sorathia,2013Hawley}. Here we present the global MHD model of CV disks with zero vertical seed field to explore the effects of seed magnetic field geometry to the steady-state behavior of MRI. Compared with previous studies, one major difference of our study is that it is the geometry of the seed magnetic field that flows in from the L1 point rather than the initial field geometry in the disk that dominates the evolution of MRI. The initial magnetic field can be washed out within a viscous timescale.

\subsection{The ``Loop-Mach10-$\beta400$-res1" Model }
\label{subsec:loop_model}

The model ``Loop-Mach10-$\beta400$-res1" has parameters as listed in Table \ref{tab:parameter}. The disk starts with uniform density and toroidal magnetic field with azimuthal wavenumber $M_c=20$ (see \S \ref{sec:IC_BC} for details of the initial and boundary conditions). No initial artificial perturbation is provided since the binary gravitational is natural perturber in the velocity and density field. The seed magnetic field flows with the gas stream into the computational domain through the L1 point as field loops in the disk plane with zero vertical component (Eq.\ref{eq:seed_loop}). The strength of the seed field is set by $\beta=400$, the same strength with the ``Bz-Mach10-$\beta400$-res1" model. The simulations runs to $t=260$ which covers $\sim 625$ Keplerian orbits at $R= 0.15$ (middle area of the disk) and $2-5$ viscous timescales.

\begin{figure*}
\centering
\includegraphics[width=0.49\textwidth]{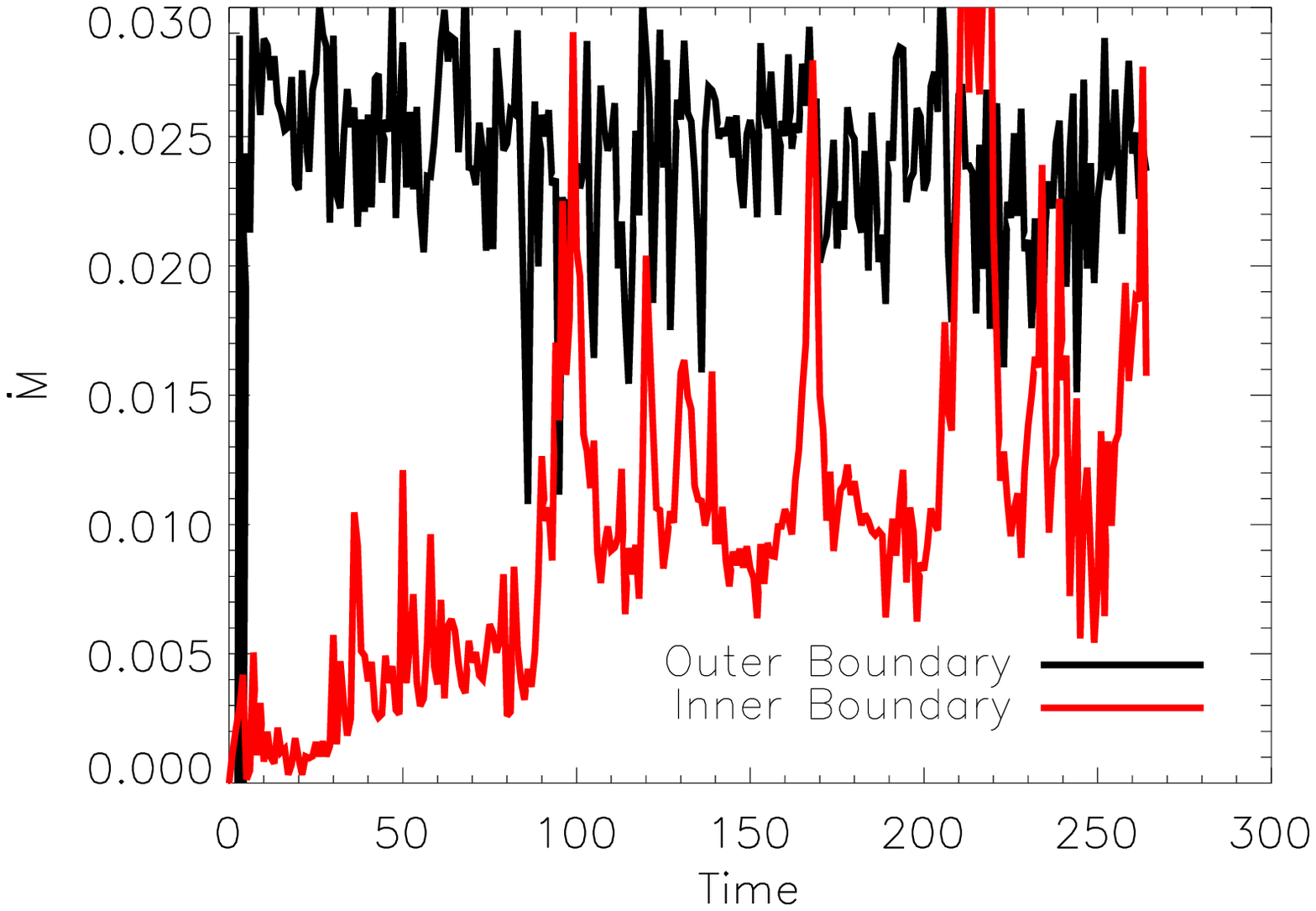}
\includegraphics[width=0.49\textwidth]{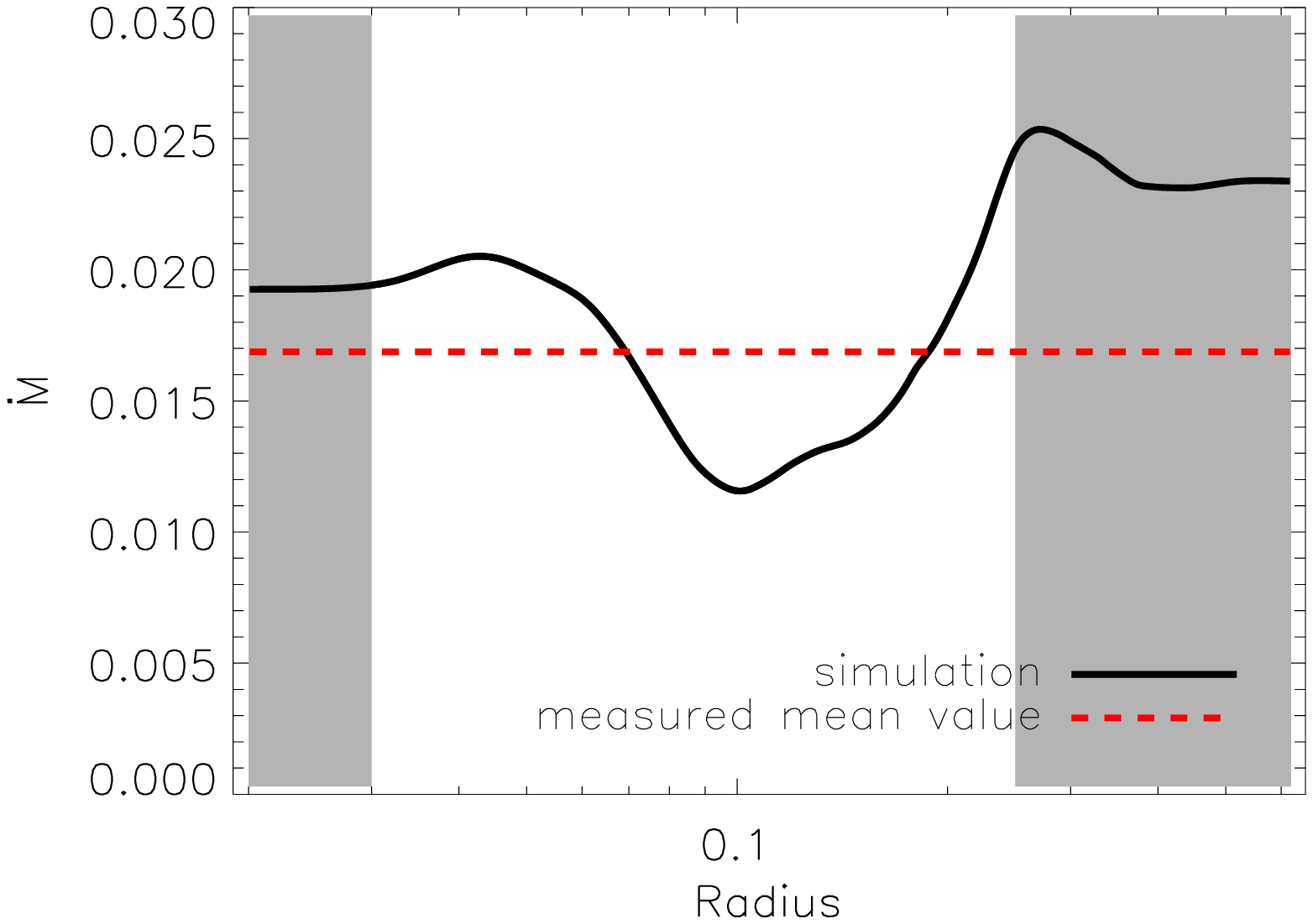}
\includegraphics[width=0.49\textwidth]{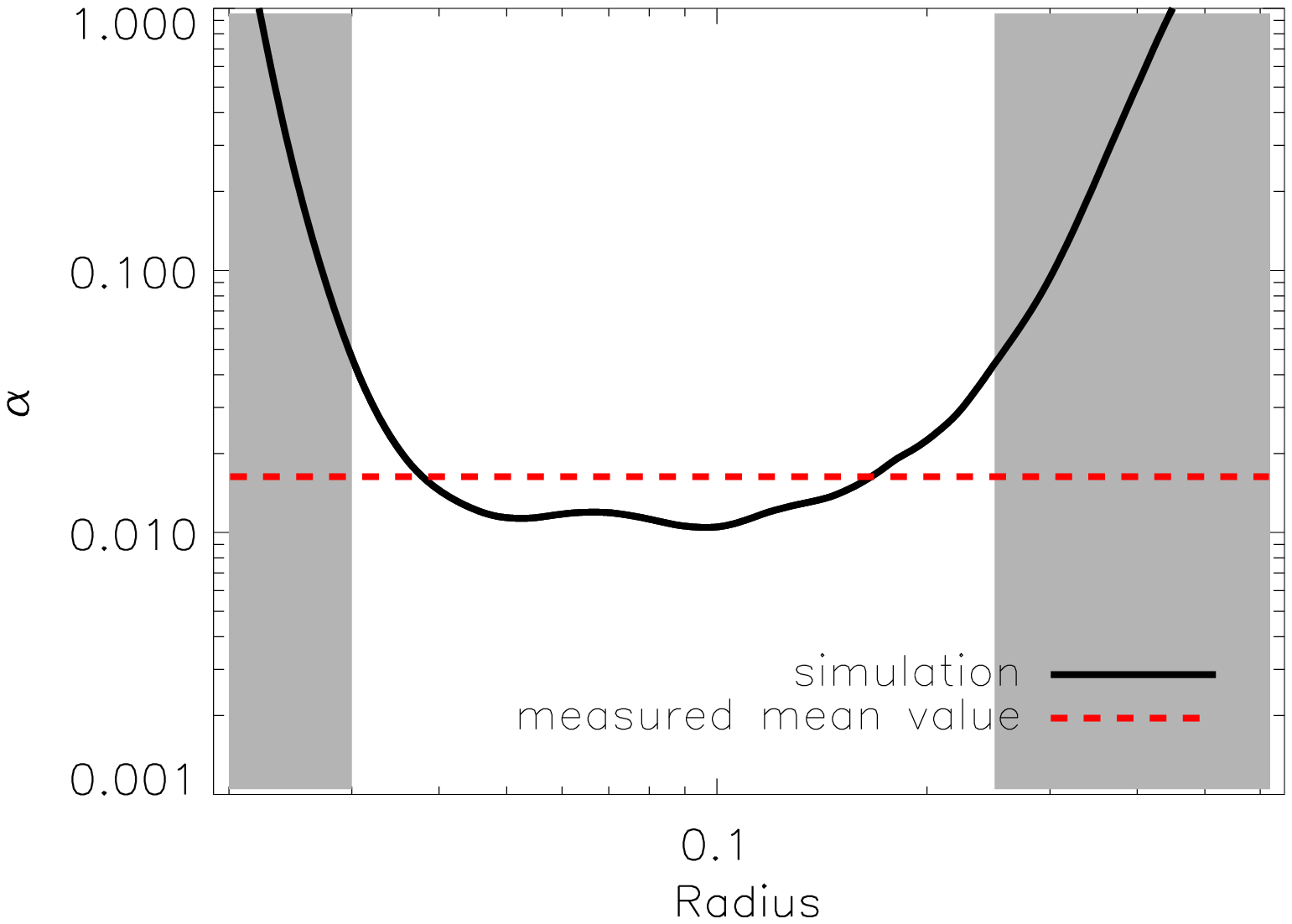}
\includegraphics[width=0.49\textwidth]{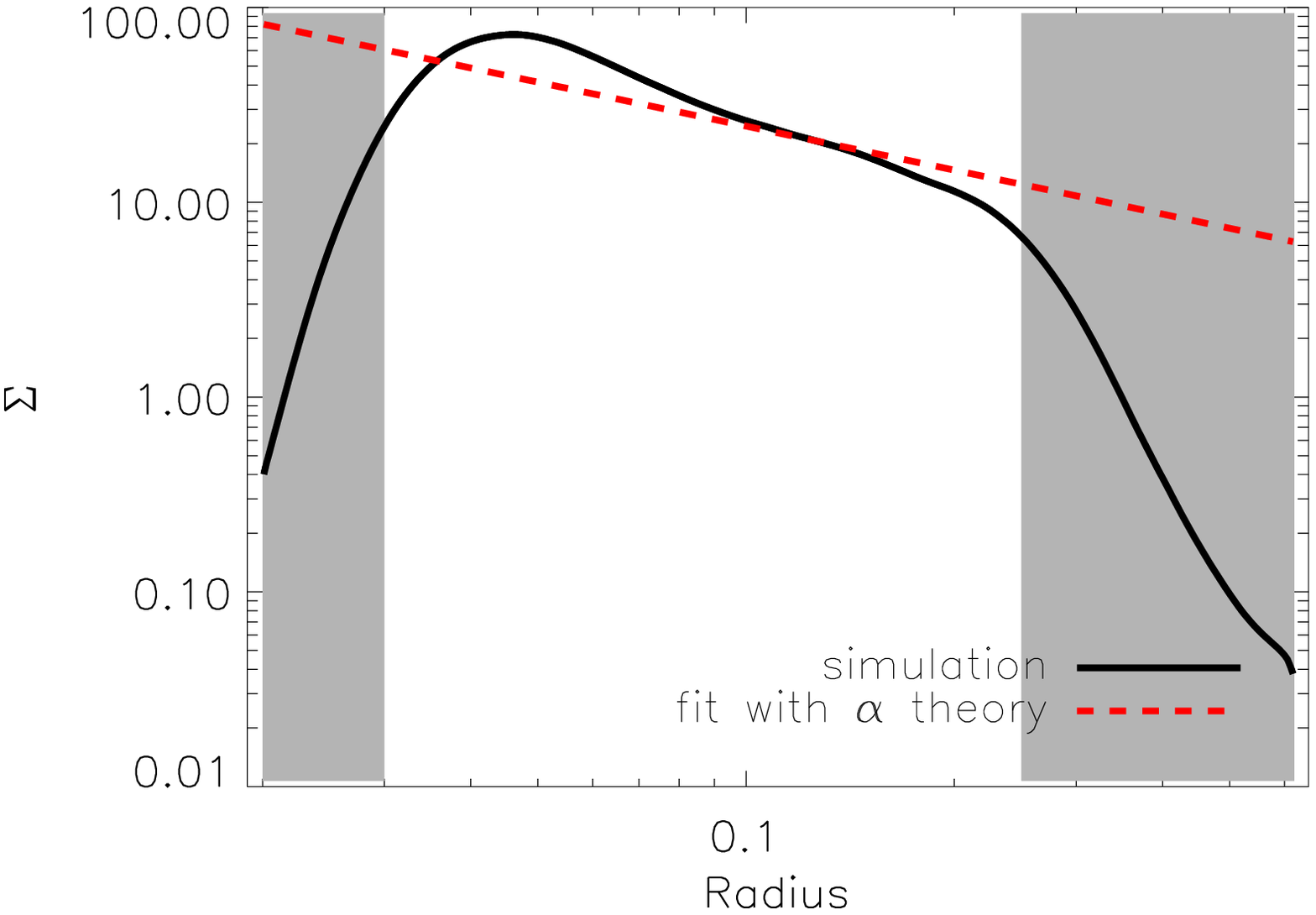}
\caption{{\it Top left}: time evolution of mass accretion rates at inner (red line) and outer (black line) boundaries for the model ``Loop-Mach10-$\beta400$-res1". {\it Other three panels}: fitting the model ``Loop-Mach10-$\beta400$-res1" using the steady-state $\alpha$ disk theory. The black lines are the radial profiles of the mass accretion rate $\dot{M}$ (top right), $\alpha_{eff}$ (bottom left), and the surface density $\Sigma$ (bottom right), all of which are volume- and time- averaged over $t=200-238$. Similar to Figure \ref{fig:Bz_Mach10_res2_fit_with_alpha_theory}, the area at $R>0.25$ is dominated by the inflow stream and the area at $R<0.03$ is affected by the inner boundary conditions, so they are both shaded for less attention. Mean values of $\dot{M}$ and $\alpha_{eff}$ (dashed red lines) are measured within the unshaded area $0.03<R<0.25$ where the majority of the disk resides. The predicted profile of $\Sigma$ according standard $\alpha$ disk theory is calculated as $\Sigma = \overline{\dot{M}} / 3 \pi \nu$ ($\nu=\overline{\alpha_{eff}} c_s H$ is the kinematic viscosity), which is the dashed red line in the bottom right panel.}
\label{fig:Loop_Mach10_res2_fit_with_alpha_theory}
\end{figure*}

In the first panel of Figure \ref{fig:Loop_Mach10_res2_fit_with_alpha_theory} we show the time-history of the mass accretion rate at the inner and outer boundaries of this model. Similar to the ``Bz-Mach10-$\beta400$-res1" model, initially the accretion rate at the inner boundary is much smaller than the mass supply rate at the outer boundary, thus the mass as well as the magnetic field piles up in the disk. As the spiral shocks and MRI are developed and grow stronger due to the increase of gas density and magnetic field strength in the disk, the inner mass accretion rate gradually catches up with the outer mass supply rate. The growth of the inner mass accretion rate is much slower than that in the ``Bz-Mach10-$\beta400$-res1" model, which is because MRI grows much slower when there is zero vertical component in the seed field.  While the outer mass supply rate is quite steady, the inner mass accretion rate has eruptive features. 

Despite the eruptive mass accretion history, the disk reaches steady state if examined over a longer period of time. Similar to the steady-state analysis of the fiducial model in \S \ref{sec:fiducial_model}, here we also fit the surface density profile using the standard $\alpha$ theory. In the last three panels of Figure \ref{fig:Loop_Mach10_res2_fit_with_alpha_theory}, we plot in solid black lines the radial profiles of the mass accretion rate $\dot{M}$, $\alpha_{eff}$ and the surface density $\Sigma$ which are time- and volume- averaged over $t=200-238$. We measure the mean values of $\dot{M}$ and $\alpha_{eff}$ within the radial range $R \in [0.03,0.25]$ where the majority of the disk resides and mark them as red dashed lines. The measurement gives $\overline{\dot{M}} = 0.017$ and $\overline{\alpha_{eff}}=0.016$. Plugging the mean values into the standard $\alpha$ theory $\Sigma = \overline{\dot{M}} / 3 \pi \nu$ where $\nu=\overline{\alpha_{eff}} c_s H$ is the kinematic viscosity, it gives the red dashed line in the last panel of Figure \ref{fig:Loop_Mach10_res2_fit_with_alpha_theory}. It is in general a good fit to the black line except the simulation profile of $\Sigma$ is slightly steeper than the theoretical prediction. The small discrepancy is because the disk is not in perfect steady state, i.e., the radial profile of the mass accretion rate is not a perfectly flat. In fact, due to the self-consistent cyclic evolution of the gas and magnetic field, a steady state is reached only in long term averages, which is equivalent to a standard thin disk with $\dot{M} = 0.017$ and $\alpha_{eff}=0.016$. Compared with the fiducial model in Figure \ref{fig:Bz_Mach10_res2_fit_with_alpha_theory}, the averaged $\dot{M}$ of the current model ``Loop-Mach10-$\beta400$-res1" is 90\% of that of the fiducial model, and the averaged $\alpha_{eff}$ of the current model is 50\% of that of the fiducial model. This indicates that a larger value of $\alpha_{eff}$ is favored when the seed magnetic field has vertical components. 

\begin{figure}
\centering
\includegraphics[width=0.48\textwidth]{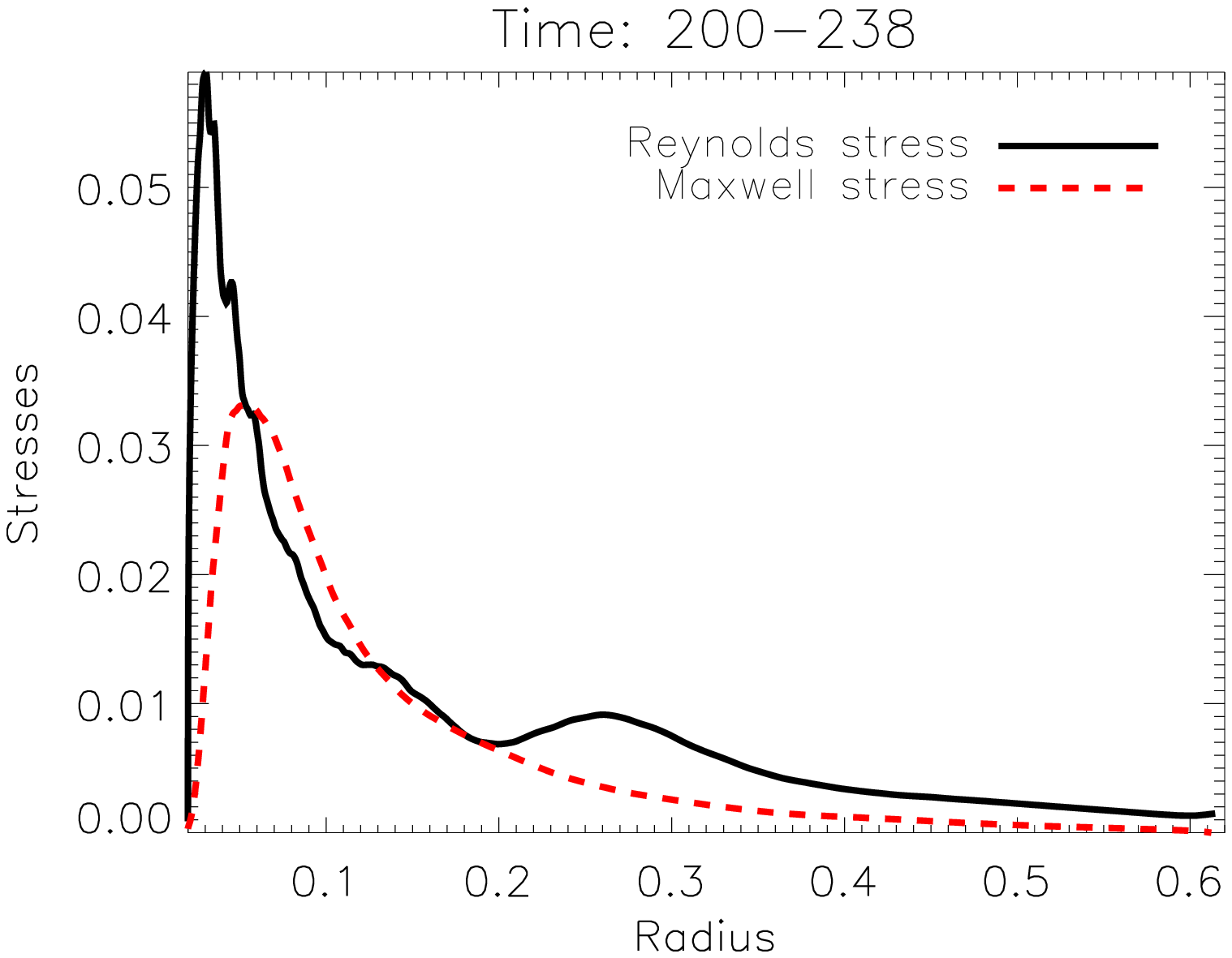}
\caption{Comparison of the Reynolds stress and the Maxwell stress in the model ``Loop-Mach10-$\beta$400-res1" as an indicator of the relative importance of spiral shocks and MRI in driving angular momentum transport. Both quantities are volume- and time- averaged over $t=200-238$.}
\label{fig:Loop_M10_res2_stress}
\end{figure}

To compare the relative importance of spiral shocks and MRI in driving angular momentum transport in this model, we plot the radial profiles of the Reynolds stress and the Maxwell stress that are volume- and time- averaged over $t=200-238$ in Figure \ref{fig:Loop_M10_res2_stress}. Similar to the ``Bz-Mach10-$\beta400$-res1" model, the Reynolds stress is comparable to the Maxwell stress in most of the disk area ($0.05<R<0.25$). Considering the Reynolds stress generated by the MRI turbulence is only $1/6$ to $1/4$ of the Maxwell stress according to previous MHD simulations, in our model, most of the Reynolds stress is due to spiral shocks. Therefore, similar to the ``Bz-Mach10-$\beta400$-res1" model, spiral shocks are as important as MRI in driving angular momentum transport in the model with zero vertical seed magnetic field. The difference of the current model from the fiducial model is that the Reynolds stress has a peak close to the inner boundary ($R<0.05$) which is higher than the Maxwell stress. This is due to the peak of surface density near the inner boundary (see \S \ref{subsec:compare_geometry} for more discussions). Due to the cyclic ``pile up - accretion" characteristic of the magnetic field, there are periods of time during the evolution when the magnetic field is so weak near the inner boundary that MRI is not resolved in that area. This results in ineffective mass accretion near the inner edge of the disk, which leads to large surface density there. 

\subsection{Comparing Seed Field Geometries}
\label{subsec:compare_geometry}

\begin{figure*}
\begin{center}
\includegraphics[width=0.3\textwidth]{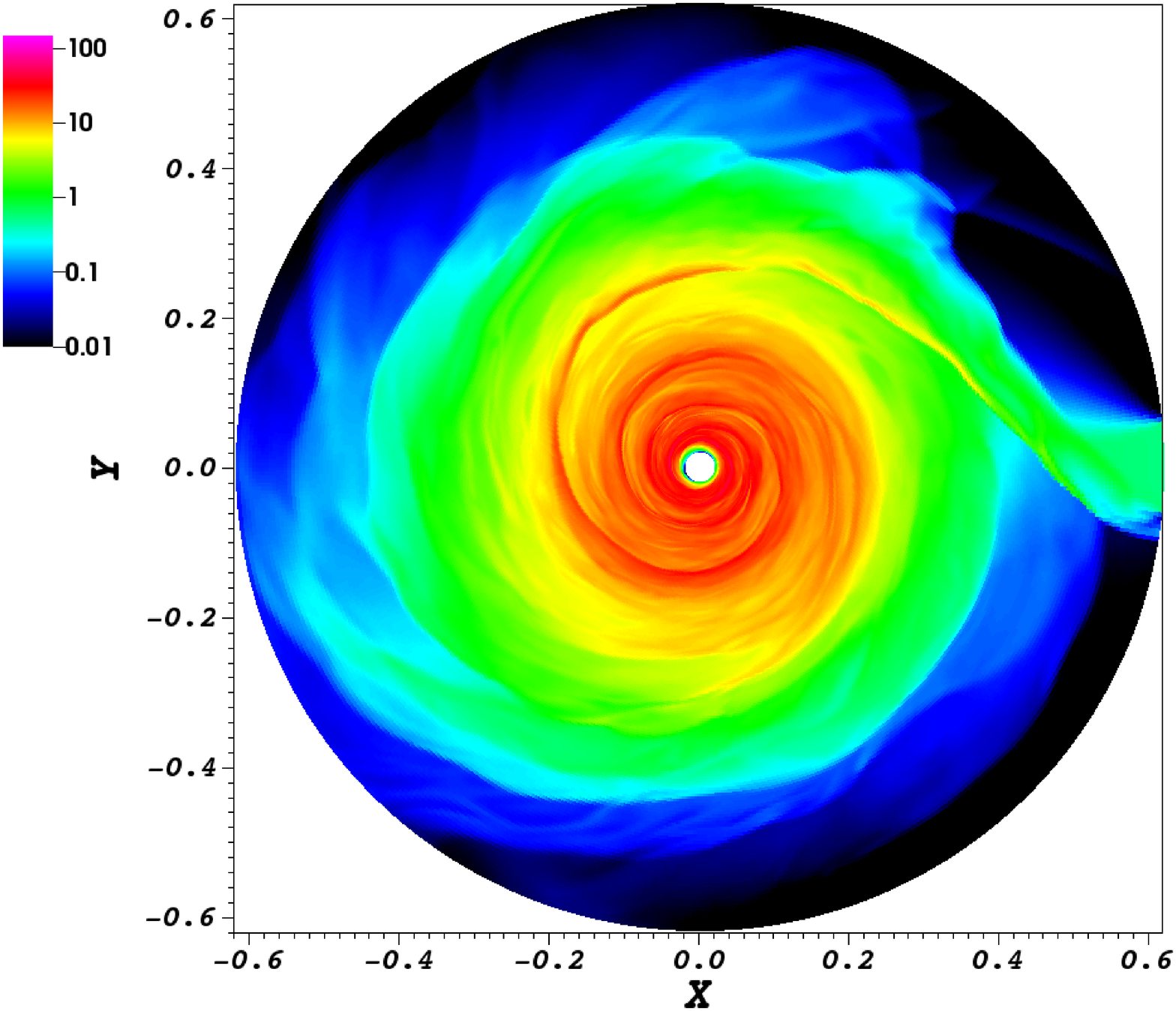}
\includegraphics[width=0.3\textwidth]{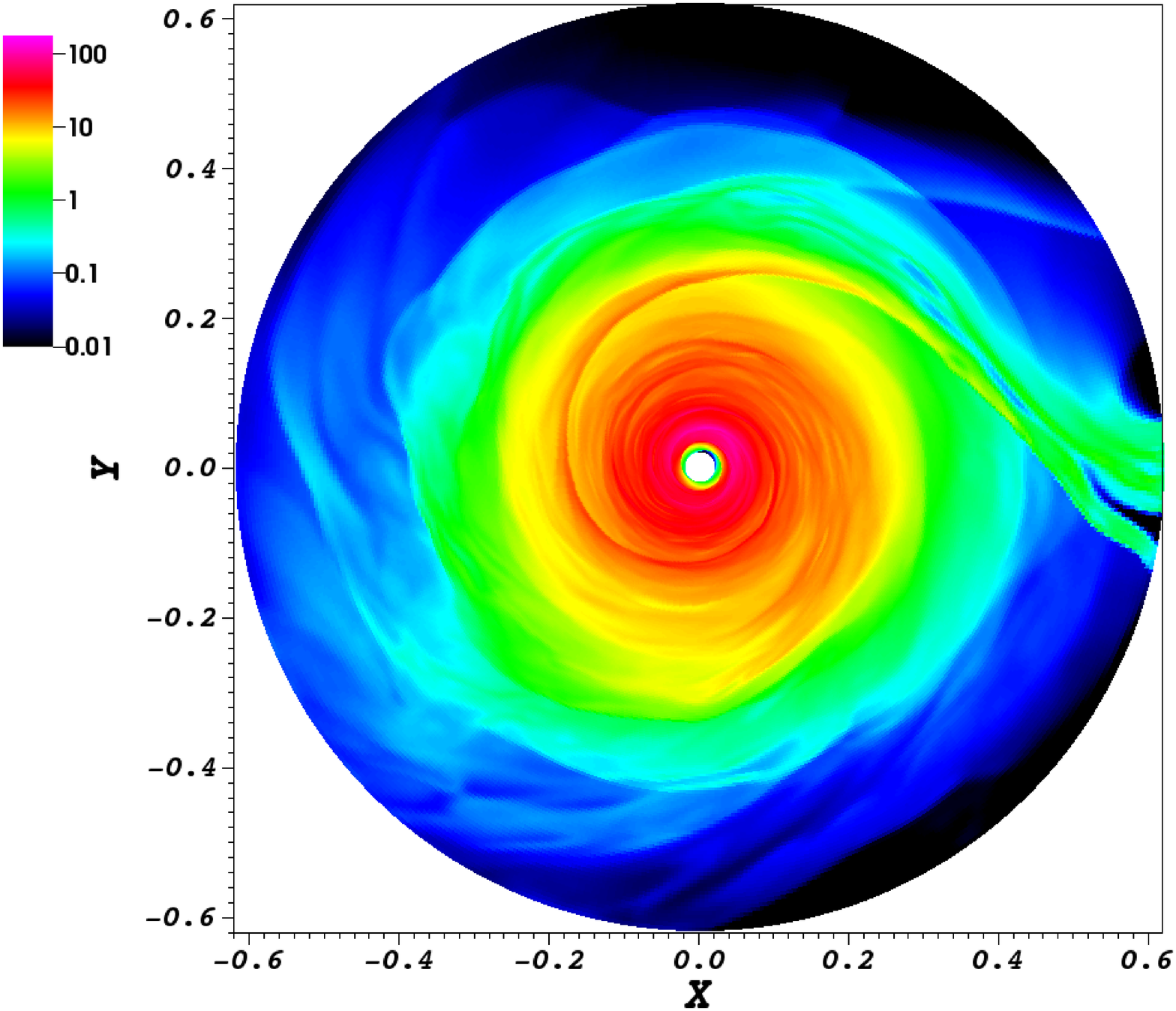}
\includegraphics[width=0.3\textwidth]{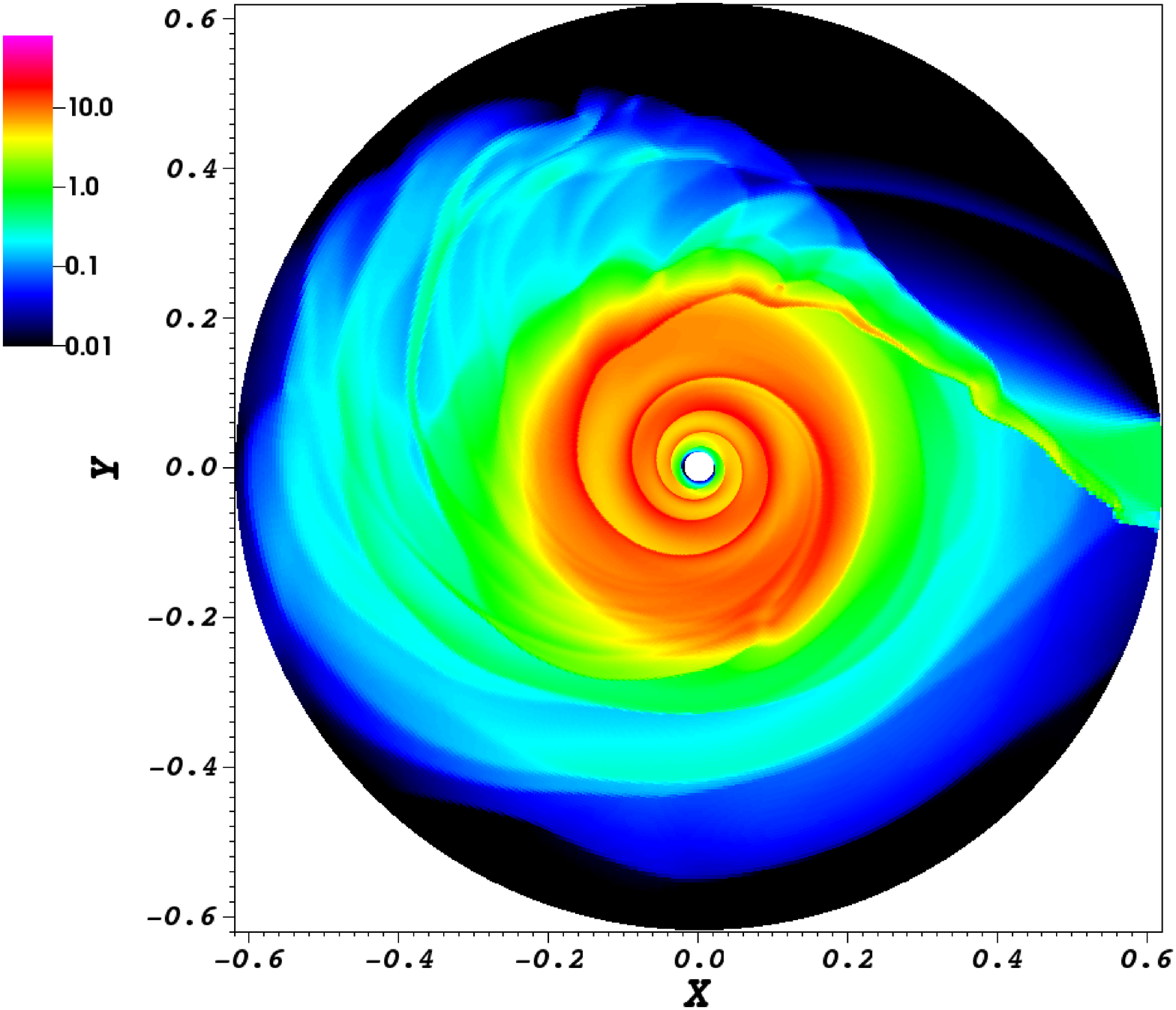}
\end{center}
\hspace{0.29in}
\includegraphics[width=0.3\textwidth]{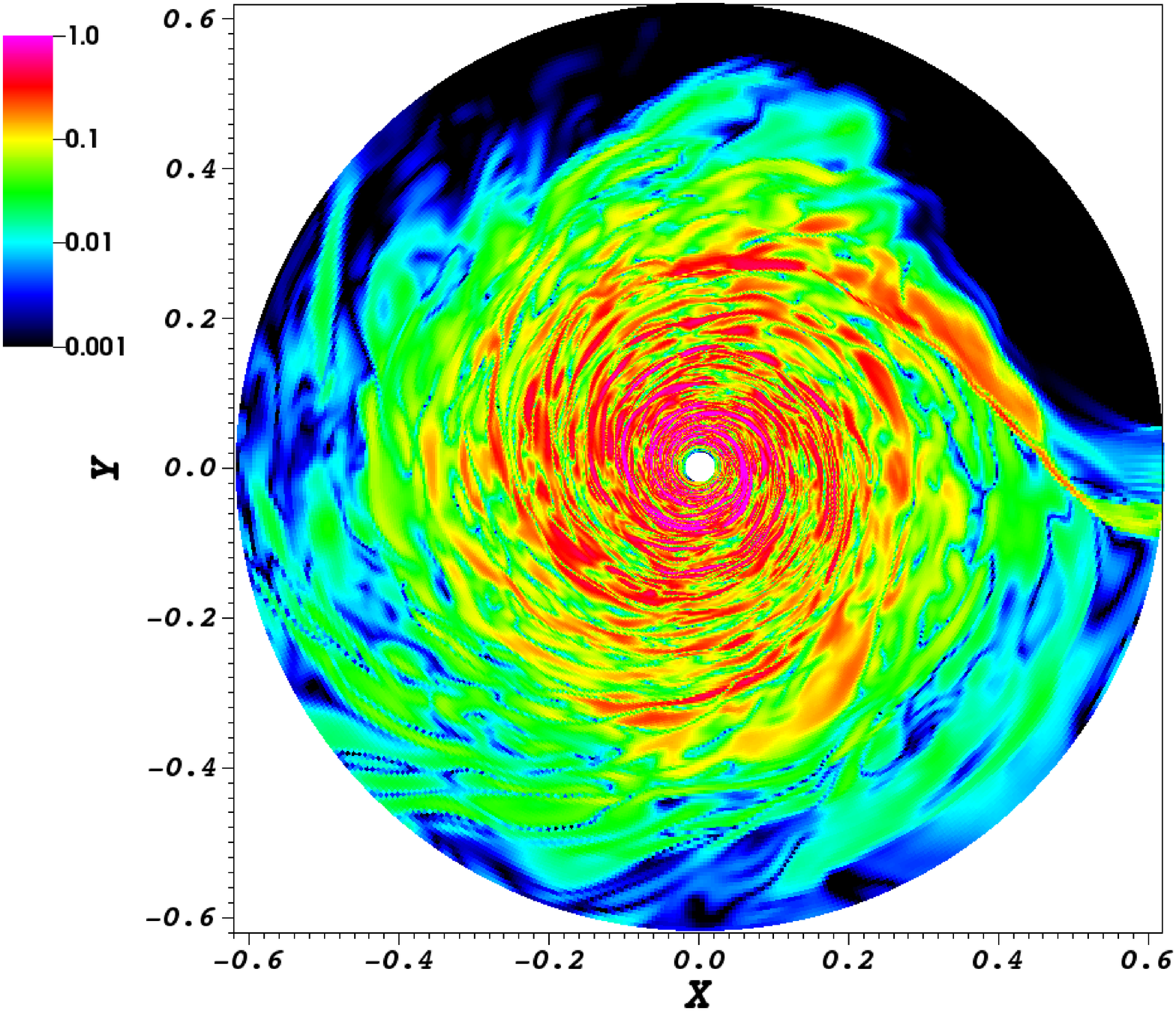}
\includegraphics[width=0.3\textwidth]{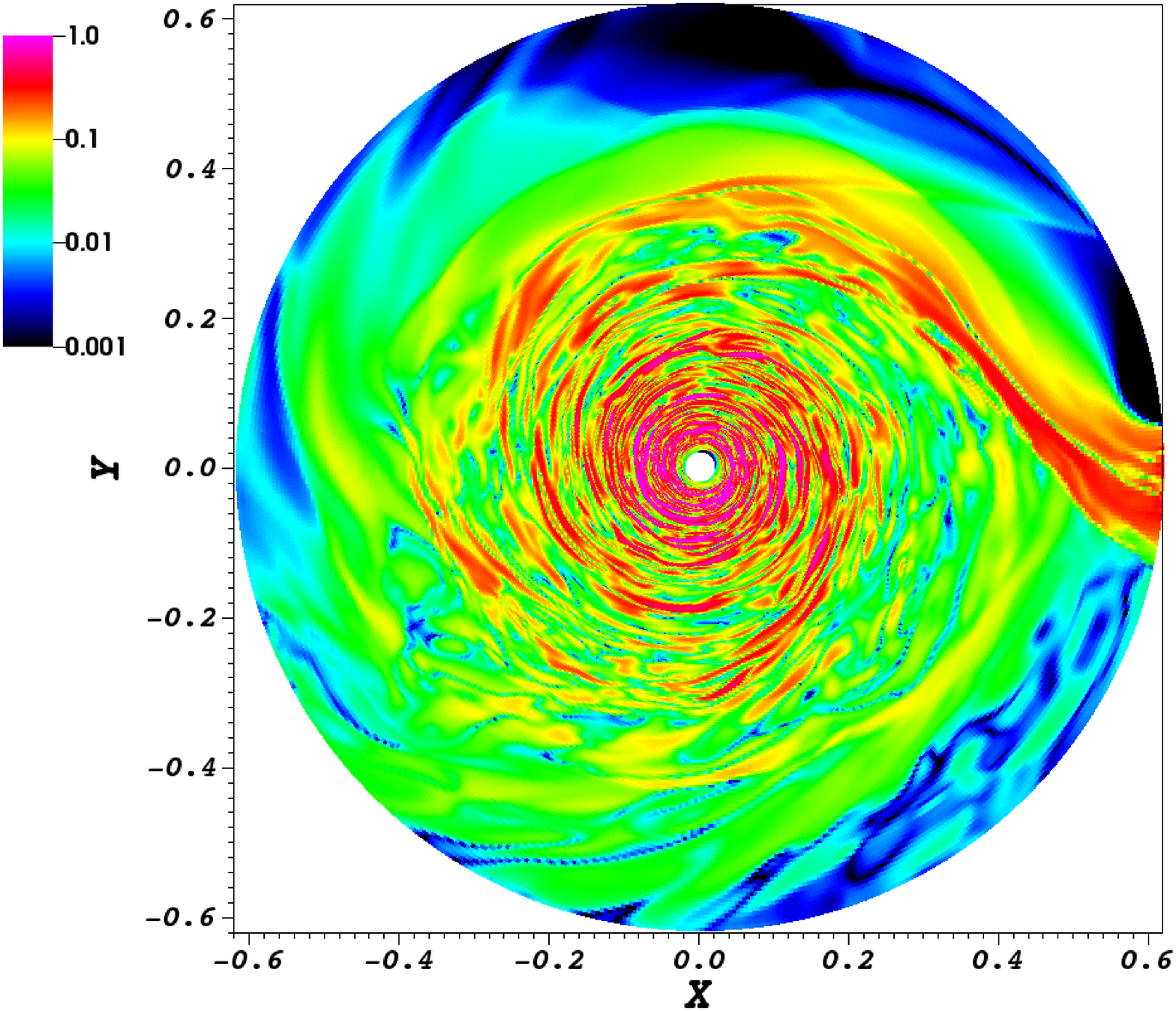}
\caption{Snapshots of the gas density (top row) and the strength of magnetic field $|{\bf B}|$ (bottom row) in the fiducial model ``Bz-Mach10-$\beta400$-res1" at $t=230$ (left column), the model ``Loop-Mach10-$\beta400$-res1" at $t=230$ (middle column), and the hydrodynamical model ``Hydro-Mach10-res1" at $t=48$ (right column).}
\label{fig:Mach10_compare_geom_2Dsnapshot_t48}
\end{figure*}

In Figure \ref{fig:Mach10_compare_geom_2Dsnapshot_t48} we show snapshots of the gas density and strength of the magnetic field for the models of ``Bz-Mach10-$\beta400$-res1" and ``Loop-Mach10-$\beta400$-res1" at $t=230$ to compare the effects of seed magnetic field geometry. From the spiral patterns in the density field and turbulence patterns in the magnetic field, there is little difference between the two models except detailed structures. In addition, to study the effects of MRI on the evolution of spiral shocks as well as the overall angular momentum transport process, we perform anther model ``Hydro-Mach10-res1" where all the initial and boundary conditions are the same with the former two MHD models except the magnetic field is zero. We show the snapshot of gas density for this hydrodynamical model at $t=48$ in the right panel of Figure \ref{fig:Mach10_compare_geom_2Dsnapshot_t48}. Compared with the MHD models, the spiral arms in the hydro model are more tightly wound. This is because the pitch angles of the spiral arms are positively correlated with the fast magnetosonic speed $\sqrt{c_s^2+v_A^2}$ where $c_s$ is the sound speed and $v_A$ is the Alfv{\'e}n speed (see \S of \citealt{2016Ju} for more discussions). While the sound speed of the hydro model is the same with the MHD models, the Alfv{\'e}n speed is zero in the hydro model, which results in smaller pitch angles of spiral arms. Since the weaker and more tightly wound spiral shocks in the hydro model drive less efficient angular momentum transport, the mass accretion rate in the hydro model is smaller than the MHD models such that gas piles up as a ring at $R \sim 0.3$ at late times of the hydro model ($t\sim 230$). 

\begin{figure*}
\centering
\includegraphics[width=0.49\textwidth]{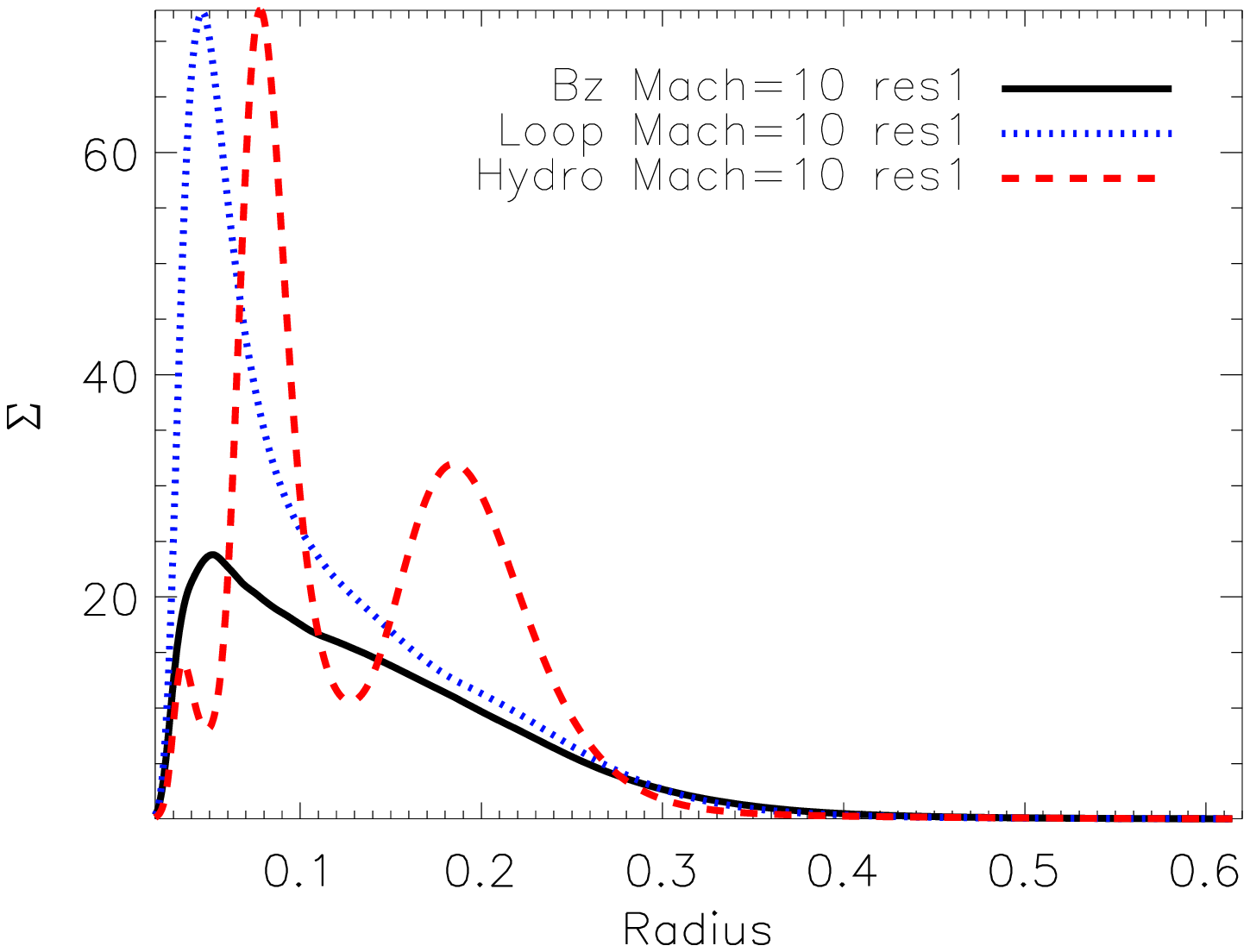}
\includegraphics[width=0.49\textwidth]{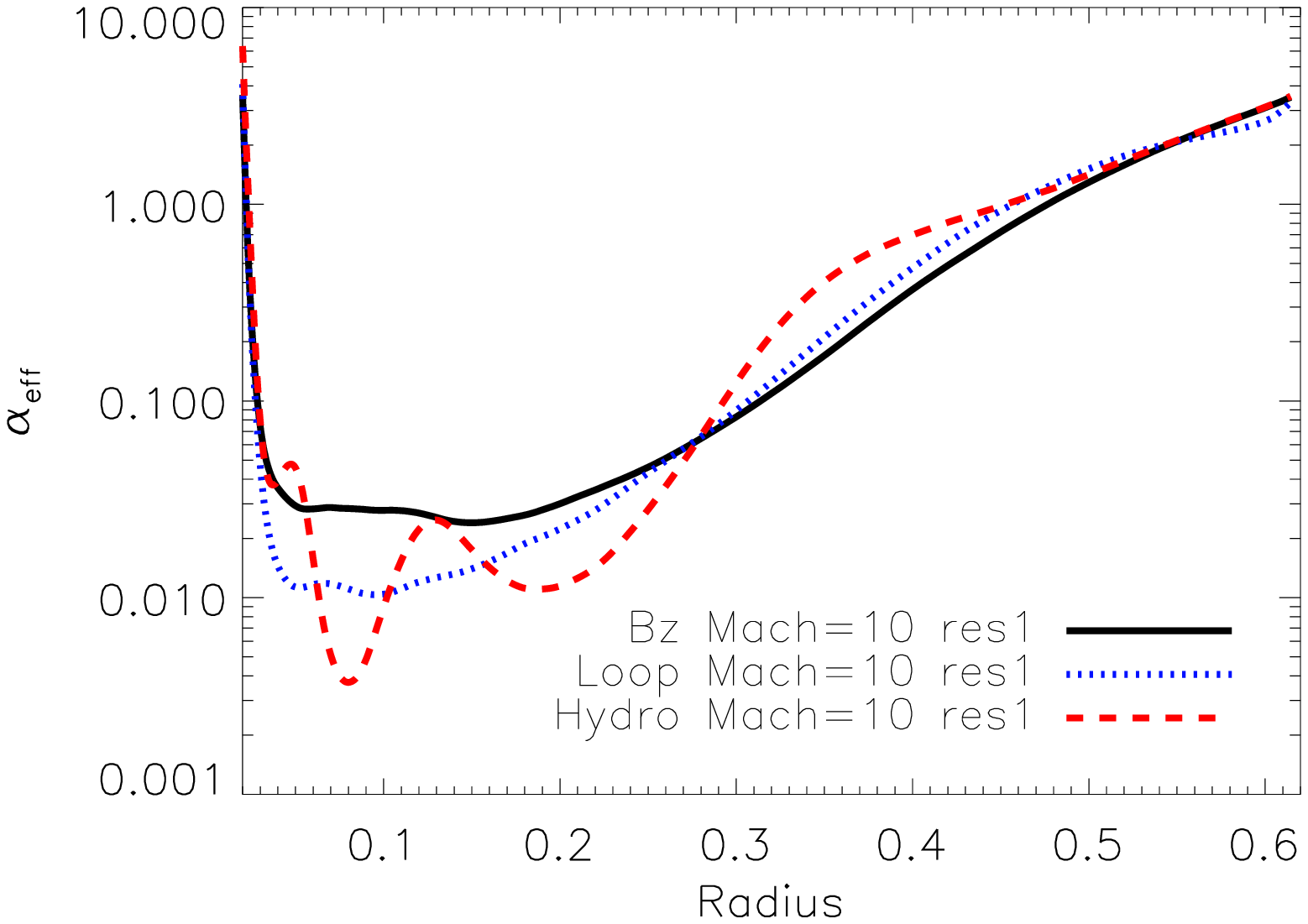}
\includegraphics[width=0.49\textwidth]{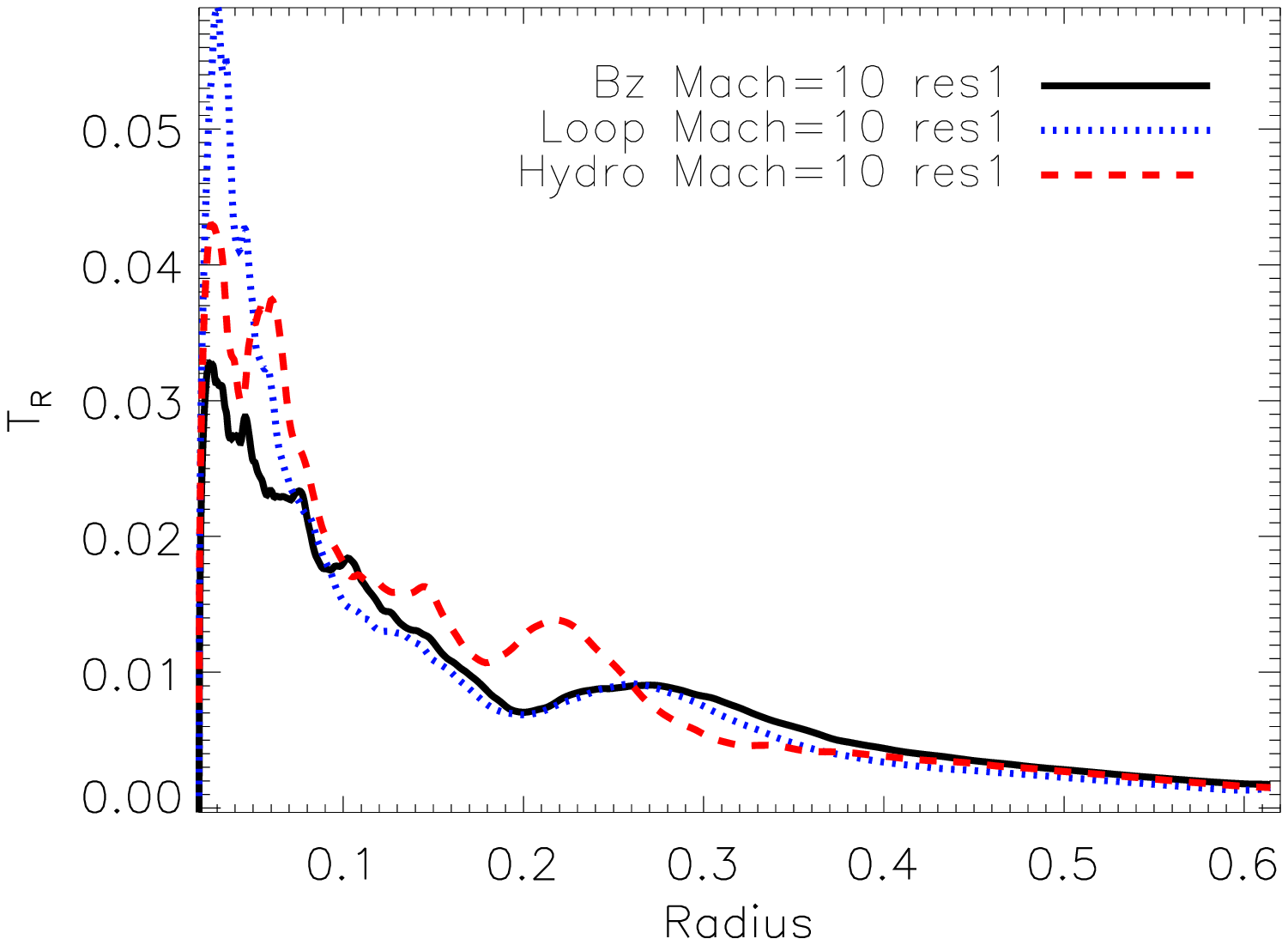}
\includegraphics[width=0.49\textwidth]{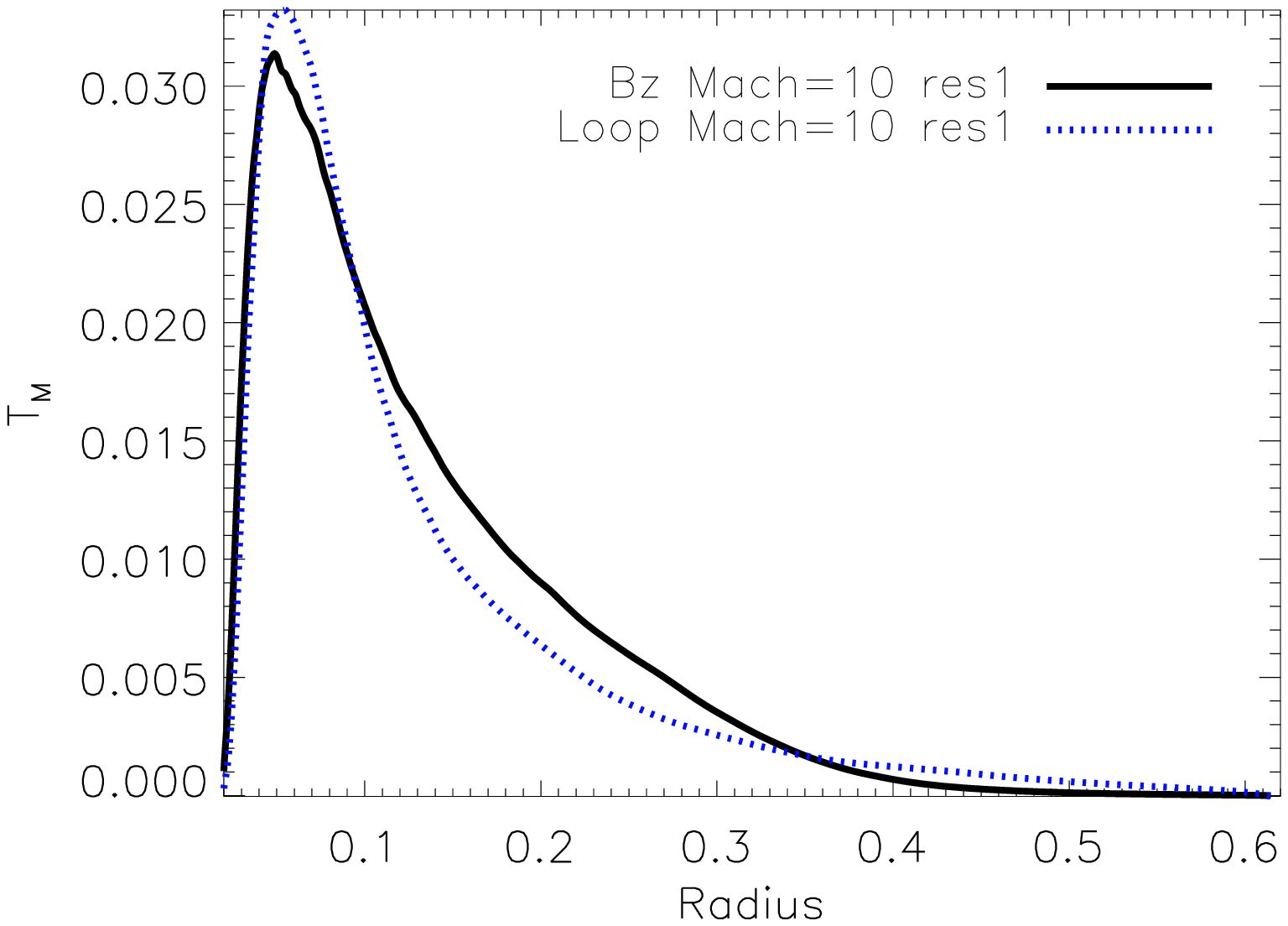}
\caption{Comparison for effects of seed magnetic field geometry. Characteristic properties of the disk in the fiducial model ``Bz-Mach10-$\beta400$-res1" (solid black lines), the model ``Loop-Mach10-$\beta400$-res1" (dotted blue lines), and the hydrodynamical model ``Hydro-Mach10-res1" (dashed red lines): the surface density $\Sigma$ (upper left panel), the effective viscosity parameter $\alpha_{eff}$ (upper right panel), the Reynolds stress $T_R$ (lower left panel) and the Maxwell stress $T_M$ (lower right panel). All quantities are volume averaged and time averaged over $t=200-238$.}
\label{fig:M10_compare_geom}
\end{figure*}

In Figure \ref{fig:M10_compare_geom} we compare more characteristic quantities of the three models with different (or no) magnetic field geometry: the surface density $\Sigma$, $\alpha_{eff}$, the Reynolds stress $T_R$, and the Maxwell stress $T_M$. All quantities are time- and volume- averaged over $t=200-238$. Although we plot the whole radial range for completeness, more attention should be paid to the region $R<0.25$ where the majority of the disk resides. Firstly, we compare the surface density in the first panel. The two MHD models ``Bz" and ``Loop" have smoother profiles than the ``Hydro" model. This is because in the ``Hydro" model when the mass inflow rate exceeds the capability of angular momentum removal driven by spiral shocks, the gas may pile up locally at certain radius which leads to density bumps. This is similar to the ``Inflow-isothermal-$c_s0.1$" hydro model in \citet{2016Ju} where the gas sound speed is so low thus the spiral waves are so weak that gas piles up as a ring at $R\sim 0.15$. However, in the MHD models, the outward transport of angular momentum driven by MRI can efficiently smear out density bumps in the disk. This further stresses that MRI is crucial for a steady-state CV disk because spiral shocks only cannot provide enough transport of angular momentum when the gas temperature is low. Comparing the two MHD models with different seed field geometry, their surface density follow the same power law profile (also see Figure \ref{fig:Bz_Mach10_res2_fit_with_alpha_theory} and Figure \ref{fig:Loop_Mach10_res2_fit_with_alpha_theory} for $\Sigma$ profiles in logarithmic scale), however, the values of surface density in the ``Loop" model is about three times of that in the ``Bz" model. This is due to their different accretion history. On the one hand, the growth rate of MRI in the ``Loop" model is much slower than in the ``Bz" model, thus the mass accretion rate in the ``Loop" model took a longer time to grow. On the other hand, since the mass accretion history of the ``Loop" model has eruptive features due to the cyclic evolution of the magnetic field strength in the disk (see \S \ref{subsec:loop_model} for detailed discussions), there are periods of time in the ``Loop" model when the magnetic field is very weak near the inner boundary thus the accretion rate is low. Therefore, although the ``Bz" model and the ``Loop" both eventually reach steady state, there is much less mass accreted in the ``Loop" model during the simulation time. Since the two models have the same mass inflow rate from the L1 point, the ``Loop" model ends up with a more massive disk.

Secondly, the profiles of the Reynolds stress $T_R$ of the three models are quite similar (lower left panel of Figure \ref{fig:M10_compare_geom}). Only within $R<0.08$ are there differences: $T_R$ of the ``Loop" model is larger than that of the ``Bz" model by a factor of 4, and $T_R$ of the ``Hydro" model is larger than that of the ``Bz" model by a factor of $\sim 1.3$. The difference between the ``Loop" model and the ``Bz" model is mostly due to the higher surface density near the inner edge of the disk in the ``Loop" model since $T_R=\Sigma v_R \delta v_\phi$ is proportional to $\Sigma$. Since the ratio of $\Sigma$ in the ``Hydro" model to that in the ``Bz" model is about $3$ but the ratio of $T_R$ is only $1.3$, this implies that the measure of perturbation $v_R \delta v_\phi$ is stronger in the ``Bz" model. This mostly comes from the extra turbulence generated by MRI. The fact that $T_R$ in the MHD models is not much greater than the hydro model implies that MRI does not enhance the strength of spiral shocks by much. Diffusive processes like MRI turbulence are expected to expand the CV disk which makes the tidal response of the disk to the companion star stronger \citep{2008Kley}. However, this effect is not obvious in our simulations since it is difficult to define the outer edge of the disk and measure its motion. 

Thirdly, the profiles of the Maxwell stress $T_M$ of the two MHD models are quite similar (lower right panel of Figure \ref{fig:M10_compare_geom}). This indicates the efficacy in driving angular momentum transport of MRI does not depend much on the seed field geometry. No matter what the geometry of seed field is, the CV disk can always reach a self-regulating steady state: when the MRI in the disk is weak and mass accretion rate is low, the gas as well as the magnetic flux piles up in the disk due to the constant supply of gas and seed magnetic field which leads to higher surface density and magnetic field strength, thus the Reynolds stress and Maxwell stress increases which makes the mass accretion rate increase until it matches the mass supply rate.

Lastly, the profiles of the effective viscosity parameter $\alpha_{eff}$ is shown in the top right panel of Figure \ref{fig:M10_compare_geom}. This parameter is defined in Eq. \ref{eq:alpha_eff} using the mass accretion rate measured from our simulations. The $\alpha_{eff}$ of the ``Bz" model ($\sim 0.03$) which is about twice that of the ``Loop" model ($\sim 0.015$). This is because the stresses that drive mass accretion in the ``Loop" model and the ``Bz" model are comparable while the ``Loop" model has higher surface density. As we discussed in \S \ref{subsec:loop_model} about Figure \ref{fig:Bz_Mach10_res2_fit_with_alpha_theory} and Figure \ref{fig:Loop_Mach10_res2_fit_with_alpha_theory}, the mass accretion rate of the two models are comparable, which may be converted to comparable luminosity in observations. Therefore, $\alpha_{eff}$ may not be the best parameter to be compared with dwarf novae observations. 
 
As a summary, by comparing the hydro model (``Hydro-Mach10-res1") and two MHD models with vertical seed magnetic field (``Bz-Mach10-$\beta400$-res1") and loop seed field with zero vertical component (``Loop-Mach10-$\beta400$-res1"), we reach the following conclusions.
\begin{itemize}
    \item MRI is crucial for a steady-state CV disk because spiral shocks only cannot provide enough transport of angular momentum when the gas temperature is low.
    \item A larger $\alpha_{eff}$ or greater accretion efficiency is favored when the seed field has vertical components.
    \item The relative importance of Reynolds stress and Maxwell stress in driving angular momentum transport does not depend on the seed field geometry.
\end{itemize}

\section{Effects of Mach Number}
\label{sec:compare_machnumber}

Thermodynamics has been realized to be crucial for spiral density waves in CV disks. For example, the pitch angles of spiral patterns are related to the Mach number of gas as $\arctan(\mathcal{M})$ \citep{1987Spruit, 2000Makita, 2016Hennebelle, 2016Ju}. In other words, the spiral arms in CV disk are more tightly wound when the gas temperature decreases (thus $\mathcal{M}$ increases), which often leads to rapid decrease in the strength of spiral shocks thus rapid decrease in the efficacy of angular momentum transport driven by spiral shocks. With the technique of Doppler tomography, spiral patterns have been reconstructed for several eclipsing CV systems \citep{1997Steeghs,1999Harlaftis,2001Groot,2004Neustroev,2005Hartley,2011Neustroev}. However, multiple follow-up numerical work found that to resemble the spiral patterns from observations, the disk needs to be much hotter than the temperature estimated from eclipse mapping \citep{1994Savonije, 1997Steeghs,1998Godon}. For example, \citet{1998Godon} did 2D hydrodynamical models with parameters appropriate for the eclipsing dwarf nova binary IP Peg using a hybrid Fourier-Chebyshev spectral method where they found that, in order to reproduce the shape of spiral patterns discovered in Doppler map of IP Peg, the temperature of CV disk in the models needs to be higher than the observed values by a factor of $\sim 30$ (although controversy exists in interpretation of the Doppler tomograms \citep{2001Smak}). \citet{1994Savonije} proposed that spiral waves cannot appear in cool CV disks at all (Mach number $\mathcal{M} > 25$) because the wavelength of tidal response is much shorter than the length-scale of the tidal force so that waves cannot successfully propagate inward without significant decay. 

Therefore, here we try to explore the MHD and the angular momentum transport process in a cooler disk. Since all the models that we discussed in previous sections adopt locally isothermal equation of state where the temperature profile follows that of a standard $\alpha$ disk and the Mach number at the inner disk edge is 10, in this section we present the model ``Bz-Mach20-$\beta100$-res2" which doubles the Mach numbers in previous models. As listed in Table \ref{tab:parameter}, the seed magnetic field is vertical and has plasma $\beta=100$ which is the same as the fiducial model. Since spiral arms are much more tightly wound in this cooler disk model, we double the resolution in the radial and vertical directions to better resolve them. Due to the challenging computational cost, we run the simulation to $t=29.7$ which takes $\sim 0.2$ million CPU~$\cdot$~hours.

\begin{figure*}
\centering
\includegraphics[width=0.46\textwidth]{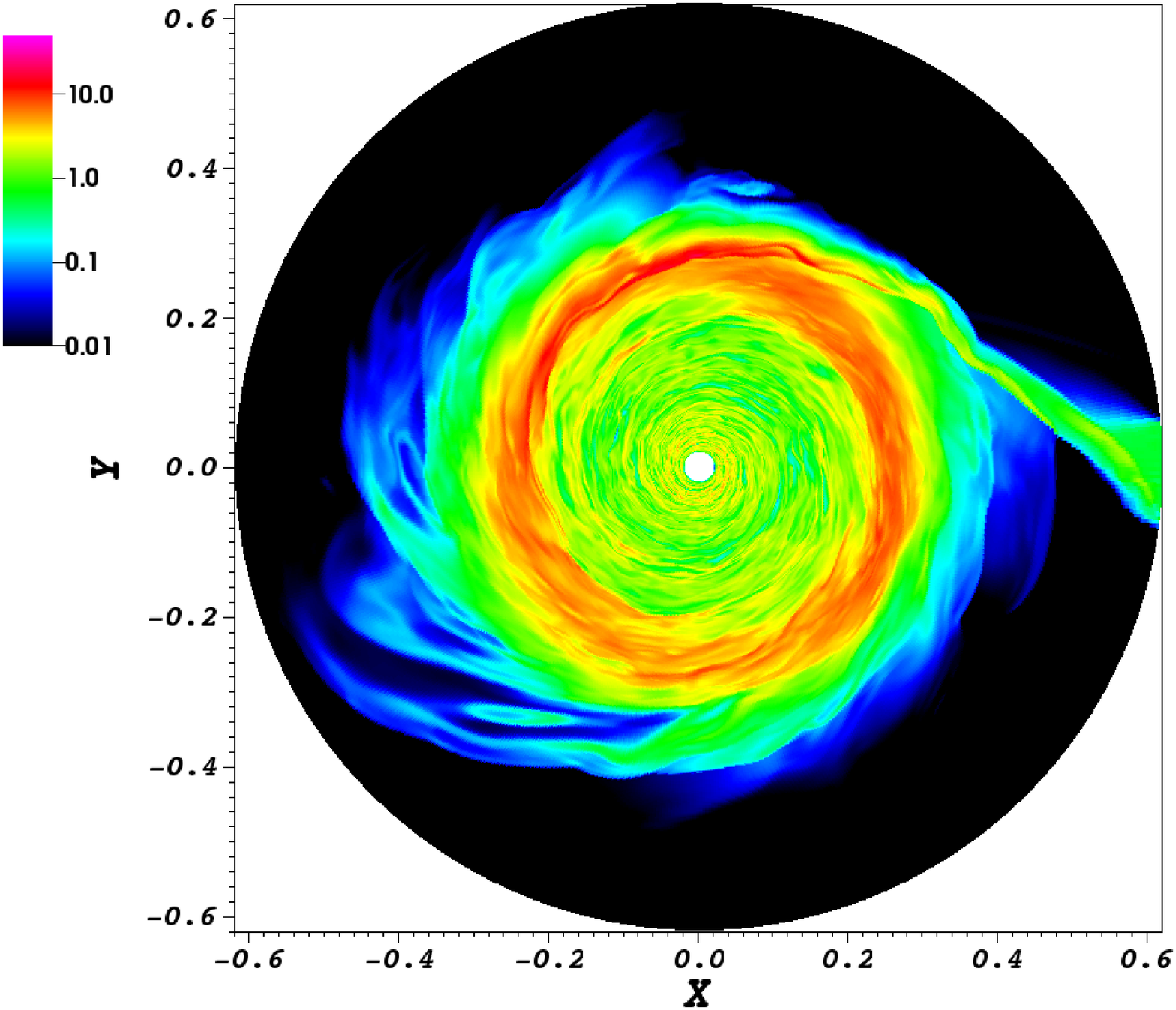}
\includegraphics[width=0.46\textwidth]{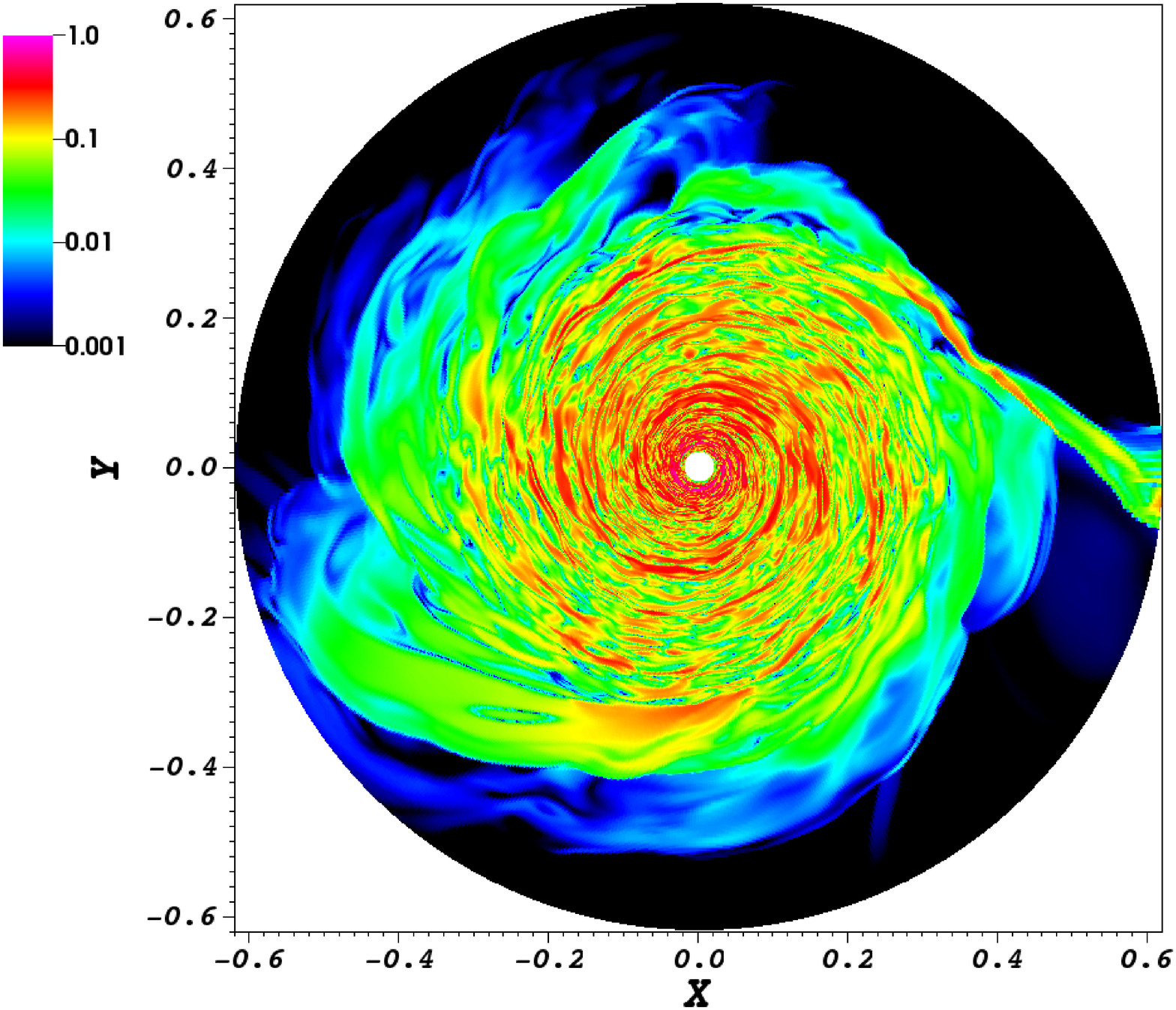}
\includegraphics[width=0.46\textwidth]{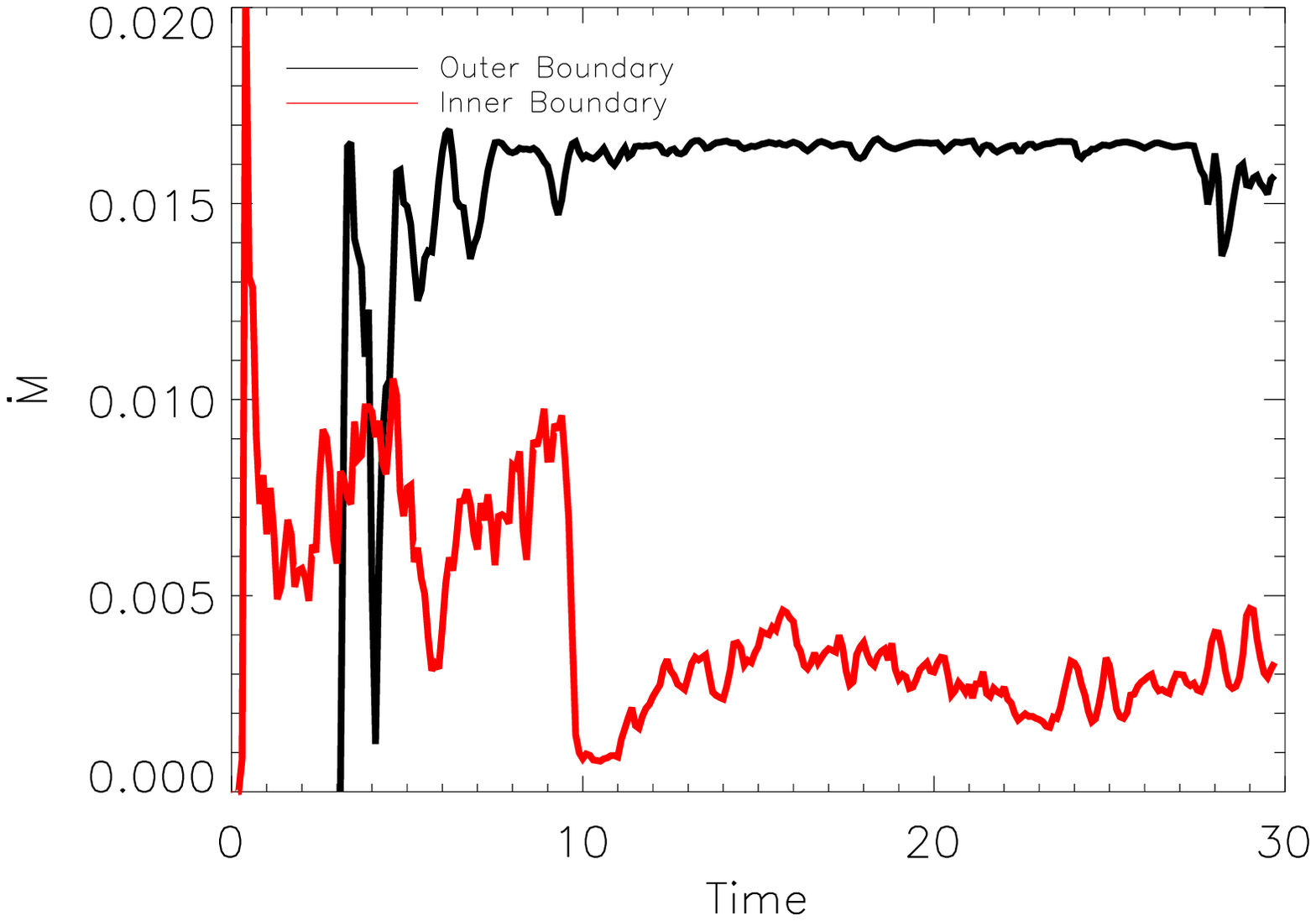}
\includegraphics[width=0.46\textwidth]{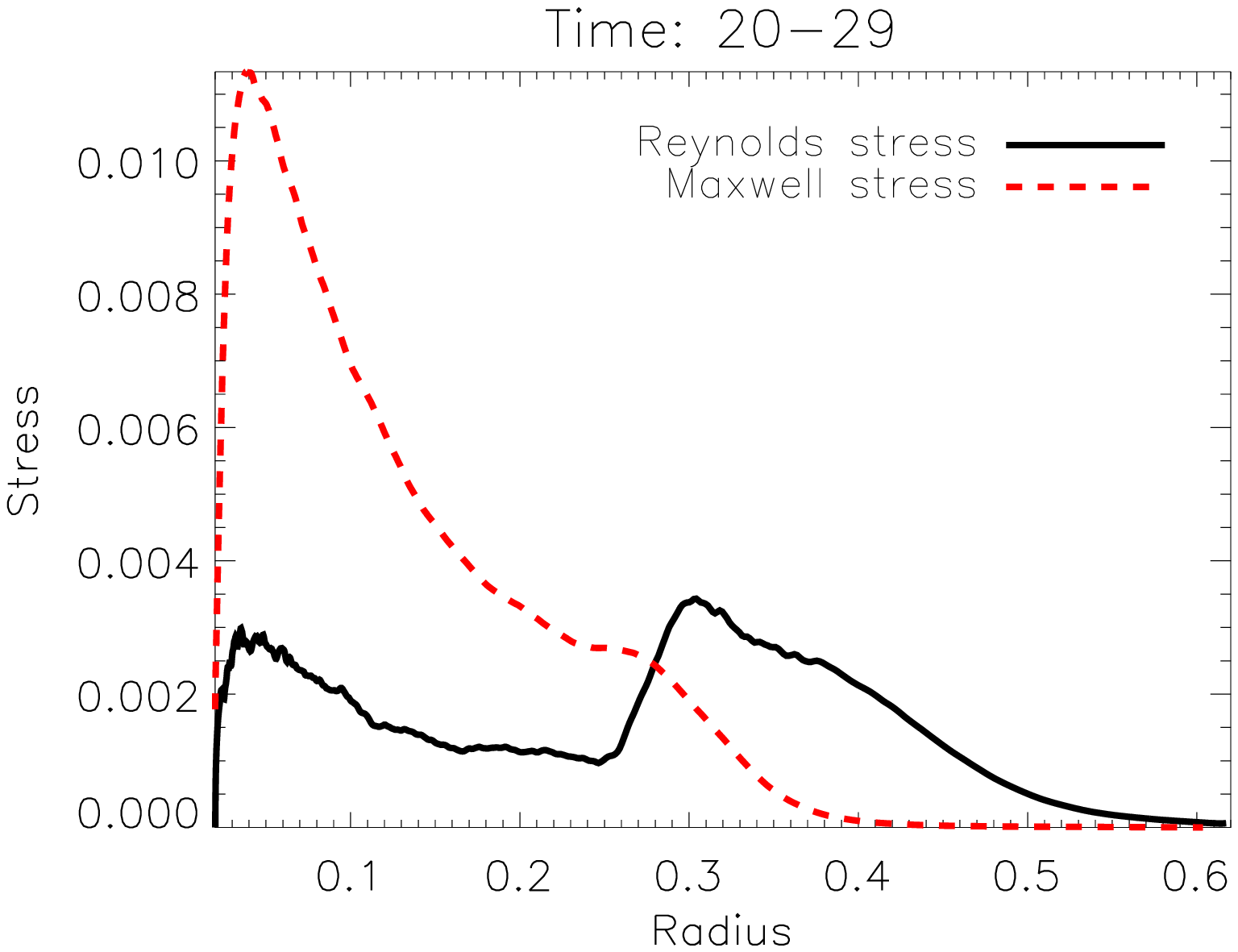}
\caption{Diagnostics of the ``Bz-Mach20-$\beta100$-res2" model. {\it Upper panels}: the snapshots of gas density (left) and the strength of magnetic field $|{\bf B}|$ (right) at $t=28$. {\it Lower left panel}: time history of the mass accretion rate $\dot{M}$ at the inner (red line) and the outer (black line) boundaries. {\it Lower right panel}: the Reynolds stress (solid black line) and the Maxwell stress (dashed red line) that are volume- and time- averaged over $t=20-29$. }
\label{fig:Bz_Mach20_res4}
\end{figure*}

In the upper panels of Figure \ref{fig:Bz_Mach20_res4} we show the snapshots of the gas density and the magnitude of magnetic field of the model ``Bz-Mach20-$\beta100$-res2" at $t=28$. Like the fiducial model, the magnetic field is turbulent indicating the presence of MRI. However, unlike the fiducial model, there are no obvious spiral structures in the snapshot of density. Gas piles up in a ring at $R \sim 0.25 - 0.3$, which is approximately the circularization radius given the initial angular momentum of the flow at the L1 point. This is because spiral density waves are more tightly wound in a cooler disk, thus are harder to propagate further inward. Actually, we also did one corresponding hydro model with the same initial and boundary conditions with the model ``Bz-Mach20-$\beta100$-res2" but zero magnetic field where we found that although gas also piles up at $R \sim 0.25 - 0.3$ as a ring, there are weak spiral density waves that propagate into the inner disk region. However, in the MHD model ``Bz-Mach20-$\beta100$-res2", these weak spiral waves may have been erased by the turbulence generated from MRI. 

In the lower left panel of Figure \ref{fig:Bz_Mach20_res4} we show the time history of the mass accretion rate $\dot{M}$ at the inner and outer boundaries. The mass accretion rate at the inner boundary is only $1/5$ of the mass supply rate at the outer boundary, which is quite steady over time and has no sign of growing. Therefore, the disk does not reach steady state and gas keeps piling up in the disk especially in the ring at $R \sim 0.25 - 0.3$. 

Given the disk is not in steady state, we are not able to do conclusive comparison of the importance of the spiral shocks and the MRI in driving angular momentum transport. But using the available data, we still plot the volume- and time- averaged radial profiles of the Reynolds stress and the Maxwell stress (averaged over $t=20-29$) in the lower right panel of Figure \ref{fig:Bz_Mach20_res4}. The Reynolds stress at the inner disk region $R<0.25$ is only $1/10$ of the values in the fiducial model (although note the values in this model are at earlier time than those in the fiducial, thus the lower density at earlier times is also a contributing factor of the lower Reynolds stress), so angular momentum transport driven by the spiral shocks is very weak. The Maxwell stress dominates the Reynolds stress by a factor of $\sim 4$, which makes the disk more like an MRI-driven turbulent disk around a single star like the previous MRI models \citep{2012Sorathia,2013Hawley}. The Reynolds stress has a bump at $R \sim 0.3$ but is not large enough to diffuse the ring efficiently. The Maxwell stress at $R \sim 0.25 - 0.3$ is as small as the Reynolds stress, thus does not induce efficient accretion of the ring either. 


\section{Convergence Study}

In order to make sure our results do not depend on numerical resolution, we perform a convergence study for the ``Bz-Mach10-$\beta400$" model with vertical seed magnetic field and inner edge Mach number $\mathcal{M}=10$. The model ``Bz-Mach10-$\beta400$-res1" has a resolution of $384(R) \times 704(\phi) \times 32 z$. For the resolution test, we double the resolution in only $R$ and $z$ directions to reduce computational cost because resolution in these two directions is the constraining factor for resolving spiral shocks and MRI. The resolution study in \citet{2013Hawley} also adopted this method. This higher resolution model is denoted with ``Bz-Mach10-$\beta400$-res2" in Table \ref{tab:parameter} which has $768(R) \times 704(\phi) \times 64(z)$. Both the lower resolution model and the higher resolution model have logarithmically spaced grids in the radial direction and evenly spaced grids in the azimuthal and vertical directions. Due to the computational cost, the higher resolution model is run to $t=22.7$ which is only one tenth of the fiducial model. But it is already long enough to study the major characteristics of spiral shocks and MRI since it covers $\sim 55$ Keplerian orbits at $R=0.15$ (middle area of the disk) and almost covers one viscous timescale (see the viscous timescales averaged over time 26 to 32 in Figure \ref{fig:Bz_Mach10_res2_time_evolution}).

\begin{figure*}
\centering
\includegraphics[width=0.49\textwidth]{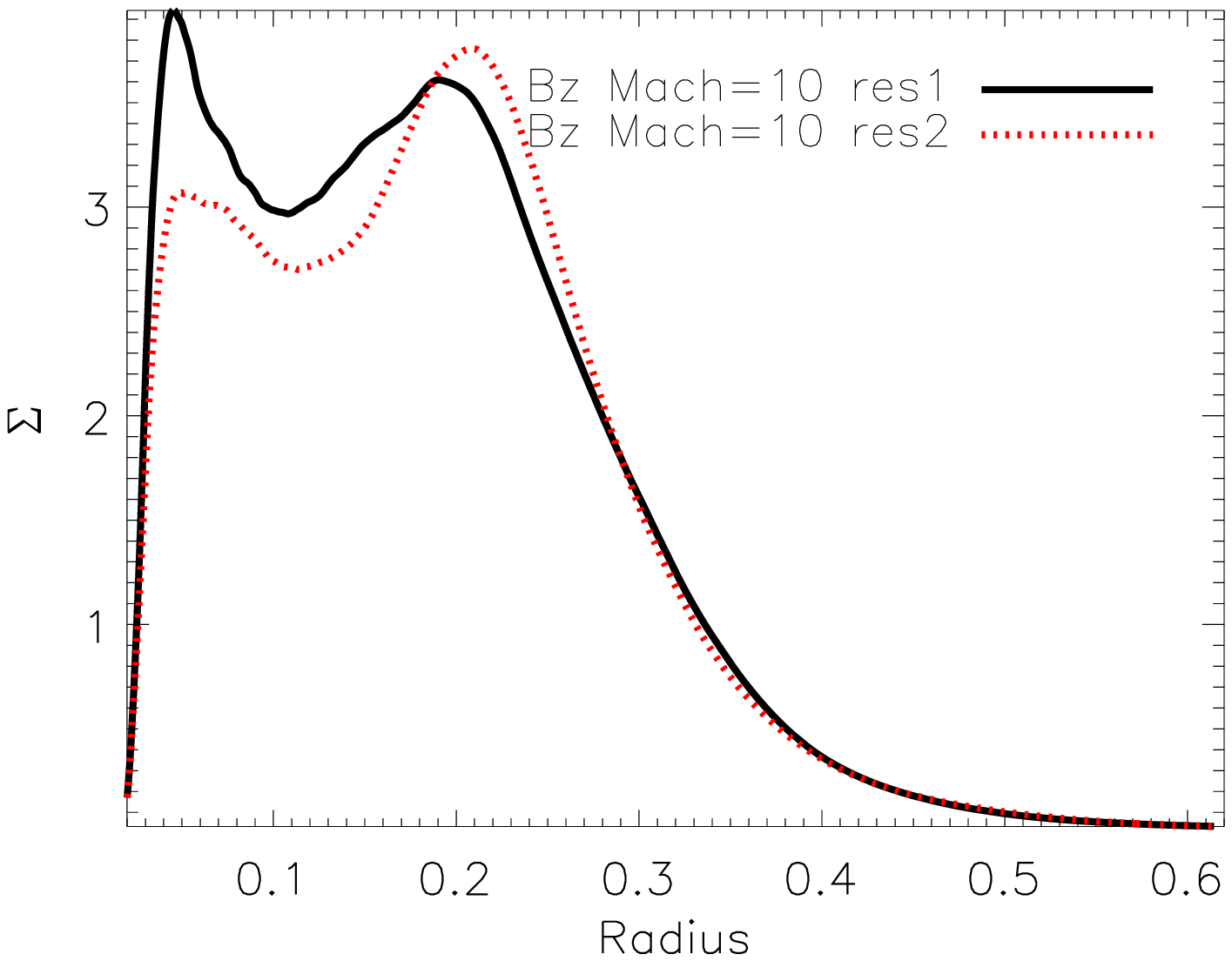}
\includegraphics[width=0.49\textwidth]{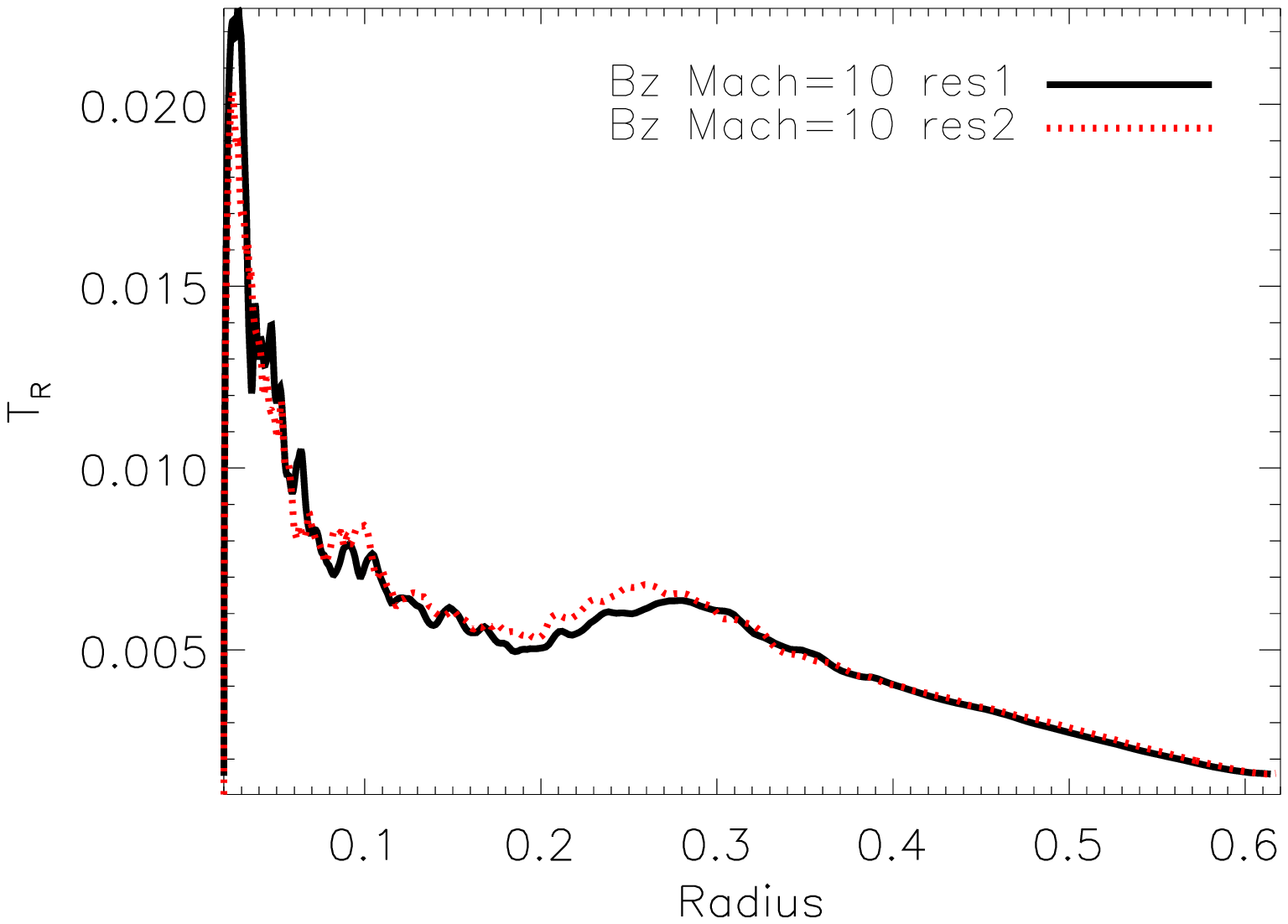}
\includegraphics[width=0.49\textwidth]{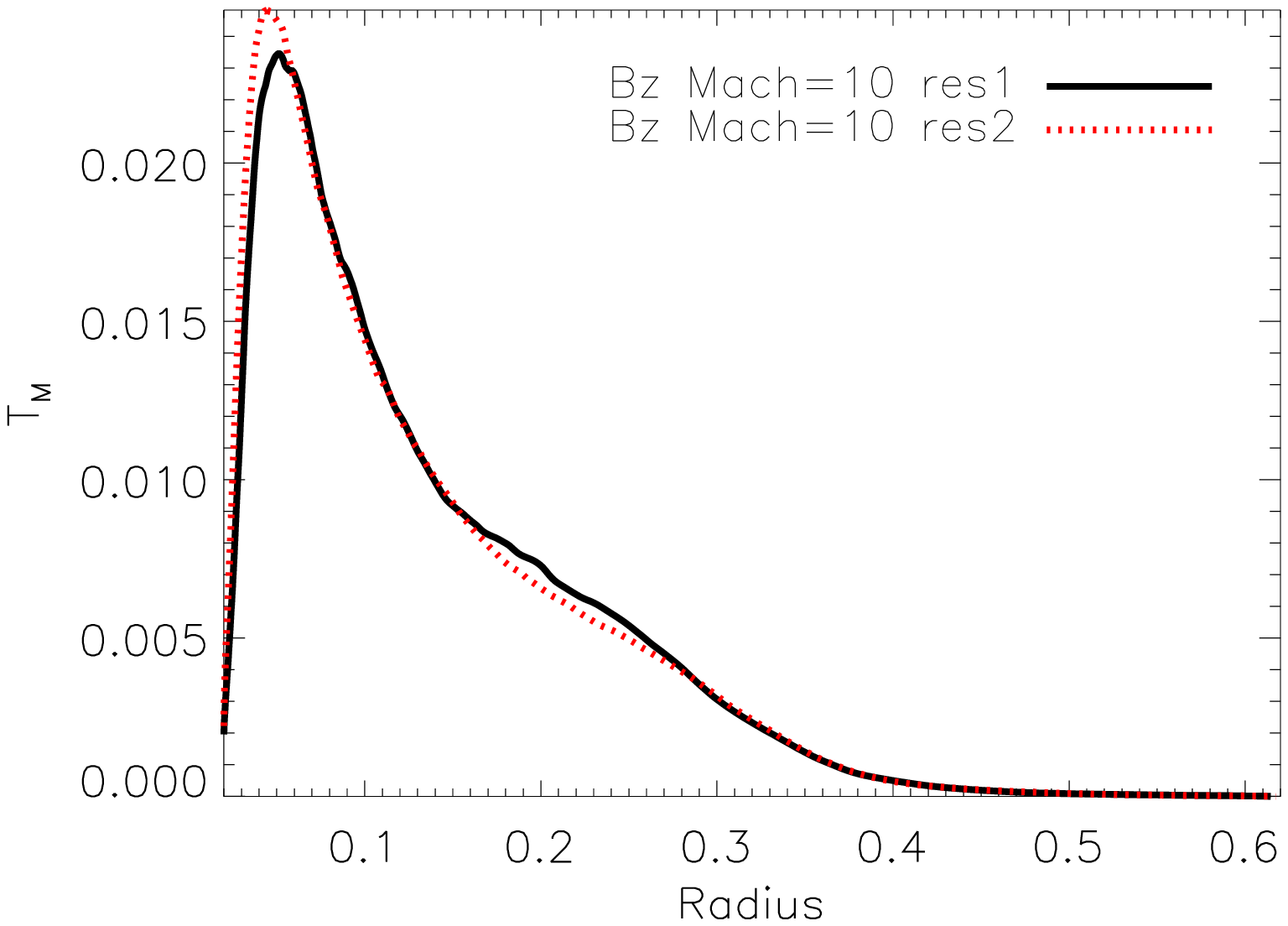}
\includegraphics[width=0.49\textwidth]{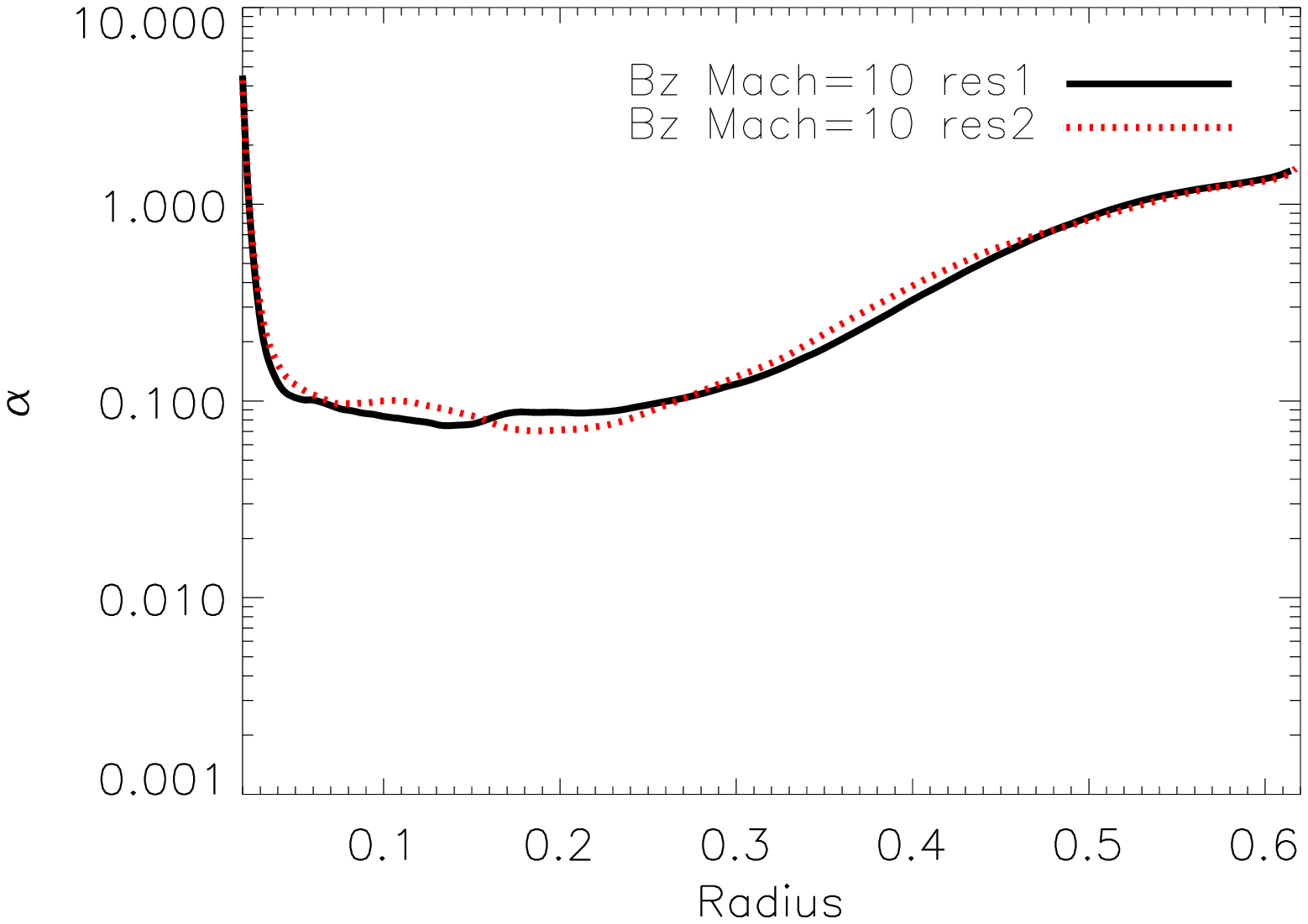}
\caption{Convergence test: comparison of the radial profiles of the surface density $\Sigma$ (upper left panel), the Reynolds stress $T_R$ (upper right panel), the Maxwell stress $T_M$ (lower left panel) and $\alpha_{eff}$ (lower right panel) for the model ``Bz-Mach10-$\beta400$-res1" (solid black lines) and the double-resolution model ``Bz-Mach10-$\beta400$-res2". All quantities are volume- and time- averaged during $t= 17-22$.}
\label{fig:Bz_M10_convergence}
\end{figure*}

In Figure \ref{fig:Bz_M10_convergence} we show the results of comparing the radial profiles of the surface density $\Sigma$, the Reynolds stress $T_R$, the Maxwell stress $T_M$ and the effective viscosity parameter $\alpha_{eff}$, all of which are volume- and time- average over $t=17-22$. For the profile of $\Sigma$, the two resolution models are in general consistent except a little difference near the inner boundary. One one hand, this is because the system is highly dynamic so there will always be transient difference in detailed structures; on the other hand, regarding the spiral shocks, as we discussed in the convergence study for our hydrodynamical models of CV disk in \S 4.5 of \citet{2016Ju}, a higher resolution could resolve more fine scale structures produced by hydrodynamical instabilities which could also contribute to differences in the distribution of surface density. For the profiles of $T_R$, $T_M$ and $\alpha_{eff}$, the two resolution models have perfect agreement. Therefore, we conclude the ``Bz-Mach10-$\beta400$-res1" model has reached convergence.

\section{Discussion and Conclusion}
\label{sec:conclusion_mri}

In this work, we conducted a series of 3D global MHD simulations of unstratified CV disks with locally isothermal equation of state in order to study the relative importance of spiral shocks and MRI in driving angular momentum transport. For the first time, we have the global MHD steady-state models of CV disks with relatively realistic temperature profiles. This is the first study of the evolution of CV disks driven by self-consistent mechanisms of angular momentum transport and supply of mass and seed magnetic field. Comparing the steady-state solutions with different seed field geometry, seed field strength, and disk Mach number, we reach the following conclusions:

\begin{itemize}
    \item When the disk Mach number is $7-10$ (the ``Mach10" models), the steady-state disk can be fit using the standard thin disk theory. The effective viscosity parameter $\alpha_{eff}$ is $\sim 0.1$ when the seed magnetic field is vertical and has $\beta=100$, is $\sim 0.032$ when the seed magnetic field is vertical and has $\beta=400$, and is $\sim 0.016$ when the seed magnetic field is field loops with zero vertical component and has $\beta=400$. This comparison indicates that a larger value of $\alpha_{eff}$ is favored when the seed magnetic field has vertical components or the flow has stronger magnetization ($1/\beta$). These results are consistent with previous local shearing-box simulations and global simulations of MRI.
    
    
    \item When the disk Mach number is $12-20$ (the ``Mach20" models), our model has not reached steady state within the simulation time which is $1/10$ in length of the ``Mach10" models due to the heavy computational cost with the higher resolution. However, apparent trends about the angular momentum transport in the disk have been observed. The spiral waves are much more tightly wound due to the lower temperature. The spiral shocks are much weaker thus provide very limited transport of angular momentum. The gas piles up near the circularization radius, which causes increase of the magnetic flux thus Maxwell stress at the over-density ring. The growing MRI would eventually diffuse the ring. Due to the weak spiral shocks, the resulting disk would behave more like a turbulent disk dominated by MRI around a single star such as previous MRI models.
    
    \item Regarding the relative importance of spiral shocks and MRI in driving angular momentum transport in CV disks, the Reynolds stress and the Maxwell stress are comparable in the ``Mach10" disks when the seed magnetic field has $\beta=400$ with or without vertical components. When the seed magnetic field has $\beta=100$, the magnetization (measured by $1/\beta$) of the disk is increased by a factor of 2, so the Maxwell stress dominates the Reynolds stress by a factor of $1.5-2$. In the ``Mach20" disks, however, the spiral shocks are much weaker due to the low temperature, so the Maxwell stress dominates the Reynolds stress by a factor of $\sim 4$ which is more like an MRI-driven turbulent disk around a single star. Therefore, the roles of spiral shocks and MRI in driving angular momentum transport have their own controls: the importance of spiral shocks is controlled by the disk temperature, and the importance of MRI is controlled by the seed magnetic field strength. 
    
\end{itemize}

Although we mostly focus on the steady-state characteristics of the disk, the evolutionary paths to the steady state as well as their dependence on the seed field geometry and strength and the Mach number are also enlightening.
\begin{itemize}
    \item The evolutionary path to the steady state follows a self-regulating manner. When angular momentum transport in the disk is inefficient, mass and magnetic flux pile up in the disk, which strengthens the spiral shocks and MRI respectively. The mass accretion rate thus keeps growing until the steady state is reached. 
    
    \item Along the path, $\alpha_{eff}$ reaches peak values of $0.1 -0.3$ in all models when MRI peaks, but decay to values of order $0.01-0.1$ after MRI saturates. However, the mass accretion rate keeps increasing during the process, which is expected to cause higher luminosity of the system.

    \item MRI plays an essential role in reaching steady state of CV disks. Without MRI, the disk may keep piling up mass due to the insufficient angular momentum transport driven by spiral shocks especially in a cool disk.
    
    \item The growth rate of the mass accretion rate depends on the disk Mach number as well as the seed magnetic field geometry and strength. Therefore, the mass accretion rate increases faster, thus the disk reaches steady state earlier, when the seed magnetic field is stronger or has vertical components, or when the disk Mach number is lower (see the MHD model in \citealt{2016Ju} for comparison).
\end{itemize}

By using a locally isothermal equation of state, we have approximated the thermodynamics of a steady disk where the cooling balances the heating. In the future, we plan to conduct global MHD simulations of stratified CV disks including radiative cooling to model convective outbursts. 


Our study may also trigger new thoughts about the boundary layer studies \citep{2013Belyaev-a,2013Belyaev-b}. In our global MHD models of CV disks (the ``Mach10" model with isothermal equation of states and the MHD model in \citealt{2016Ju} with adiabatic equation of states), the spiral waves propagate all the way to the inner boundary even with the existence of strong MRI. Therefore, the global $m=2$ mode excited by tidal forces may interfere with the acoustic waves excited by the supersonic shearing in the boundary layer. The ultimate solution to the angular momentum transport process at the boundary layer of CV disks can only be found in global MHD models with inner boundary conditions and resolution appropriate for the boundary layer region.

\begin{figure}
\centering
\includegraphics[width=0.48\textwidth]{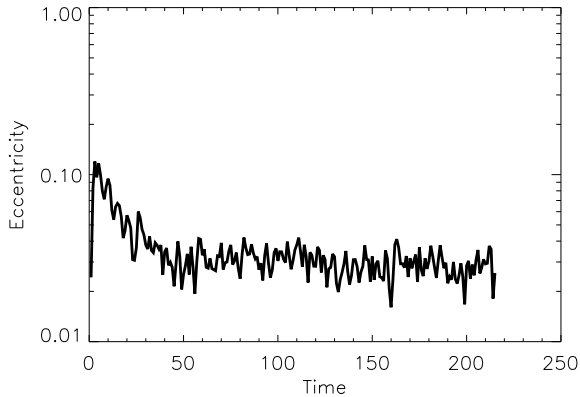}
\caption{ Time evolution of mass-averaged eccentricity in the ``Bz-Mach10-$\beta$100-res2" model. No growth of eccentricity is observed. }
\label{fig:ecc_Bz_Mach10_beta100}
\end{figure}

Lastly, as we discussed in the first paper of this series \citep{2016Ju}, we examine the eccentricity growth driven by elliptical instability which is potentially related to the observed superhumps in CVs \citep{1991Lubow, 2008Kley}. \citet{2008Kley} observed development of eccentric disks in close binary systems with large viscosity. Since MRI turbulence is often suggested to act as a ``viscosity", here we pick our model with the strongest MRI, the ``Bz-Mach10-$\beta$100-res2" model, and measure the mass averaged eccentricity. In Figure \ref{fig:ecc_Bz_Mach10_beta100} we show the time evolution of the mass-averaged eccentricity, where we find that although the eccentricity rises to $\sim 0.1$ at the beginning of the simulation due to excitation of spiral arms, it decays and becomes stable at $e \sim 0.03$ after the disk reaches steady state. No sign of eccentricity growth is observed. Adopting $\alpha_{eff} \sim 0.1$ (see Fig. \ref{fig:M10_compare_beta}) and $\mathcal{M} \sim 10$ in the ``Bz-Mach10-$\beta$100-res2" model, the corresponding dimensionless viscosity is $\nu = (\alpha / \mathcal{M}^2) R v_K \sim 3 - 6 \times 10^{-4}$.  According to the model with the largest viscosity $\nu = 10^{-4}$ and mass ratio $q=0.3$ in \citet{2008Kley} (see their Figure 18), the disk eccentricity is expected to start growing from $e \sim 0.08$ at $\sim 10$ binary orbits ($t \sim 60$ in our time units) to $e \sim 0.5$ at $\sim 30$ binary orbits ($t \sim 188$ in our time units). This discrepancy may originate from the lack of disk expansion due to MRI. Since the specific angular momentum is proportional to $\sqrt{R}$, little material that is pushed outward can carry away the angular momentum of a larger amount of material that is accreted inward. In our simulations, due to the efficient outward transport of angular momentum, most of the disk material moves inward. Although there is more matter being pushed outward as the whole disk becomes more massive and the mass accretion rate increases, there is a much larger amount of material moving inward, which stabilizes the disk.

\acknowledgements We thank J. Goodman, E. Ostriker and G. Bakos for insightful discussions and suggestions. This project is supported by NSF grant AST-1312203. Numerical simulations in this work are carried out using computational resources supported by the Princeton Institute of Computational Science and Engineering, and the Texas Advanced Computing Center (TACC) at The University of Texas at Austin through XSEDE grant TG-AST130002. Z.Z. acknowledges support by NASA through Hubble Fellowship grant HST HF- 51333.01-A awarded by the Space Telescope Science Institute, which is operated by the Association of Universities for Research in Astronomy, Inc., for NASA, under contract NAS 5-26555.


\end{document}